\documentclass[aps,prb,reprint,amsmath,amssymb,superscriptaddress,nofootinbib]{revtex4-2}

\usepackage{graphicx}
\usepackage{booktabs}
\usepackage{bm}
\usepackage{amsthm}
\usepackage{hyperref}
\usepackage{xcolor}
\usepackage{physics}
\usepackage{mathtools}
\usepackage{microtype}

\hypersetup{colorlinks = true,
	linkcolor = blue,
	urlcolor = blue,
	citecolor = blue,
	pdftitle={Coherence controls the visibility of frozen disorder in an open quantum chain},
	pdfsubject={Coherence, frozen routing disorder, and complete terminal populations},
	pdfauthor={S. Y. Li, M. H. Yu, and X. Z. Zhang}}

\newcommand{\E}{\mathbb E}
\newcommand{\Pp}{\mathbb P}
\newcommand{\R}{\mathbb R}
\newcommand{\Z}{\mathbb Z}
\newcommand{\Var}{\operatorname{Var}}
\newcommand{\Cov}{\operatorname{Cov}}
\newcommand{\ee}{\mathrm e}
\newcommand{\ind}{\mathbf 1}
\newcommand{\HH}{\mathcal H}

\begin{document}
\title{Coherence controls the visibility of frozen disorder in an open quantum chain}
\author{S. Y. Li}
\affiliation{College of Physics and Materials Science, Tianjin Normal University, Tianjin 300387, China}
\author{M. H. Yu}
\affiliation{College of Physics and Materials Science, Tianjin Normal University, Tianjin 300387, China}
\author{X. Z. Zhang}
\email{zhangxz@tjnu.edu.cn}
\affiliation{College of Physics and Materials Science, Tianjin Normal University, Tianjin 300387, China}
\affiliation{Interdisciplinary Center, Tianjin Normal University, Tianjin 300387, China}
\date{22 October 2026}
\begin{abstract}
A Gaussian position distribution need not imply that frozen disorder is irrelevant. We study a resonant quantum chain where neighboring sites couple with opposite signs to auxiliary states that decay with frozen, random return probabilities. Lowering dephasing spreads the effect of each local return bias over more sites. We take dephasing to zero at a fixed, sufficiently weak disorder amplitude. On the scale of a coherent wave packet, the position distribution approaches the same Gaussian for almost every frozen sample. At larger distances and later times, it remains sample dependent and is described by diffusion in a random potential with Brownian spatial statistics. The covariance of bounded, direction-sensitive position readouts from independent trials in the same field distinguishes these regimes. Simulations resolve their separation at finite dephasing. Taking weak disorder before vanishing dephasing also reveals a different packet-scale response from a memoryless random-force model matched in clean diffusion and leading potential strength. The coherent response develops an integrable logarithmic peak near the initial site, where the reference response stays finite. It reflects the correlated duration and displacement of motion between jumps, together with the unfinished packet at measurement.
\end{abstract}
\maketitle
\fussy
\raggedbottom

\section{Introduction}\label{sec:intro}

Dephasing sets the length scale over which a particle explores the lattice coherently~\cite{HakenStrobl}. Static energy disorder can localize coherent motion~\cite{Anderson}. Dephasing can assist transport in disordered quantum networks or turn coherent spreading into diffusion~\cite{Plenio,Rebentrost,Moix,ZnidaricHorvat}. Other combinations of noise and disorder produce subdiffusion, rare-region effects, or localized modes with slow decay~\cite{Gopalakrishnan,Taylor,Longhi2024}. A nearly Gaussian population profile leaves an important distinction unresolved. Disorder may have become irrelevant to the measured motion, or its effect may remain visible only at larger lengths and later times.

Random left and right transition rates make this question concrete. In one dimension, local directional biases accumulate into a random potential~\cite{Solomon,BouchaudGeorges}. When the potential increments have zero mean, random barriers produce the slow spreading known as Sinai transport~\cite{Sinai}. In a weak-disorder scaling limit, the potential becomes a Brownian function of position and the particle diffuses in this frozen potential~\cite{Brox,Seignourel}. Dissipative quantum chains with reciprocal coherent hopping and asymmetric random jumps already realize this mechanism~\cite{Shapira,Avnit,Longhi2026}. In Longhi's model, biased transitions accumulate population near minima of the random potential. Slow crossings between these minima produce Sinai-type spreading in the dissipative regime \cite{Longhi2026}. We ask how changing the coherence time changes the measured effect of one fixed set of return biases. For each frozen sample, we average a position readout at a specified time and spatial resolution. We then compare these conditional means across samples.

A quantum jump fixes the starting site of the next coherent segment, but it does not determine the position measured later. Between jumps, the particle spreads as a coherent wave packet. The segment duration is correlated with its displacement. If the measurement occurs before the next jump, it also records the spread of this unfinished packet. Replacing a coherent segment by a nearest-neighbor hop can therefore change the final position distribution. Replacing its random duration by the mean can do so as well. The effect is most direct when dephasing is weak and the unfinished packet is as wide as the observed region. The reservoir sets the probabilities of returning left or right, as in the related local return mechanism of Ref.~\cite{NC}. Here we retain both ground and auxiliary populations in the final position measurement.

In our model, neighboring ground states couple to the same auxiliary state with opposite signs. Their contributions nearly cancel when the ground-state amplitudes vary slowly in space, suppressing auxiliary excitation at small momentum. Ground-state dephasing at rate $\gamma$ limits the lifetime of these coherent modes. We lower $\gamma$ while keeping the coherent coupling, auxiliary decay, and local return biases fixed. Long coherent excursions become less strongly suppressed, spreading the effect of a local bias over more sites. On the packet scale, the accumulated fluctuations vanish and the complete position distribution approaches the same Gaussian for almost every frozen sample. On the long scale defined below, those fluctuations remain finite, and different fields produce different distributions described by diffusion in a Brownian potential.

The difference between these regimes can be measured without observing the reservoir. We use the covariance of two position readouts from trials that are independent once the frozen field is fixed. This covariance is the variance, across fields, of the mean readout in each field. At the three simulated rates, the packet-scale covariance decreases while the long-scale covariance remains statistically resolved. To identify what coherent motion adds, we compare the weak-disorder response with a memoryless model that matches the disorder-free diffusion constant and the leading random-potential strength. The two models share the clean Gaussian limit but respond differently to a local change in the return probabilities. In the coherent chain, a perturbation closer to the initial site produces a logarithmically larger response, although its integrated effect remains finite. Matching the clean spreading law therefore does not fix disorder sensitivity.

Sections~\ref{sec:model} and~\ref{sec:scales} define the model, terminal measurement, and two limits; Sec.~\ref{sec:mechanism} explains their mechanism. Sections~\ref{sec:data} and~\ref{sec:response} present the finite-rate results and response comparison, and Sec.~\ref{sec:discussion} discusses their scope. Appendices~\ref{app:renewal}--\ref{app:statistics} contain the renewal derivation, proofs of the two limits, response estimates, and numerical error bounds. The remainder of this paper is organized as follows.

\section{Model and position measurement}\label{sec:model}
We set the lattice spacing and $\hbar$ equal to one. The ground state $\ket{j}$ is located at the integer position $j\in\Z$. The auxiliary state $\ket{e_j}$ lies halfway between $\ket{j}$ and $\ket{j+1}$, at position $j+1/2$. We consider the infinite chain and prepare the particle in the ground state $\ket0$. The coherent coupling and rates have units of inverse time. Numerical results below use $\kappa$ as the rate unit and $\kappa^{-1}$ as the time unit. At resonance, the auxiliary states are degenerate with the ground states, and the Hamiltonian is
\begin{equation}
	H=g\sum_{j\in\Z}\left[|e_j\rangle(\langle j|-\langle j+1|)+\mathrm{h.c.}\right],\qquad g>0.
	\label{eq:model}
\end{equation} 
The alternating ground and decaying sites resemble lossy bipartite quantum walks~\cite{RudnerLevitov}. Here each auxiliary decay returns the particle to the ground chain. The two auxiliary-return channels and the ground-state dephasing are described by the jump operators
\begin{align}
	R_{j,L}&=\sqrt{\frac{\kappa(1-b_j)}2}\,|j\rangle\langle e_j|,\nonumber\\
	R_{j,R}&=\sqrt{\frac{\kappa(1+b_j)}2}\,|j+1\rangle\langle e_j|,\qquad
	L_j=\sqrt\gamma P_j.\label{eq:jumps}
\end{align}
Here $P_j=|j\rangle\langle j|$ projects onto ground site $j$. The total auxiliary return rate is $\kappa>0$, and the ground-state dephasing rate is $\gamma>0$. The density operator $\rho$ obeys the Lindblad equation~\cite{GKS,Lindblad,BreuerPetruccione},
\begin{equation}
	\partial_t\rho=\mathcal L_b\rho=-i[H,\rho]+\sum_\mu\left(J_\mu\rho J_\mu^\dagger-\tfrac12\{J_\mu^\dagger J_\mu,\rho\}\right),
	\label{eq:master}
\end{equation}
where $J_\mu$ runs over all jump operators in Eq.~\eqref{eq:jumps}. The two returns are separate decay channels that end in different states; they do not add coherently.

The variable $b_j$ sets the left--right return bias on bond $j$. We write $b_j=\varepsilon\xi_j$ and refer to the collection as the frozen field. The variables $\xi_j$ are independent and identically distributed (i.i.d.), with a distribution symmetric about zero. They satisfy $|\xi_j|\le M$ and $\E_\xi\xi_j^2=\sigma^2>0$. Binary disorder means $\xi_j=\pm1$ with equal probabilities, so $b_j=\pm\varepsilon$. We require $\varepsilon M<1$ so that the rates in Eq.~\eqref{eq:jumps} remain positive. Each realization of the field is drawn once and reused at every revisit. We take $\gamma\downarrow0$ with $g$, $\kappa$, and the disorder amplitude fixed. The general two-scale theorem requires $0<\varepsilon M<d_*$. At fixed $g$ and $\kappa$, the same positive threshold applies to every sufficiently small $\gamma$; Eq.~\eqref{eq:Bthreshold} gives a sufficient choice. A separate long-scale result applies to binary disorder with $\varepsilon=1/5$ and $g/\kappa=1/10$.

Let $P_G=\sum_jP_j$ and $P_E=\sum_j|e_j\rangle\langle e_j|$ be the orthogonal projectors onto the ground and the auxiliary subspaces. The total jump-rate operator is
\begin{equation}
\sum_\mu J_\mu^\dagger J_\mu=\gamma P_G+\kappa P_E.\label{eq:effect}
\end{equation}
Thus the no-jump propagator
\begin{equation}
\mathcal U_\gamma(u)=\exp\!\left[-iuH-\frac u2(\gamma P_G+\kappa P_E)\right]
\label{eq:nojump}
\end{equation}
is independent of the frozen field. Every jump prepares a ground basis state. The next coherent segment starts from that state, and the next jump selects a new starting site. Quantum-jump evolution~\cite{Dalibard,Molmer,PlenioKnight} therefore gives an exact description as a sequence of such segments. We refer to this description as the renewal representation. A ground-state dephasing event counts as a reset even when it leaves the particle at the previous starting site. Dephasing suppresses coherence between the two ground states coupled to the same auxiliary state. It does not distinguish the decay channels or add an energy potential to the Hamiltonian.

For the preparation $\rho_0=P_0$, the state at time $T$ is $\rho_T^b=\ee^{T\mathcal L_b}P_0$. This state averages over the jump record while holding $b$ fixed. We measure the position operator
\begin{equation}
X=\sum_jjP_j+\sum_j(j+\tfrac12)|e_j\rangle\langle e_j|.
\end{equation}
 The terminal measurement records position at the final time $T$ and counts both ground and auxiliary populations. A bounded function $f$ converts the measured position into the readout $Y=f(cX)$, where $c>0$ sets the spatial resolution. Its mean for a frozen field is
\begin{equation}
F_b(T;c)=\Tr[f(cX)\ee^{T\mathcal L_b}P_0].\label{eq:terminal}
\end{equation}
This expectation averages over the jump histories and the final Born measurement. Auxiliary population contributes at the bond positions; it is neither discarded nor counted as a loss.

For two terminal trials $A$ and $B$ that are conditionally independent given the frozen field $b$,
\begin{equation}
	\Cov(Y_A,Y_B)=\Var_\xi F_{\varepsilon\xi}(T;c).\label{eq:pair}
\end{equation}

\begin{figure*}[t]
\includegraphics[width=\textwidth]{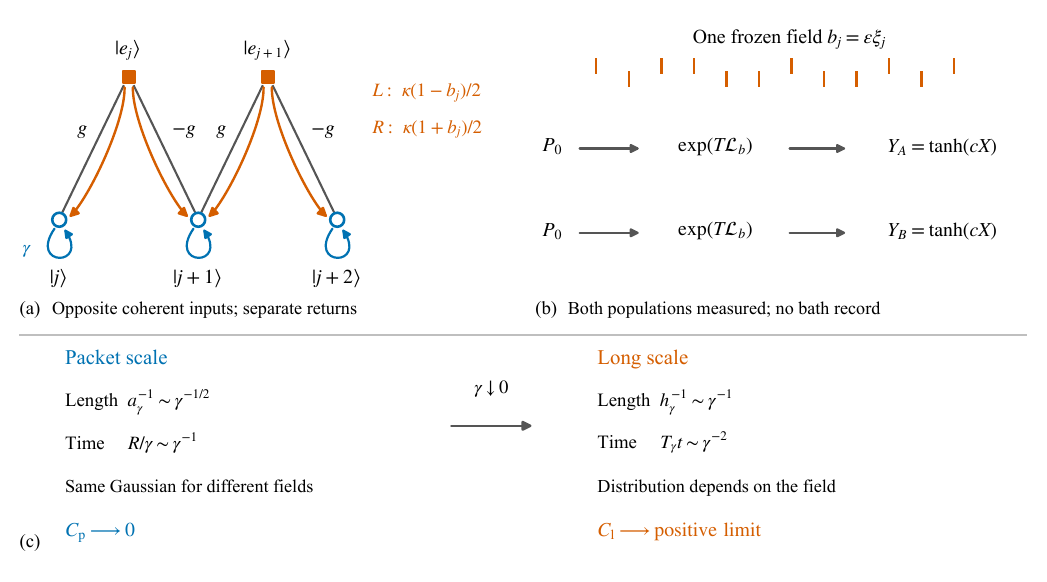}
\caption{Coherent spreading, frozen return biases, and the two observation scales. (a) The auxiliary state on bond $(j,j+1)$ couples to the neighboring ground states with amplitudes $g$ and $-g$. It returns left or right at rates $\kappa(1-b_j)/2$ and $\kappa(1+b_j)/2$. The bias $b_j=\varepsilon\xi_j$ is unchanged on every revisit; ground-state dephasing has rate $\gamma$. (b) Trials $A$ and $B$ share one frozen field and are conditionally independent. Each position measurement counts both populations at their physical positions and gives $Y=\tanh(cX)$. Averaging the paired readouts over fields yields $\Cov(Y_A,Y_B)=\Var_bF_b$, without a bath record. (c) The two length and time scales separate as $\gamma$ decreases. Their exact factors are defined in Eq.~\eqref{eq:scales}. The limiting comparison holds at fixed rescaled times, fixed $g,\kappa$, and a frozen field $b_j=\varepsilon\xi_j$ with bounded symmetric i.i.d. $\xi_j$ and $0<\varepsilon M<d_*$. This is a schematic of the model and measurement, not a plot of simulated populations.}
\label{fig:mechanism}
\end{figure*}

For each fixed field, independence gives $\E[Y_A Y_B\mid\xi]=F_{\varepsilon\xi}(T;c)^2$. Averaging these pairs over fields gives Eq.~\eqref{eq:pair}. The covariance measures variation of the conditional mean between samples, rather than the Born variance within one sample. It requires only the two final position measurements, not a bath record. We use $f(x)=\tanh x$ for the quantitative readout: it is bounded, smooth, and odd, and gives a nonzero response to directional redistribution. Figure~\ref{fig:mechanism} summarizes the chain and the two observation regimes.

\section{Exact segment dynamics and two limiting position distributions}\label{sec:scales}
\subsection{Joint duration and endpoint of a coherent segment}

Start a coherent segment at ground site $0$. Before the next jump, the unnormalized populations in the two subspaces are
\begin{equation}
p^G_j(u)=|\langle j|\mathcal U_\gamma(u)|0\rangle|^2,\qquad
p^E_j(u)=|\langle e_j|\mathcal U_\gamma(u)|0\rangle|^2.
\end{equation}
Here $u$ is the elapsed laboratory time since the last jump. At $b=0$, the clean case, the joint duration--endpoint density is
\begin{equation}
	\mathcal K_\gamma(j,u)=\gamma p^G_j(u)+\frac\kappa2[p^E_j(u)+p^E_{j-1}(u)].
	\label{eq:jointlaw}
\end{equation}
 Integrating $\mathcal K_\gamma(j,u)$ over $u$ gives the clean probability $p_j$ of resetting to site $j$. Summing both no-jump populations gives the survival probability, $S_\gamma(u)=\sum_j[p^G_j(u)+p^E_j(u)]$. It satisfies
\begin{equation}
	-S_\gamma'(u)=\gamma\sum_jp^G_j(u)+\kappa\sum_jp^E_j(u).
\end{equation}
For the initial state $\ket0$, define the integrated no-jump ground occupation $Q_z$ and the auxiliary-return probability $q_j$ by
\begin{align}
	Q_z&=\int_0^\infty|\langle z|\mathcal U_\gamma(u)|0\rangle|^2\dd u,\nonumber\\
	q_j&=\kappa\int_0^\infty|\langle e_j|\mathcal U_\gamma(u)|0\rangle|^2\dd u.
\end{align}
The clean reset probability is $p_j=\gamma Q_j+(q_j+q_{j-1})/2$. Three properties of a segment set the observation scales. The probability of an auxiliary return is $\eta=\sum_jq_j$, the mean duration is $\tau=\sum_jQ_j+\eta/\kappa$, and the endpoint variance is $\nu=\sum_j j^2p_j$.

Equation~\eqref{eq:effect} bounds the total jump rate below by $\min(\gamma,\kappa)$. The survival probability therefore decays at least exponentially, and a segment ends in a jump with probability one. Hence $\sum_jp_j=1$, or $\gamma\sum_jQ_j+\eta=1$. The mean duration is $\int_0^\infty S_\gamma(u)\,\dd u$, which includes the time spent in both subspaces.

 For a segment starting at site $j$, an auxiliary decay on bond $j+n$ returns the particle to one of that bond's endpoints. The left and right probabilities are $(1-b_{j+n})/2$ and $(1+b_{j+n})/2$. A reset at site $j+n$ can therefore follow decay on bond $j+n-1$ or bond $j+n$, as well as a ground-state dephasing event. Adding these contributions gives the joint density
\begin{align}
	\mathcal K_b(j,j+n;u)={}&\gamma p^G_n(u)+\frac\kappa2\big[(1+b_{j+n-1})p^E_{n-1}(u)\nonumber\\
	&+(1-b_{j+n})p^E_n(u)\big].
	\label{eq:biasedjointlaw}
\end{align}
 Integrating over time gives the reset kernel exactly,
\begin{equation}
	K_b(j,j+n)=p_n+\frac12\left[b_{j+n-1}q_{n-1}-b_{j+n}q_n\right].
	\label{eq:reset}
\end{equation}
In particular, the reset kernel stays within a factor $(1\pm d)$ of the clean one:
\begin{equation}
	(1-d)p_n\le K_b(j,j+n)\le(1+d)p_n,\qquad d=\varepsilon M<1.
	\label{eq:kernelcomparison}
\end{equation}
The bound in Eq.~\eqref{eq:kernelcomparison} makes the reset chain irreducible, with strictly positive transition probabilities. Let $J_n$ be the ground site after the $n$th jump, with $J_0=0$. This count includes dephasing events that leave the reset position unchanged. Summing Eq.~\eqref{eq:biasedjointlaw} over all endpoints gives $-S_\gamma'(u)$, independent of $b$. The joint distribution of duration and endpoint still depends on the field.

\subsection{Finite-rate coefficients and observation scales}\label{sec:renewal}
The observation scales are set by how far the particle moves in a segment and how long that segment lasts. We therefore calculate the mean duration and endpoint variance in the clean chain. The duration includes time spent in both subspaces, and the endpoint variance includes the choice of a ground site after auxiliary decay. To evaluate these quantities, we use the Fourier convention centered on the auxiliary positions. The amplitude generator is
\begin{equation}
	A_\gamma(k)=\begin{pmatrix}-\ell&c(k)\\-c(k)&-r\end{pmatrix},\qquad
	c(k)=2g\sin(k/2),
	\label{eq:bloch}
\end{equation}
where $\ell=\gamma/2$, $r=\kappa/2$, and $\Gamma=\ell+r$. For $e_1=(1,0)^{\mathsf T}$, the first unit vector, put
\begin{equation}
	\mathsf M(k)=\int_0^\infty\ee^{uA_\gamma(k)}e_1e_1^{\mathsf T}
	\ee^{uA_\gamma(k)^{\mathsf T}}\dd u.
\end{equation}
The diagonal entries of $\mathsf M(k)$ are the time-integrated ground and auxiliary occupations of mode $k$. Write $x=\mathsf M_{11}$, $y=\mathsf M_{12}$, and $z=\mathsf M_{22}$. They obey the Sylvester equation $A_\gamma\mathsf M+\mathsf M A_\gamma^{\mathsf T}=-e_1e_1^{\mathsf T}$, whose component equations are
\begin{align}
	2\ell x-2cy&=1,\quad \Gamma y=c(z-x),\quad 2rz=-2cy,\nonumber\\
	x&=\frac{c^2+r\Gamma}{2\Gamma(\ell r+c^2)},\qquad
	z=\frac{c^2}{2\Gamma(\ell r+c^2)}.
\end{align}
 Using $\int_{-\pi}^{\pi}(a-b\cos k)^{-1}\dd k/(2\pi)=(a^2-b^2)^{-1/2}$ and defining $\mathcal D_\gamma=\sqrt{\gamma\kappa(\gamma\kappa+16g^2)}$, we obtain
\begin{equation}
	\sum_jQ_j=\frac{1+\kappa^2/\mathcal D_\gamma}{\gamma+\kappa}.
\end{equation}
The normalization $\gamma\sum_jQ_j+\eta=1$ and the mean segment duration $\tau=\sum_jQ_j+\eta/\kappa$ now fix $\eta$ and $\tau$. The endpoint variance requires one further contribution. In a clean auxiliary return, the two endpoints lie at distances $-1/2$ and $1/2$ from the auxiliary center and have equal probability. Choosing between them adds $1/4$ to the variance about that center, or $\eta/4$ per segment. Appendix~\ref{app:renewal} includes this term in the endpoint characteristic function. We obtain
\begin{align}
\eta&=\frac\kappa{\gamma+\kappa}\left(1-\frac{\gamma\kappa}{\mathcal D_\gamma}\right),\nonumber\\
\tau&=\frac{2+\kappa(\kappa-\gamma)/\mathcal D_\gamma}{\gamma+\kappa},\nonumber\\
\nu&=\frac{4\kappa g^2}{(\gamma+\kappa)\mathcal D_\gamma}
+\frac{32g^4}{(\gamma+\kappa)^2(\gamma\kappa+8g^2+\mathcal D_\gamma)}.
\label{eq:constants}
\end{align}
 For small $\gamma$, with $g$ and $\kappa$ fixed, they behave as
\begin{equation}
	\tau\sim\frac{\sqrt\kappa}{4g\sqrt\gamma},\quad
	\nu\sim\frac g{\sqrt{\kappa\gamma}},\quad
	\frac\nu{2\tau}\longrightarrow D_*:=\frac{2g^2}\kappa.
	\label{eq:asymptotic}
\end{equation}
To make the rate dependence explicit, we write $\eta_\gamma,\tau_\gamma,\nu_\gamma$. The spatial factors below convert lattice positions into coordinates that remain finite in each limit. The quantity $n_\gamma$ sets the number of jumps on the long scale. Multiplying by the mean segment duration gives the corresponding laboratory time, $T_\gamma=\tau_\gamma n_\gamma$:
\begin{align}
	a_\gamma&=\frac{\sqrt{\kappa\gamma}}g,\qquad
	\alpha_\gamma=\frac{\eta_\gamma}{\nu_\gamma}\sim a_\gamma,\nonumber\\
	h_\gamma&=\alpha_\gamma^2,\qquad
	n_\gamma=\frac1{\nu_\gamma h_\gamma^2},\qquad
	T_\gamma=\tau_\gamma n_\gamma\sim\frac{g^2}{4\kappa\gamma^2}.
	\label{eq:scales}
\end{align}
The choice of $n_\gamma$ gives unit clean reset variance in the long-scale coordinate, since $h_\gamma^2\nu_\gamma n_\gamma=1$. The packet readout is $f(a_\gamma X)$ at time $R/\gamma$, with $R>0$ fixed. The long-scale readout is $f(h_\gamma X)$ at time $T_\gamma t$, with $t>0$ fixed. At finite $\gamma$, we use the exact ratio $\alpha_\gamma=\eta_\gamma/\nu_\gamma$. Although $\alpha_\gamma\sim a_\gamma$, the two factors are not identical at finite rate. For the readout $f=\tanh$, write $F_{\rm p}=F_{\varepsilon\xi}(R/\gamma;a_\gamma)$ and $F_{\rm l}=F_{\varepsilon\xi}(T_\gamma t;h_\gamma)$, with $C_{\rm p}=\Var_\xi F_{\rm p}$ and $C_{\rm l}=\Var_\xi F_{\rm l}$. These two covariances refer to complete (ground plus auxiliary) terminal measurements, compared at their respective rate-dependent resolutions and physical times.

\subsection{Gaussian packet and random-potential limits}
There is a threshold $d_*>0$ that can be chosen independently of $\gamma$ for all sufficiently small $\gamma$. Fix a nonzero disorder amplitude with $0<\varepsilon M<d_*$. The following two limits then hold as $\gamma\downarrow0$. For every fixed bounded continuous $f$ and fixed $R>0$, almost every original field realization satisfies
\begin{equation}
	\Tr[f(a_\gamma X)\rho^{\varepsilon\xi}_{R/\gamma}]
	\longrightarrow\int_\R f(x)\frac{\ee^{-x^2/(8R)}}{\sqrt{8\pi R}}\dd x.
	\label{eq:packetlaw}
\end{equation}
The limiting variance of $a_\gamma X$ is $4R$. Equivalently, the limiting variance of $\sqrt\gamma X$ is $2D_*R$. For $f=\tanh$, the limiting value of $F_{\rm p}$ vanishes. Since the readout is bounded, the frozen-sample variance also tends to zero by bounded convergence.

On the long scale, different frozen samples retain different position distributions. Regard the conditional position distribution as a random measure, with randomness supplied by the frozen field. This measure converges in distribution to the time-$t$ law of a diffusion $Z_t^V$ started at the origin. Its potential is
\begin{equation}
	V(x)=-\Sigma(\varepsilon)W(x),\qquad
	\Sigma^2(\varepsilon)=\varepsilon^2\sigma^2+O(\varepsilon^4)>0.
	\label{eq:potential}
\end{equation}
Here $W$ is a two-sided standard Brownian motion with $W(0)=0$. Its argument is position, not laboratory time: each realization gives one fixed spatial potential. The limiting process is specified by two quantities. The scale density determines probabilities of exiting an interval through either endpoint. Together with it, the speed measure determines residence times. Here these quantities are
\begin{equation}
	s_V'(x)=\ee^{V(x)},\qquad m_V(\dd x)=2\ee^{-V(x)}\dd x.
	\label{eq:scale-speed}
\end{equation}
For a smooth potential, the generator of the diffusion is $\tfrac12\ee^V\partial_x(\ee^{-V}\partial_x)$; Eq.~\eqref{eq:scale-speed} defines the diffusion for the continuous, nondifferentiable Brownian potential~\cite{Brox}. With $V=0$, the diffusion reduces to standard Brownian motion with variance $t$ at time $t$. The normalization here differs from that of the packet coordinate in Eq.~\eqref{eq:packetlaw}.

The limiting law retains sample dependence detectable by the same bounded readout:
\begin{equation}
	\lim_{\gamma\downarrow0}\Var_\xi
	\Tr[\tanh(h_\gamma X)\rho^{\varepsilon\xi}_{T_\gamma t}]
	=\Var_W\E_0^V\tanh Z_t^V>0.
	\label{eq:positive}
\end{equation}
To see why this variance is positive, compare the two linear potentials $\varphi_\pm(x)=\pm x$. They give $Z_t^V=B_t\mp t/2$, so the two means of $\tanh Z_t^V$ have opposite signs for every $t>0$. Choose a compact interval large enough that exiting it changes either bounded mean by less than one quarter of their separation. Potentials uniformly close to $\varphi_+$ or $\varphi_-$ on this interval have close stopped scale and speed laws, and retain separated terminal means. Gaussian increments and Brownian bridges give positive probability for a Brownian potential of nonzero amplitude to lie in either neighborhood. If their probabilities are $p_\pm$ and the separation is $m_t>0$, then $\Var_W\E_0^V\tanh Z_t^V\ge p_-p_+m_t^2>0$, where the variance is over the limiting Brownian potential. Boundedness transfers terminal-law convergence to this variance; it supplies no numerical finite-rate covariance floor.

\subsection{Explicit binary amplitude and scope}
For $g/\kappa=1/10$ and independent $b_j=\pm1/5$ each with probability $1/2$, the long-scale conditional-distribution limit in Eqs.~\eqref{eq:scale-speed} and~\eqref{eq:positive} holds with the same $h_\gamma$ and $T_\gamma$. The potential coefficient satisfies
\begin{equation}
	\Sigma^2(1/5)\ge\frac{529}{14400}>0.
	\label{eq:explicit-amplitude}
\end{equation}
This result applies at the disorder amplitude and coupling ratio used in the numerical study below. Equation~\eqref{eq:explicit-amplitude} bounds the variance of the limiting potential, not the value of the terminal covariance. The proof requires sufficiently small $\gamma$, but does not specify how small. Appendix~\ref{app:potential} gives the argument for this explicit amplitude. The general two-scale statement uses the common amplitude range $0<\varepsilon M<d_*$. Appendix~\ref{app:clock} treats the packet limit separately in the original frozen field. At resonance, its sufficient condition is $dC_*/2<1$, with $C_*=2$.

The two limits describe sample dependence in different ways. Equation~\eqref{eq:packetlaw} holds for almost every fixed realization of the original frozen field. The long-scale result concerns the distribution of conditional position laws across samples, jointly with the rescaled potential. Both are weak limits: they determine expectations of bounded continuous position functions at the stated spatial resolution. They do not make all finer measurements identical. We call disorder visible when the mean of a smooth position readout still varies between frozen samples.

\section{How coherent segments separate the scales}\label{sec:mechanism}
\subsection{From endpoint bias to a smoothed potential}
A spatially uniform ground-state amplitude excites an auxiliary state through two equal and opposite contributions, which cancel [Fig.~\ref{fig:mechanism}(a)]. For a mode of momentum $k$, the remaining coupling has magnitude $2g|\sin(k/2)|\simeq g|k|$. The coupling therefore vanishes at long wavelength. At $\gamma=0$, the corresponding slow no-jump eigenvalue is $-(2g^2/\kappa)k^2+O(k^4)$. Ground-state dephasing limits the lifetime of these long-wavelength modes and cuts off their propagation at a length of order $a_\gamma^{-1}$. The frozen field does not enter this no-jump dispersion. It acts when an auxiliary excitation returns, so its effect is averaged over the spatial region explored before that return.

The bias in the exact reset kernel, Eq.~\eqref{eq:reset}, is evaluated at the bond from which the auxiliary state decays. A single segment can cross many sites, and the reset probabilities generally violate detailed balance. At zero dephasing, the displacement tail is proportional to $|n|^{-2}$. Positive dephasing cuts off this tail at a distance of order $\alpha_\gamma^{-1}\sim\gamma^{-1/2}$. This cutoff distance differs from the root-mean-square displacement of one reset, $\sqrt{\nu_\gamma}=O(\gamma^{-1/4})$. Rare long coherent excursions account for the difference.

To determine the accumulated bias, we look for positive site weights $r_j$ that are unchanged after one reset. The net flow across each cut between neighboring sites is zero. These weights grow subexponentially, but their sum on the infinite line need not be finite, so they are not necessarily a normalized steady state. Their spatial variation encodes the accumulated bias. We extract its large-scale profile by smoothing with the positive kernel
\begin{align}
A_n&=\sum_{m>|n|}p_m(m-|n|),\nonumber\\
v_j&=\frac{(A*r)_j+(A*r)_{j+1}}\nu,\qquad v_0=1.
\label{eq:envelope}
\end{align}
Convolution is defined by $(A*r)_j=\sum_nA_nr_{j-n}$. Here $v_j$ is a time-independent smoothed weight, with its normalization fixed at the origin. It is distinct from the no-jump auxiliary population $p^E_j(u)$ in Sec.~\ref{sec:scales}. Place the values $-\log v_j$ at the rescaled positions $x=h_\gamma j$ and interpolate linearly between them. As $\gamma\downarrow0$, this interpolated potential converges in distribution to Eq.~\eqref{eq:potential}. The construction uses the smoothed weights $v_j$, not the unsmoothed weights $r_j$.

Here and below, $\alpha=\alpha_\gamma$ at the fixed rate $\gamma$ under consideration. The stationary flow across every cut vanishes. Summing all jumps that cross the cut between $j$ and $j+1$ gives
\begin{equation}
	(A*r)_{j+1}-(A*r)_j=\frac12b_j(q*r)_j.
	\label{eq:zerocurrent}
\end{equation}
This relation retains the full nonlocal kernel, including jumps that cross several bonds. Appendix~\ref{app:potential} proves the recurrence and growth bounds needed to construct the invariant weight on the infinite chain.

We now express the zero-current relation as a potential increment. The quantity $x_j=2(A*r)_j/\nu$ is the normalized weight associated with a cut, and $w_j=(q*r)_j/\eta$ is the local auxiliary-return activity. The bias weighted by this local activity is $a_j=b_jw_j/v_j$, so $x_{j+1}-x_j=\alpha a_jv_j$. With the relative increment $u_j=\alpha a_j/2$, the two neighboring values of $x$ become
\begin{equation}
	x_j=v_j(1-u_j),\qquad x_{j+1}=v_j(1+u_j).
	\label{eq:relativeincrement}
\end{equation}
Positivity of the invariant weight and $q_n\le2(A_n+A_{n+1})$ imply $|u_j|\le d<1$. Comparing the two expressions for $x_{j+1}$ obtained at neighboring cuts gives $v_{j+1}/v_j=(1+u_j)/(1-u_{j+1})$. Telescoping therefore yields
\begin{equation}
\log\frac{v_n}{v_0}=\sum_{j=0}^{n-1}2\operatorname{artanh}u_j
+\log(1-u_0)-\log(1-u_n).
\label{eq:logincrement}
\end{equation}
 For $n<0$ we use the oriented-sum convention $\sum_{j=0}^{n-1}f_j:=-\sum_{j=n}^{-1}f_j$. The distributions of the local ratios are invariant under lattice translations, although the anchored weights themselves are not. Reflection about a bond sends $a_j$ to $-a_{-j-1}$. This symmetry makes the mean logarithmic increment vanish, including its nonlinear terms. The increment in an individual sample can still be nonzero.

The two lengths follow from how the local biases accumulate. Coherent propagation spreads the effect of one return bias over many bonds. The resulting potential increment is of order $\alpha_\gamma$ times a zero-mean field with spatial correlations. Its fluctuations over $L$ sites are of order $\alpha_\gamma\sqrt L$; the statistics of this field set their amplitude. These fluctuations vanish for $L\sim\alpha_\gamma^{-1}$, but remain finite for $L\sim\alpha_\gamma^{-2}$. The particle samples the same field on repeated visits. The smoothed local contributions are therefore treated as correlated quantities. Appendix~\ref{app:potential} constructs the source and controls these correlations by replacing one field variable at a time.

The potential describes how return biases accumulate in space. To fix the rate of motion, we also need the variance acquired per reset. We evaluate it using a coordinate whose expected change in one reset is zero. This harmonic coordinate is obtained from the invariant weight $r^-_j$ for the opposite field $-b$:
\begin{align}
	H_\gamma(j+1)-H_\gamma(j)&=\frac{(q*r^-)_j}{\eta},\nonumber\\
	H_\gamma(0)&=0,\qquad K_bH_\gamma=H_\gamma.
	\label{eq:harmonic}
\end{align}
The smoothed weights for $b$ and $-b$ become reciprocal in the same limiting potential: $v_j^+v_j^-\to1$. Averaging the invariant weight over microscopic distances gives the limiting spatial occupation measure. The harmonic coordinate gives the scale density, which determines exit probabilities. Its accumulated variance fixes the rate of motion in that coordinate. Together these limits give the speed measure in Eq.~\eqref{eq:scale-speed}. The time scale $T_\gamma$ therefore needs no additional factor depending on the disorder amplitude. Appendix~\ref{app:potential} derives these limits and treats exits from compact intervals.

\subsection{The physical clock and the unfinished packet}\label{sec:clock}

The reset-chain limit counts completed jumps, whereas the measurement is made at a fixed laboratory time. This distinction also arises in continuous-time random walks~\cite{MontrollWeiss}. Correlations between travel time and displacement can affect the resulting transport~\cite{Zaburdaev}. On the long scale, we connect the two clocks by controlling the fluctuations in elapsed time and the spreading after the last reset.

Let $W_i$ be the duration of the $i$th segment, $S_n=\sum_{i=1}^nW_i$, and $N(T)=\max\{n:S_n\le T\}$. At each fixed $\gamma$, the segment durations are independent and identically distributed for every frozen field. Each reset starts from a ground basis state, and Eq.~\eqref{eq:effect} is independent of the starting site and the field. A duration is nevertheless correlated with the displacement produced by that same segment.

To bound fluctuations of the elapsed time, we first estimate how long a segment can survive. In the following estimates, a second subscript records the auxiliary-decay rate. Near zero momentum, the zero-dephasing matrix in Eq.~\eqref{eq:bloch} has the eigenvalue $-(2g^2/\kappa)k^2+O(k^4)$. As $k\to0$, the slow mode has ground-state component tending to one and an auxiliary component that is odd in $k$ and of order $k$. Let $M_{2,G}$ and $M_{2,E}$ denote the unnormalized position second moments about the starting site. They use the integer ground positions and half-integer auxiliary positions, respectively. Parseval's identity gives
\begin{align}
	S_{0,\kappa}(u)&\asymp(1+u)^{-1/2},\nonumber\\
	M_{2,G}(u)&\le C(1+u)^{1/2},&
	M_{2,E}(u)&\le C(1+u)^{-1/2}.
	\label{eq:sectorbounds}
\end{align}
The estimates are uniform as the auxiliary decay rate varies over a compact neighborhood of $\kappa$. Modes away from zero momentum have decay rates bounded away from zero uniformly in the small-dephasing limit. Near a degeneracy of the two modes, we use the full matrix exponential $\ee^{uA_\gamma(k)}$ rather than a single-mode approximation.

For $0<\gamma\le\kappa/2$ the exact identity
\begin{equation}
	\mathcal U_{\gamma,\kappa}(u)=\ee^{-\gamma u/2}\mathcal U_{0,\kappa-\gamma}(u)
\end{equation}
and Eq.~\eqref{eq:sectorbounds} imply
\begin{equation}
\E W^2=2\int_0^\infty uS_\gamma(u)\dd u\le C\gamma^{-3/2}.
\end{equation}
For every fixed field, the centered duration sums have zero conditional mean increments and therefore form a martingale. For fixed $U>0$ and $n_\gamma=(\nu h_\gamma^2)^{-1}$, Doob's maximal inequality gives, uniformly in $b$,
\begin{align}
\E_b\sup_{0\le s\le U}\left|\frac{S_{\lfloor n_\gamma s\rfloor}}{\tau n_\gamma}-s\right|^2
&\le\frac{C_U\Var W}{\tau^2n_\gamma}+\frac2{n_\gamma^2}
=O(\gamma).\label{eq:clock}
\end{align}
The jump count $N(T)$ increases with $T$. Equation~\eqref{eq:clock} therefore gives $N(T_\gamma t)/n_\gamma\to t$ uniformly on compact time intervals. Thus the conversion from jump count to physical time becomes deterministic on this scale. Its limit can be combined with the spatial path limit without assuming independence between a segment's duration and displacement.

At the measurement time, the particle has evolved since its last reset. Let $J=J_{N(T)}$ be that reset site and $A(T)=T-S_{N(T)}$ the elapsed time since the reset. Conditional on these quantities, the state is $\mathcal U_\gamma(A)\ket{J}/S_\gamma(A)^{1/2}$. The terminal Born position $X_B(T)$ is drawn from this state and includes both subspaces. Equation~\eqref{eq:sectorbounds} bounds its position variance about $J$ by $C(1+A)$, after division by $S_\gamma(A)$. The total jump rate is at least $\gamma$, so
\begin{equation}
	\Pp_b(A(T)>a)\le\ee^{-\gamma a},\qquad
	\E_bA(T)\le\min(T,\gamma^{-1}).
\end{equation}
 The case of no jump by age $T$ is included. Hence
\begin{equation}
	\E_b|h_\gamma[X_B(T)-J]|^2
	\le Ch_\gamma^2[1+\min(T,\gamma^{-1})]=O(\gamma).
	\label{eq:born}
\end{equation}
The bound is uniform in the frozen field and in time. On every finite scaled-time interval, the expectation of a bounded Lipschitz readout of $h_\gamma X_B(T)$ differs from its expectation at $h_\gamma J_{N(T)}$ by $O(\sqrt\gamma)$. Uniform control of the position tails extends this convergence to bounded continuous readouts. The estimate includes both ground and auxiliary populations.

\begin{table*}[t]
\caption{Observation scales and covariance estimates at three dephasing rates on the infinite resonant chain. We use $\kappa=1$, $g=0.1$, binary disorder $b_j=\pm0.2$, and $R=t=1$. The rates are $\gamma_H=2.5\times10^{-4}$, $\gamma_M=\gamma_H/4$, and $\gamma_L=\gamma_H/16$. Times are measured in $\kappa^{-1}$ and lengths in lattice units. Covariances are in units of $10^{-3}$; parentheses give one standard error across fields in the last two displayed digits. The bounds on numerical bias $B_{\rm p,l}$ for each target are in units of $10^{-6}$ and rounded upward. The simulation uses unrounded scale factors.}
\label{tab:scandata}
\begin{ruledtabular}
\begin{tabular}{ccccccccc}
Rate & $1/\gamma$ & $T_\gamma$ & $1/a_\gamma$ & $1/h_\gamma$ & $10^3\widehat C_{\rm p}$ & $10^3\widehat C_{\rm l}$ & $10^6B_{\rm p}$ & $10^6B_{\rm l}$\\
$H$ & $4.0\times10^3$ & $4.83\times10^4$ & $6.32$ & $43.8$ & $2.07(21)$ & $2.00(17)$ & $1.7$ & $11$\\
$M$ & $1.6\times10^4$ & $7.03\times10^5$ & $12.6$ & $167$ & $0.91(14)$ & $1.97(16)$ & $3.3$ & $72$\\
$L$ & $6.4\times10^4$ & $1.07\times10^7$ & $25.3$ & $655$ & $0.31(15)$ & $1.89(18)$ & $0.45$ & $38$\\
\end{tabular}
\end{ruledtabular}
\end{table*}

\subsection{Gaussian reconstruction on the packet scale}\label{sec:packet}

On the packet scale, neither the random segment durations nor the spreading after the last reset can be discarded. At time $O(\gamma^{-1})$, elapsed-time fluctuations at a fixed jump count remain correlated with displacement. The unfinished packet also has width $O(\gamma^{-1/2})$, comparable with the observation length. We therefore keep duration and displacement together for each segment and include the unfinished packet at measurement. We use the rescaled displacement $z=\sqrt\gamma\,\Delta J$ and the duration $u=\gamma W$ for one segment. Set $D_*=2g^2/\kappa$ and $c_*=\sqrt\kappa/(4g)$. After rescaling the jump count as $n\sqrt\gamma$, the limiting joint process is described by a L\'evy measure. This measure specifies the frequency, per unit rescaled jump count, of segments in each range of durations and displacements:
\begin{align}
	\Pi(\dd z,\dd u)={}&\frac{c_*}{\sqrt\pi}\ee^{-u}
	\left[u^{-1/2}+\frac{z^2}{2D_*u^{5/2}}\right]\nonumber\\
	&\times\mathcal N(0,D_*u)(\dd z)\dd u.
	\label{eq:levymarks}
\end{align}
Here $\mathcal N(0,D_*u)$ is the centered Gaussian measure of variance $D_*u$. The two terms in brackets come from ground-state resets and auxiliary returns, respectively. Appendix~\ref{app:clock} derives them from the slow coherent mode and controls the contribution of short segments. Let $s>0$ be conjugate to scaled time in the Laplace transform, and let $k$ be conjugate to displacement in the Fourier transform. The joint exponent is $\psi(s,k)=\int(1-\ee^{-su+ikz})\Pi(\dd z,\dd u)$, which gives
\begin{equation}
\psi(s,k)=c_*\frac{s+D_*k^2}{\sqrt{1+s+D_*k^2/2}}.
\label{eq:packetexponent}
\end{equation}
The unfinished packet at scaled age $a$ is Gaussian with variance $D_*a$ about its last center. Let $\widehat{\widetilde p}(s,k)$ denote the Laplace--Fourier transform of the complete terminal law in the $\sqrt\gamma X$ coordinate and scaled time $R$. To obtain the measured position, we combine the last reset position with the unfinished packet and its age. In the transform below, the reciprocal of the denominator accounts for all completed segments. The numerator accounts for the survival and spatial spread of the segment present at measurement. Together they give
\begin{equation}
	\widehat{\widetilde p}(s,k)=
	\frac{c_*/\sqrt{1+s+D_*k^2/2}}{\psi(s,k)}
	=\frac1{s+D_*k^2}.
	\label{eq:packet-transform}
\end{equation}
The number of completed jumps, their total displacement, and the elapsed time in the unfinished segment must be treated together~\cite{StrakaHenry}. Replacing segment durations by their mean would fail to reproduce Eq.~\eqref{eq:packet-transform}.

In Eq.~\eqref{eq:packet-transform}, the square-root factors from the joint reset process and the unfinished packet cancel. The remaining transform is that of a Gaussian with variance $2D_*R$ in $\sqrt\gamma X$, or $4R$ in $a_\gamma X$. The frozen field must also produce no surviving drift on this scale. Appendix~\ref{app:clock} proves this centering for almost every original field, with the same return biases used on every revisit. The result describes a position measurement at one time; it does not assign an unmeasured classical position path to a coherent segment.

\section{Separation of the terminal covariances at finite rates}\label{sec:data}
\subsection{Observation scales and terminal sampling}

The finite-rate calculation holds the microscopic frozen disorder fixed while changing the ground-state dephasing rate. We use $\kappa=1$ as the rate unit, $g=0.1$, binary $b_j=\pm0.2$ with equal probabilities, and $R=t=1$. The three rates are $\gamma_H=2.5\times10^{-4}$, $\gamma_M=\gamma_H/4$, and $\gamma_L=\gamma_H/16$. For each rate we use the exact $a_\gamma$, $h_\gamma$, and $T_\gamma$ from Eq.~\eqref{eq:scales}. As $\gamma$ decreases, the measurement time and spatial resolution change according to the two observation scales. The long-scale theorem applies to this amplitude and coupling ratio in the limit $\gamma\downarrow0$. The simulations probe three finite rates.

We use the same 512 frozen fields at all three rates. For each field and rate, two trajectory batches are independent conditional on the field. Each batch contains 256 paths at $H$ and $M$, and 192 paths at $L$, for a total of 720\,896 paths. We evaluate both terminal times along each path. Table~\ref{tab:scandata} lists the measurement scales, covariance estimates, and separate bounds on numerical bias.

Starting from a ground state, the joint densities for a dephasing event at $z$ and an auxiliary decay at bond $a$ are
\begin{equation}
	f_G(z,u)=\gamma p^G_z(u),\qquad f_E(a,u)=\kappa p^E_a(u).
\end{equation}
For numerical sampling, we tabulate these joint densities once per rate. Fourier inversion of $\ee^{uA_\gamma(k)}e_1$ yields the site amplitudes. The auxiliary amplitude indexed by bond $a$ differs from the centered convention of Eq.~\eqref{eq:bloch} by the phase $\ee^{ik/2}$. After sampling an auxiliary decay on bond $J+a$, we choose its left or right endpoint using the frozen bias $b_{J+a}$. The sampling includes dephasing resets that leave the site unchanged. It uses the infinite chain without a reflecting boundary, a periodically repeated field, or sample-dependent recentering.

For the last reset site $J$ and the elapsed time $A$ since that reset, the conditional mean of the readout is
\begin{align}
	\mathcal F_f(J,A;c)&=\frac{N_G+N_E}{S_\gamma(A)},\nonumber\\
	N_G&=\sum_zf(c(J+z))p^G_z(A),\nonumber\\
	N_E&=\sum_af(c(J+a+1/2))p^E_a(A).
	\label{eq:conditional-terminal}
\end{align}
Here $c=a_\gamma$ or $h_\gamma$, and $f=\tanh$. For each simulated jump record, Eq.~\eqref{eq:conditional-terminal} averages the final Born measurement analytically. The record includes jump times, channels, and reset sites. This conditional average reduces Monte Carlo noise without changing the terminal mean, and both populations remain in $S_\gamma$. An experimental trial instead supplies one sampled final position. The path counts here specify the numerical sampling.

The sampler is based on the exact renewal representation of the full Lindblad evolution. Numerical approximation enters through the tabulated joint distribution and terminal readout. Duration, channel, and position are sampled jointly, so duration--displacement correlations are retained. Appendix~\ref{app:statistics} bounds the approximation error over the full laboratory-time interval and converts it into a bound on covariance bias.

\subsection{Covariance estimates and statistical uncertainty}

For a frozen sample $\xi^{(i)}$, rate $r\in\{H,M,L\}$, and readout $e\in\{\mathrm p,\mathrm l\}$, write $A_{ire}$ and $B_{ire}$ for the two batch means and $F_{ire}$ for their common conditional expectation. With $n=512$ fields, we define
\begin{align}
	w_{ire}&=(A_{ire}-\bar A_{re})(B_{ire}-\bar B_{re}),\nonumber\\
	\widehat C_{re}&=\frac1{n-1}\sum_iw_{ire},\nonumber\\
	\zeta_{ire}&=\frac n{n-1}(w_{ire}-\bar w_{re}).
	\label{eq:estimator}
\end{align}
Overbars denote averages over the $n$ fields. Conditional independence of batches $A$ and $B$ gives $\E[A_{ire}B_{ire}\mid\xi^{(i)}]=F_{ire}^2$. Their cross-covariance therefore estimates $\Var_bF_{re}$ without adding the trajectory variance within a field. Statistical independence applies to complete fields, not to individual trajectories pooled across the study.

\begin{figure*}[t]
	\includegraphics[width=\textwidth]{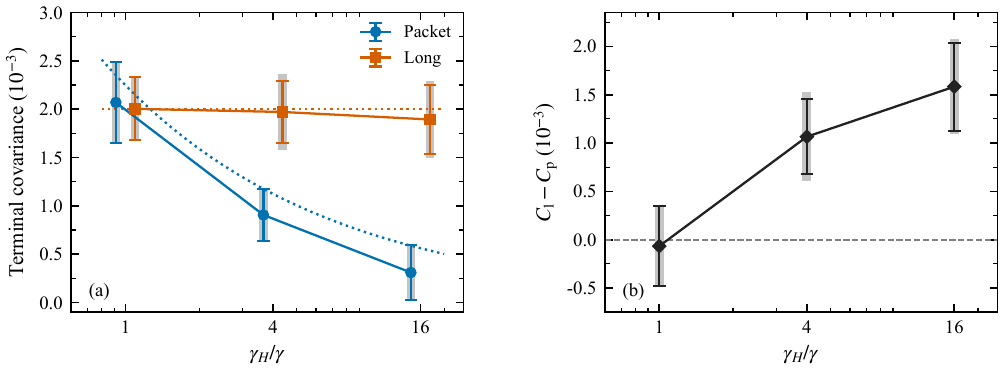}
\caption{Terminal covariances respond differently to decreasing dephasing at fixed return biases. The resonant infinite chain starts from $P_0$, with $\kappa=1$, $g=0.1$, binary $b_j=\pm0.2$, and $R=t=1$. (a) Covariances of $\tanh(a_\gamma X)$ at $1/\gamma$ and $\tanh(h_\gamma X)$ at $T_\gamma$, plotted against $\gamma_H/\gamma$. Thin intervals are approximate pointwise 95\% $t$ intervals computed across independent frozen fields. The thicker gray intervals include the separate bounds on numerical bias. Colored lines connect estimates; small horizontal offsets separate the readouts. Dotted curves show the ordered weak-disorder references $\varepsilon^2a_\gamma B_1$ and $\varepsilon^2A_1$, with $\sigma^2=1$. They are not finite-amplitude fits. (b) The paired difference between the long-scale and packet covariances, evaluated on the same fields. The renewal sampler retains duration--displacement correlations and both populations of the unfinished packet. Section~\ref{sec:data} and Table~\ref{tab:design} specify the numerical sampling design.}
	\label{fig:finite}
\end{figure*}

To compare rates and readouts, consider a linear combination $\theta=\sum_{r,e}c_{re}C_{re}$. The contribution of field $i$ to this contrast is $\zeta_{i\theta}=\sum_{r,e}c_{re}\zeta_{ire}$. Its standard error is $\operatorname{sd}(\zeta_{i\theta})/\sqrt n$. Keeping all observations from a field together preserves the correlations between rates and readouts. Figure~\ref{fig:finite} shows approximate pointwise $t$ intervals computed across fields. The bounds on numerical bias widen these intervals separately.

The two covariances change differently as the rate decreases [Fig.~\ref{fig:finite}(a)]. The packet covariance falls from $2.07(21)\times10^{-3}$ to $0.31(15)\times10^{-3}$, while the long-scale covariance changes from $2.00(17)\times10^{-3}$ to $1.89(18)\times10^{-3}$. Parentheses give one standard error across fields in the last displayed digits. The packet point estimate is reduced by about $85\%$ over this sixteenfold rate decrease. The long-scale signal stays near $2\times10^{-3}$, and the gap between the two covariances grows from a value consistent with zero to $1.58(23)\times10^{-3}$ (Fig.~\ref{fig:finite}(b)).

The three contrasts used in the joint analysis are
\begin{align}
	S_{\rm p}&=C_{\rm p}(H)-C_{\rm p}(L),\nonumber\\
	\Delta_L&=C_{\rm l}(L)-C_{\rm p}(L),\nonumber\\
	\mathcal I&=\Delta_L-[C_{\rm l}(H)-C_{\rm p}(H)].
	\label{eq:primarycontrasts}
\end{align}
The contrasts distinguish suppression of the packet signal, separation at the lowest rate, and growth of that separation. 

For these three comparisons, we draw 30\,000 bootstrap resamples of the 512 fields. All measurements from a selected field are kept together, and all six covariances are recomputed in each resample. In each resample, we center the three contrast estimates and divide them by their original standard errors across fields. Taking the largest standardized value accounts for the three comparisons together. Its 95th percentile is $q_{0.95}\simeq2.03$, giving the common one-sided lower bound
\begin{equation}
	\widehat\theta-q_{0.95}\widehat{\mathrm{SE}}_\theta-B_\theta,\qquad
	B_\theta=\sum_{r,e}|c_{re}|B_{re}.
\end{equation}
Table~\ref{tab:primarytests} rounds the lower bounds downward. Coverage of both the $t$ intervals and the bootstrap intervals is approximate. We subtract the bound on numerical bias from the statistical lower endpoint, rather than combine it in quadrature with the standard error.

\begin{table}[t]
	\caption{Paired contrasts with standard errors across fields in parentheses and simultaneous one-sided 95\% lower bounds. Estimates and lower bounds are in units of $10^{-3}$. Bounds on numerical bias are in units of $10^{-6}$ and rounded upward, while lower bounds are rounded downward. The rows test three changes: suppression of the packet covariance ($S_{\rm p}$), separation of the two covariances at the lowest rate ($\Delta_L$), and growth of that separation across the rate window ($\mathcal I$).}
	\label{tab:primarytests}
	\begin{ruledtabular}
		\begin{tabular}{cccc}
			Target & Estimate & $10^6B_\theta$ & Lower bound\\
			$S_{\rm p}$ & $1.76(25)$ & $2.1$ & $1.25$\\
			$\Delta_L$ & $1.58(23)$ & $38$ & $1.07$\\
			$\mathcal I$ & $1.65(31)$ & $50$ & $0.96$\\
		\end{tabular}
	\end{ruledtabular}
\end{table}

All three simultaneous lower bounds are positive. The difference of differences, $\widehat{\mathcal I}=1.65(31)\times10^{-3}$, directly tests whether the two covariances change differently across the rate window. Its positive joint lower bound supports that difference. This comparison does not rely on whether separate endpoint intervals overlap.

For the long-scale covariance, a separate descriptive comparison gives $C_{\rm l}(L)-C_{\rm l}(H)=-0.11(24)\times10^{-3}$. Its approximate $90\%$ interval, widened by numerical bias, is $[-0.56,0.34]\times10^{-3}$, inside a tolerance of $\pm0.8\times10^{-3}$. The tolerance is coarse; the observed change of about 6\% does not establish constancy to that precision. These comparisons are exploratory. The intervals describe uncertainty for the chosen contrasts, tolerance, and bootstrap settings. They do not account for possible selection of these choices after inspecting the data. The three rates distinguish the trends over this window, but do not determine an asymptotic exponent.

\section{Covariance response on the two physical scales}\label{sec:response}
\subsection{Exact response of the full Lindblad evolution}
The fixed-amplitude data show that the two covariances change differently over the simulated rate window. To understand how coherent motion affects their sensitivity to disorder, we now calculate the initial response of each covariance as disorder is introduced. At every fixed positive $\gamma$, we take the disorder amplitude $\varepsilon$ to zero. The quadratic coefficients are
\begin{equation}
\mathsf C_{\rm p}(\gamma;R)=\lim_{\varepsilon\to0}\frac{\Var_\xi F_{\rm p}}{\varepsilon^2},\qquad
\mathsf C_{\rm l}(\gamma;t)=\lim_{\varepsilon\to0}\frac{\Var_\xi F_{\rm l}}{\varepsilon^2}.
\label{eq:C2}
\end{equation}
Below, $\mathsf C$ denotes the coefficient for a chosen $(T,c)$.

In the full generator, each auxiliary-return term depends on the square of its jump amplitude. The square roots in Eq.~\eqref{eq:jumps} therefore cancel. The frozen field enters only linearly:
\begin{equation}
	\mathcal L_b=\mathcal L_0+\sum_jb_j\mathcal B_j,\qquad
	\mathcal B_j\rho=\frac\kappa2\rho_{e_j,e_j}(P_{j+1}-P_j).
	\label{eq:insertion}
\end{equation}
Let $\mathsf E_j(s)=\langle e_j|\ee^{s\mathcal L_0}P_0|e_j\rangle$ be the auxiliary population at time $s$ under the full clean evolution. It includes every earlier reset. Let $U_f(v,j)=\Tr[f(cX)\ee^{v\mathcal L_0}P_j]$ be the mean terminal readout after evolving for time $v$ from ground site $j$. Duhamel's formula gives the response to a bias on bond $j$:
\begin{align}
G_j(T;c)&=\frac\kappa2\int_0^T\mathsf E_j(s)d_j(T-s;c)\dd s,\nonumber\\
d_j(v;c)&=U_f(v,j+1)-U_f(v,j),\qquad
\mathsf C=\sigma^2\sum_jG_j^2.
\label{eq:gradient}
\end{align}
Here $G_j=\left.\partial F_b(T;c)/\partial b_j\right|_{b=0}$ measures the change caused by a weak bias on bond $j$. The first factor gives the population on the perturbed auxiliary state at time $s$. The second gives the difference between the later readouts after a right return and a left return. Their product, integrated over $s$, is the response to that bond. For an increasing readout, $G_j\ge0$. A single no-jump occupation cannot replace $\mathsf E_j$, because it excludes all earlier resets.

The perturbation in Eq.~\eqref{eq:insertion} is bounded in trace-class norm: $\|\sum_j\xi_j\mathcal B_j\|\le\kappa M$. The derivative is therefore defined on the infinite chain for every fixed positive $\gamma$ and finite $T$. For $f=\tanh$, translation invariance and $0<f'\le1$ give $0\le d_j(v;c)\le c$. Hence $G_j\ge0$ and $\sum_jG_j\le\kappa cT/2$. The random derivative $\sum_j\xi_jG_j$ is summable. Independence and centering of the $\xi_j$ then give $\mathsf C=\sigma^2\sum_jG_j^2$.

\subsection{Spatial response on the two observation scales}

To compare the bond responses at different dephasing rates, we express them as functions of rescaled position. The limiting response kernel measures how strongly a weak bias at each position changes the final mean readout. We first determine the auxiliary population on each observation scale. The two scales use Gaussian propagators with different variance conventions. Let
\begin{equation}
	\mathsf p_v(x)=\frac{\ee^{-x^2/(4v)}}{\sqrt{4\pi v}},\qquad
	p_r(x)=\frac{\ee^{-x^2/(2r)}}{\sqrt{2\pi r}}.
\end{equation}
We denote convolution with these kernels by $\mathsf P_v$ and $P_r$, respectively. Their variances are $2v$ and $r$.

The small-$\gamma$ response depends on the full auxiliary occupation $\mathsf E_j(s)$, including every earlier reset. Each reset starts another segment, so summing over their number produces repeated convolutions in time and space. Laplace and Fourier transforms turn this sum into a geometric series. We denote the Laplace transform in physical time by a tilde, with parameter $\lambda>0$. In space, we use the auxiliary Fourier phase $\ee^{-ik(j+1/2)}$. Summing all clean resets before time $s$ then gives
\begin{align}
\widetilde{\mathsf E}_c(k,\lambda)&=\frac{\widetilde p^E_c(k,\lambda)}{1-\widetilde{\mathcal K}_\gamma(k,\lambda)},\nonumber\\
\widetilde{\mathcal K}_\gamma&=\gamma\widetilde p^G+\kappa\cos(k/2)\widetilde p^E_c.
\label{eq:renewalquotient}
\end{align}
The endpoint cosine occurs only in the denominator. Multiplying the clean duration--endpoint density by $\ee^{-\lambda u}$ and integrating over $u$ preserves positivity and spatial symmetry. Hence
\begin{align}
1-\widetilde{\mathcal K}_\gamma(k,\lambda)
={}&1-\widetilde{\mathcal K}_\gamma(0,\lambda)\nonumber\\
&+\sum_j\widetilde{\mathcal K}_\gamma(j,\lambda)[1-\cos(kj)]>0.
\label{eq:positivedeficit}
\end{align}
At fixed $\gamma>0$ and $\lambda>0$, Eq.~\eqref{eq:positivedeficit} keeps the denominator positive and bounded away from zero over all lattice momenta. This remains true when the transformed numerator changes sign. The lower bound need not be uniform as $\gamma$ and $\lambda$ vanish; Appendix~\ref{app:response} treats that limit. The mean auxiliary residence time per segment is $\eta/\kappa$. Dividing it by the mean segment duration gives $f_{E,\gamma}=\eta/(\kappa\tau)$, the long-time fraction of laboratory time spent in the auxiliary subspace. Define positive step densities on cells of the indicated widths:
\begin{align}
	e^p_\gamma(r,x)&=\frac{\kappa \mathsf E_j(r/\gamma)}{4\gamma},&x&\in[a_\gamma j,a_\gamma(j+1)),\nonumber\\
	e^l_\gamma(r,x)&=\frac{\mathsf E_j(T_\gamma r)}{h_\gamma f_{E,\gamma}},&x&\in[h_\gamma j,h_\gamma(j+1)).
\end{align}
Let $z>0$ be the Laplace parameter for scaled time and $p$ the Fourier variable conjugate to the rescaled position $x$. We retain the full lattice momentum range $|k|\le\pi$. The centered lattice Fourier amplitudes on this momentum interval are
\begin{align}
	\mathcal A^p_{\gamma,z}(p)&=\frac{a_\gamma\kappa}{4}\widetilde{\mathsf E}_c(a_\gamma p,\gamma z)
	\longrightarrow\frac{\sqrt{1+z+p^2}}{z+2p^2},\nonumber\\
	\mathcal A^l_{\gamma,z}(p)&=\frac{\widetilde{\mathsf E}_c(h_\gamma p,z/T_\gamma)}{T_\gamma f_{E,\gamma}}
	\longrightarrow\frac1{z+p^2/2}.
	\label{eq:occupation-symbols}
\end{align}
The Laplace--Fourier transforms of the step densities are $\operatorname{sinc}(a_\gamma p/2)\mathcal A^p_{\gamma,z}(p)$ and $\operatorname{sinc}(h_\gamma p/2)\mathcal A^l_{\gamma,z}(p)$, respectively, where $\operatorname{sinc}u=\sin u/u$ and $\operatorname{sinc}0=1$. Integrating over a cell supplies this factor. The auxiliary-center phase is already included in $\widetilde{\mathsf E}_c$. At each fixed $p$, both cell factors tend to one. On the packet scale, the square root depends on both transformed time and momentum, retaining the correlation between segment duration and displacement. On the long scale, this factor disappears and the transform reduces to that of ordinary diffusion. Inverting the packet limit in Eq.~\eqref{eq:occupation-symbols} gives
\begin{align}
	e_r(x)={}&\frac{\ee^{-r}\mathsf p_r(x)}{\sqrt{\pi r}}\nonumber\\
	&+\int_0^r\frac{\ee^{-u}}{\sqrt{\pi u}}
	[\mathsf p_{2r-u}(x)+\partial_x^2\mathsf p_{2r-u}(x)]\dd u,\nonumber\\
	H_R(x)&=2\int_0^Re_r(x)(\mathsf P_{2(R-r)}f')(x)\dd r,\nonumber\\
	B_R&=\int_\R H_R(x)^2\dd x.
	\label{eq:packetkernel}
\end{align}
The kernel $e_r$ is nonnegative, although the terms in its displayed expression need not be separately positive. It follows from the ratio in Eq.~\eqref{eq:renewalquotient}, which includes all earlier resets. For the long scale,
\begin{align}
K_t(x)&=\int_0^tp_r(x)(P_{t-r}f)'(x)\dd r,\nonumber\\
A_t&=\frac14\int_\R K_t(x)^2\dd x.
\label{eq:longkernel}
\end{align}
Both $B_R$ and $A_t$ are finite and strictly positive. The factor $1/4$ in $A_t$ comes from squaring the factor $1/2$ in the first-order response, Eq.~\eqref{eq:roughresponse}.

To relate these continuum kernels to the covariance, each bond is represented by a cell in rescaled space. Its contribution depends on both the response amplitude and the cell width. We represent each exact gradient as a constant on its cell. On cells of width $a_\gamma$, $G_j(R/\gamma;a_\gamma)/a_\gamma$ converges to $H_R$ in spatial $L^2$. On cells of width $h_\gamma$, $G_j(T_\gamma t;h_\gamma)/\alpha_\gamma$ converges to $K_t/2$ in the same norm. Thus the integral of the squared difference tends to zero in each case. The exact definitions of $f_{E,\gamma}$ and $T_\gamma$ give
\begin{equation}
	\frac\kappa2T_\gamma f_{E,\gamma}h_\gamma^2=\frac\alpha2.
\end{equation}
Equation~\eqref{eq:gradient} sums the squared response over bonds. On the packet scale, each gradient carries one cell-width factor; squaring it and summing over cells leaves one such factor. On the long scale, the squared gradient scale equals the cell width, so no vanishing factor remains. The two covariance coefficients therefore satisfy
\begin{equation}
\frac{\mathsf C_{\rm p}(\gamma;R)}{a_\gamma}\longrightarrow\sigma^2B_R,\qquad
\mathsf C_{\rm l}(\gamma;t)\longrightarrow\sigma^2A_t.
\label{eq:ordered}
\end{equation}
When the weak-disorder limit is taken first, the packet response coefficient $\mathsf C_{\rm p}$ is asymptotically proportional to $\sqrt\gamma$, whereas the long-scale coefficient $\mathsf C_{\rm l}$ approaches a positive constant. This $L^2$ limit includes the bonds nearest the initial site.

The covariance involves the squared response, so a weak limit of the terminal distribution is not enough to obtain Eq.~\eqref{eq:ordered}. We must also control how much of the squared response comes from bonds near the initial site. Appendix~\ref{app:response} establishes this control over the full momentum range, including the extra Fourier copies from the spatial cells. It also bounds the contributions near both ends of the time integral. These estimates are needed because the packet response diverges logarithmically at the origin:
\begin{equation}
	H_R(x)=\frac{(\mathsf P_{2R}f')(0)}\pi\log\frac1{|x|}+O_R(1),\qquad x\to0.
\label{eq:cusp}
\end{equation}
The logarithmic peak comes from the short-time occupation kernel. It is unbounded at the origin, but its square is integrable. The coefficient remains $(\mathsf P_{2R}f')(0)/\pi$ for any bounded smooth readout with bounded first and second derivatives. It is positive for a nonconstant increasing readout and may vanish for other choices of $f$. Appendix~\ref{app:response} establishes the required norm convergence by combining the full-frequency bounds with positivity of the occupations. 

\begin{figure*}[t]
\includegraphics[width=\textwidth]{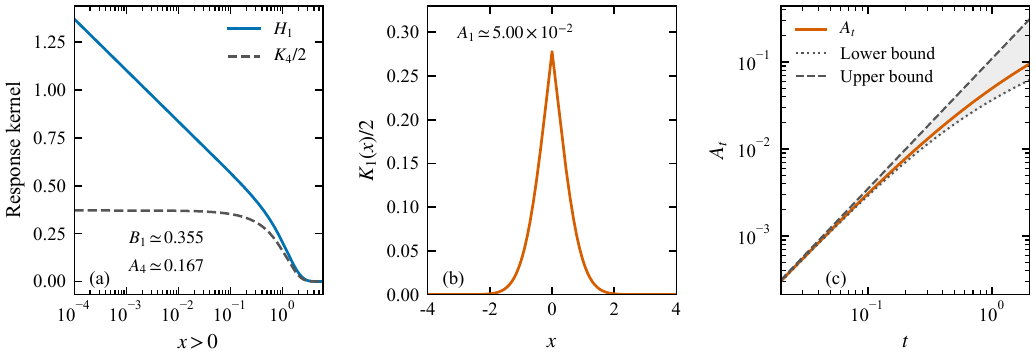}
\caption{Spatial responses of the terminal readout $f(x)=\tanh x$. (a) The coherent-chain packet kernel $H_1$ (solid) and the matched memoryless kernel $K_4/2$ (dashed) have different spatial profiles. The two models are compared at the same clean terminal variance, 4. Both kernels are even; their squared norms are $B_1$ and $A_4$. The logarithmic peak of $H_1$ is given in Eq.~\eqref{eq:cusp}. (b) The long-scale kernel $K_1/2$, whose squared norm is $A_1$. Panels (a) and (b) use their respective rescaled position coordinates. (c) The solid curve is $A_t$ from Eq.~\eqref{eq:longkernel}. Dotted and dashed curves give the lower and upper bounds $C_0t^{3/2}/(1+2t)$ and $C_0t^{3/2}$, with $C_0=2(\sqrt2-1)/(3\sqrt{2\pi})$. The shaded region lies between them. All curves are deterministic numerical quadratures, not trajectory fits. Appendix~\ref{app:response} gives convergence estimates and a separate enclosure for $A_1$.}
\label{fig:response}
\end{figure*}

Figure~\ref{fig:response} displays the two response kernels and the long coefficient. At unit rescaled times, deterministic quadrature gives $B_1\simeq0.355$ and $A_1\simeq5.00\times10^{-2}$, with $A_1/B_1\simeq0.141$. We use these values for the weak-disorder reference curves in Fig.~\ref{fig:finite}. For the response coefficients defined by the ordered limits in Eq.~\eqref{eq:C2},
\begin{equation}
	\lim_{\gamma\downarrow0}a_\gamma
	\frac{\mathsf C_{\rm l}(\gamma;1)}{\mathsf C_{\rm p}(\gamma;1)}=\frac{A_1}{B_1}.
	\label{eq:ratio}
\end{equation}

\subsection{A memoryless random-force comparison}\label{sec:classical}
 To isolate the effect of coherent propagation, we compare the response to a local change in the return probabilities. The reference model matches both the clean spreading and the leading random-potential strength. We use a nearest-neighbor continuous-time chain with bond rates
\begin{equation}
	k_{j\to j+1}=D_\gamma \ee^{\alpha_\gamma b_j/2},\qquad
	k_{j+1\to j}=D_\gamma \ee^{-\alpha_\gamma b_j/2}.
	\label{eq:classicalreference}
\end{equation}
Here $D_\gamma=\nu_\gamma/(2\tau_\gamma)$ matches the clean diffusion constant of the coherent chain. The rates also match the leading accumulated potential of the coherent chain. To first order in the biases, Eq.~\eqref{eq:logincrement} gives $-\alpha_\gamma\sum_{j=0}^{n-1} b_j$ across an interval, up to endpoint terms. The reference assigns $-\alpha_\gamma b_j$ to each bond.

 In Eq.~\eqref{eq:classicalreference} the invariant ratio across bond $j$ is exactly $\ee^{\alpha_\gamma b_j}$. The potential increment is therefore $-\alpha_\gamma b_j$, and the clean generator is $D_\gamma\Delta_{\rm lat}$, where $\Delta_{\rm lat}$ is the nearest-neighbor lattice Laplacian. In the coordinate $x=a_\gamma j$, the clean variance at time $R/\gamma$ is
\begin{equation}
	\sigma^2_{{\rm cl},\gamma}=2D_\gamma a_\gamma^2R/\gamma\longrightarrow4R.
\end{equation}
Let $p_j^{\rm cl}(u)$ be the clean probability of reaching site $j$ from the origin at time $u$. Let $U^{\rm cl}(v,j)$ be the mean readout after a further time $v$ from site $j$. Differentiating both rates on bond $j$ gives
\begin{align}
	G_j^{\rm cl}(T;c)={}&\frac{D_\gamma\alpha_\gamma}{2}\int_0^T[p_j^{\rm cl}(u)+p_{j+1}^{\rm cl}(u)]\nonumber\\
	&\times[U^{\rm cl}(T-u,j+1)-U^{\rm cl}(T-u,j)]\dd u.
	\label{eq:exactclassicalgradient}
\end{align}
Both forward occupations enter with the same sign. Taking the clean heat-kernel limit gives $K_{\sigma^2_{{\rm cl},\gamma}}/2$, with environmental potential variance per unit $x$ equal to $\varepsilon^2\sigma^2\alpha_\gamma^2/a_\gamma$. Therefore
\begin{align}
\mathsf C_{\rm p}^{\rm cl}(\gamma;R)
&=\sigma^2\frac{\alpha_\gamma^2}{a_\gamma}[A_{4R}+o(1)],\nonumber\\
\frac{\mathsf C_{\rm p}^{\rm cl}(\gamma;R)}{a_\gamma}
&\longrightarrow\sigma^2A_{4R}.
\label{eq:classicalresponse}
\end{align}
In the coordinate $h_\gamma j$, the clean reference variance at $T_\gamma t$ is exactly $t$. Since $\alpha_\gamma^2/h_\gamma=1$, its long-scale response coefficient is $\sigma^2A_t$. These comparisons take the weak-disorder limit first. The reference has no unfinished packet, yet matches the clean limiting variance on both observation scales.

At $R=1$, $A_4\simeq0.167$ and $B_1\simeq0.355$, so that $B_1/A_4\simeq2.13$. The kernels in Fig.~\ref{fig:response}(a) also differ in shape: $H_1$ has the integrable logarithmic peak in Eq.~\eqref{eq:cusp}, while $K_4/2$ stays finite at the origin. The two kernels therefore cannot be made identical by an overall amplitude rescaling.

 We can probe the kernel difference by changing the return probabilities only in a chosen spatial region. Let $v$ be a deterministic, bounded, piecewise continuous profile that vanishes outside a finite interval. Set $b_j^{(\theta)}=\theta v(a_\gamma j)$, with $|\theta|\|v\|_\infty<1$. For fixed $R>0$, write $\Phi_{\gamma,v}(\theta)=F_{b^{(\theta)}}(R/\gamma;a_\gamma)$. We differentiate at $\theta=0$ at each fixed $\gamma$, with no environment average. Equation~\eqref{eq:gradient} gives
\begin{equation}
	\Phi_{\gamma,v}'(0)
	=\sum_jv(a_\gamma j)G_j(R/\gamma;a_\gamma)
	\longrightarrow\int_\R v(x)H_R(x)\dd x
	\label{eq:spatialprobe}
\end{equation}
as $\gamma\downarrow0$. In contrast, the reference response tends to $\tfrac12\int vK_{4R}$. For $v=\ind_{[-r,r]}$, take this limit at fixed $r$ before sending $r$ to zero. The coherent-chain response then has order $r\log(1/r)$, whereas the reference response has order $r$. At fixed $r$, the perturbation covers $O(r/a_\gamma)$ microscopic bonds. Both responses vanish as $r\downarrow0$, but with different small-$r$ dependence. This terminal measurement therefore distinguishes the two spatial susceptibilities in the stated order of limits. It distinguishes the specified memoryless reference, not all classical descriptions that retain age and displacement correlations.

\subsection{Connection to the Brownian-potential response}
The coefficient $A_t$ also describes the variation of the mean readout between weak Brownian potentials. This relation follows directly from the limiting diffusion. Let $U_\lambda(t)=\E_0^{\lambda W}\tanh Z_t$. As the potential amplitude $\lambda$ tends to zero, its first-order response converges in $L^2$ over Brownian environments:
\begin{align}
\frac{U_\lambda(t)}\lambda&\longrightarrow-\frac12\int_\R K_t(x)\dd W(x),\nonumber\\
\Var_W U_\lambda(t)&=\lambda^2A_t+o(\lambda^2).
\label{eq:roughresponse}
\end{align}
The kernel of the stochastic integral is deterministic and square integrable. Its variance is given by Eq.~\eqref{eq:longkernel}. A Brownian potential is not differentiable. We therefore derive the response from the scale function and speed measure, first for bounded potentials and then by controlling Brownian paths on growing finite intervals; see Appendix~\ref{app:response}.

 We can bound $A_t$ directly by expanding the square of $K_t$. With $G$ a standard normal variable and $q=uv/(u+v)$, this gives
\begin{align}
	A_t={}&\frac14\int_0^t\!\int_0^t p_{u+v}(0)\nonumber\\
	&\times\E[(P_{t-u}f')(\sqrt qG)(P_{t-v}f')(\sqrt qG)]\dd u\dd v.
\end{align}
For $f'=\sech^2$, the inequality $\sech^2x\ge\ee^{-x^2}$ and Gaussian integration give the lower bound $(1+2t)^{-1}$ for the expectation. Its upper bound is one. The remaining double Gaussian integral is elementary, giving
\begin{equation}
	\frac{C_0t^{3/2}}{1+2t}\le A_t\le C_0t^{3/2},\qquad
	C_0=\frac{2(\sqrt2-1)}{3\sqrt{2\pi}}.
	\label{eq:Abounds}
\end{equation}

The lattice response calculation first takes weak disorder at fixed dephasing. In the reverse order, we first take the low-dephasing limit at fixed admissible amplitude, and then take weak disorder in the resulting Brownian potential. Combining Eq.~\eqref{eq:roughresponse} with the fixed-amplitude law and Eq.~\eqref{eq:potential} gives
\begin{equation}
\lim_{\varepsilon\to0}\frac{\lim_{\gamma\downarrow0}\Var_\xi F_{\rm l}}{\varepsilon^2}
=\sigma^2A_t.
\label{eq:reverse}
\end{equation}
The two iterated limits, Eqs.~\eqref{eq:ordered} and~\eqref{eq:reverse}, give the same coefficient. This equality does not control an arbitrary simultaneous approach of $\varepsilon$ and $\gamma$ to zero. Such a result would require a finite-amplitude remainder bound uniform in $\gamma$, which is not established here.

The length and time scales used so far keep the disorder amplitude fixed. For weaker disorder, the particle must travel farther and evolve longer to probe a potential of order-one strength. Write $Z_t^\Sigma$ for diffusion in the potential $V=-\Sigma W$. Brownian scaling gives $Z_t^\Sigma\overset{d}{=}\Sigma^{-2}Z_{\Sigma^4t}^{1}$, where the equality includes both the environment and diffusion laws. The physical length and time corresponding to unit potential amplitude are therefore
\begin{equation}
\ell_{\gamma,\Sigma}=\frac1{h_\gamma\Sigma^2},\qquad
 t_{\gamma,\Sigma}=T_\gamma\Sigma^{-4}.
\label{eq:cost}
\end{equation}
For weak disorder, the additional length and time factors have orders $\varepsilon^{-2}$ and $\varepsilon^{-4}$. These factors are obtained by taking $\gamma\downarrow0$ at each fixed admissible amplitude; they do not specify a finite-size onset or an experimental lifetime. The covariance at unit potential amplitude is $C(s)=\Var_W\E_0^{-W}\tanh Z_s$. This differs from the weak-potential coefficient $A_s$. Appendix~\ref{app:response} uses the positivity of $C(s)$ and proves $C(s)\sim C_0s^{3/2}$ at short times. The numerical value of $A_1$ does not determine $C(1)$.

\section{Physical scope and conclusions}\label{sec:discussion}
The same frozen return biases can become invisible in one scaled terminal distribution and remain visible in another. In this model, neighboring ground states couple to the same auxiliary state with opposite signs, suppressing its excitation when the ground-state amplitudes vary slowly in space. Reducing ground-state dephasing allows longer coherent excursions, distributing a local bias over a wider region. The accumulated fluctuations vanish on the packet scale but survive on the long scale. On the packet scale, the correlated segment durations and displacements, together with the unfinished packet, combine to give a Gaussian distribution. On the long scale, elapsed-time fluctuations become small relative to the total observation time, and the unfinished packet is negligible at the observation length. The remaining motion is described by diffusion in the accumulated random potential.

Conditionally independent terminal measurements in the same frozen field provide an observable measure of this sample dependence. Their covariance is the variance of the conditional mean across fields. At fixed binary disorder, the three-rate calculation resolves a decrease in the packet covariance while the long-scale covariance remains statistically resolved. Changing the diffusion constant by the same amount in every field would preserve a centered distribution and a zero mean for the odd readout. It therefore cannot produce this covariance. Sample-dependent directional classical motion can produce it. The observable measures sensitivity to frozen disorder, not a uniquely quantum signature or a fixed-rate Sinai asymptotic.

Dissipation can select coherent many-body states~\cite{Diehl,Kraus}, including aligned spins~\cite{ZhangMagnetization} and a steady state with long-range pair coherence~\cite{ZhangSingleSite}. Reservoir coupling can also select slowly decaying modes in a correlated lattice~\cite{ZhangSlowSectors}. These examples concern state or mode selection. Here coherence controls how frozen return biases enter a spreading position distribution.

The long-scale process belongs to the established class of diffusion in a random potential. Dephasing-dependent transport~\cite{Plenio,Moix,Gopalakrishnan,Taylor,Longhi2024} and coherent changes to random transition rates~\cite{Shapira,Avnit} have earlier precedents. References \cite{Shapira,Avnit} also include independent site dephasing. Longhi's model has kinetic valleys and slow barrier crossings. In its dissipative expansion, coherent hopping adds a symmetric rate $2\mathcal J^2/\Gamma_n$ [Eqs. (37) and (42)--(44) of Ref. \cite{Longhi2026}]. Here $\mathcal J$ is the coherent hopping amplitude of that model, and $\Gamma_n$ is fixed by its dissipative hopping rates.

The additional result concerns the response to a spatially confined change of the return probabilities. At fixed return biases, destructive interference preserves a long-wavelength coherent mode as ground-state dephasing tends to zero. The correlations between segment duration and displacement, together with the unfinished packet, determine the kernel $H_R$, whose profile differs from the matched memoryless kernel $K_{4R}/2$. This changes the terminal response without changing the Sinai--Brox class of the long-scale limit \cite{Sinai,Brox,Seignourel}. The local input and return structure is related to the reciprocal Mott construction~\cite{NC}. Effective-operator methods can eliminate excited states in a weakly driven dissipative system~\cite{ReiterSorensen}. The present calculation instead retains both the ground and auxiliary populations in the measured distribution. A distinction between Born spread and sample variation also appears with dynamical noise \cite{LonghiOptics}. Here the field is frozen, and all dynamical histories are averaged before samples are compared.

The terminal-distribution results use the assumptions and convergence modes of Sec.~\ref{sec:scales}. The explicit binary result still requires sufficiently small dephasing, without a quantitative onset. The response comparison in Sec.~\ref{sec:response} takes the weak-disorder limit before the low-dephasing limit; no uniform remainder for an arbitrary joint limit is established. Within these conditions, two models can share the same clean Gaussian spreading while responding differently to local changes of the return probabilities. The complete terminal population retains this measurable effect of coherent motion.

\sloppy
\begin{acknowledgments}
X. Z. Zhang acknowledges support from the National Natural Science Foundation of China (Grant Nos. 12675024 and 12275193).
\end{acknowledgments}

\section*{Data availability}
The numerical data and code supporting the findings of this article are available from the corresponding author upon reasonable request.

\appendix
\fussy
\numberwithin{equation}{section}
\renewcommand{\theequation}{\thesection\arabic{equation}}
\section{Auxiliary-center correction to the endpoint variance}\label{app:renewal}
This appendix completes the derivation of the finite-rate coefficients in Sec.~\ref{sec:renewal}. We use the centered Fourier generator of Eq.~\eqref{eq:bloch}, with $\ell=\gamma/2$, $r=\kappa/2$, and $\Gamma=\ell+r$.

The spatial second moment involves two momenta, $q_\pm=q\pm k/2$. Define $c_\pm=2g\sin(q_\pm/2)$ and the integrated cross-mode matrix
\begin{equation}
\mathsf C(q,k)=\int_0^\infty\ee^{uA_\gamma(q_+)}e_1e_1^{\mathsf T}
\ee^{uA_\gamma(q_-)^{\mathsf T}}\dd u.
\end{equation}
Integration of the time derivative gives
\begin{equation}
A_\gamma(q_+)\mathsf C+\mathsf C A_\gamma(q_-)^{\mathsf T}=-e_1e_1^{\mathsf T}.
\end{equation}
In components,
\begin{align}
2\ell C_{11}-c_+C_{21}-c_-C_{12}&=1,\nonumber\\
\Gamma C_{12}&=c_+C_{22}-c_-C_{11},\nonumber\\
\Gamma C_{21}&=c_-C_{22}-c_+C_{11},\nonumber\\
2rC_{22}&=-c_+C_{12}-c_-C_{21}.
\label{eq:cross-mode-components}
\end{align}
Eliminating $C_{12}$ and $C_{21}$ leaves
\begin{align}
(2\ell\Gamma+c_+^2+c_-^2)C_{11}-2c_+c_-C_{22}&=\Gamma,\nonumber\\
-2c_+c_-C_{11}+(2r\Gamma+c_+^2+c_-^2)C_{22}&=0.
\end{align}
The determinant of this two-by-two system is $4\mathcal Z(q,k)$, where
\begin{align}
\mathcal Z(q,k)={}&\ell r\Gamma^2+2\Gamma^2g^2[1-\cos(k/2)\cos q]\nonumber\\
&+4g^4\sin^2(k/2)\sin^2q.
\end{align}
Solving gives
\begin{align}
C_{11}&=\frac{\Gamma\{r\Gamma+2g^2[1-\cos(k/2)\cos q]\}}{2\mathcal Z(q,k)},\nonumber\\
C_{22}&=\frac{\Gamma g^2[\cos(k/2)-\cos q]}{\mathcal Z(q,k)}.
\end{align}
Ground resets have rate $\gamma=2\ell$. Auxiliary returns have rate $\kappa=2r$ and choose the two neighboring endpoints with equal probability. Their characteristic function is therefore
\begin{equation}
\widehat p(k)=\int_{-\pi}^{\pi}
[\gamma C_{11}(q,k)+\kappa\cos(k/2)C_{22}(q,k)]\frac{\dd q}{2\pi}.
\end{equation}
Substitution gives the deficit of the endpoint characteristic function,
\begin{align}
	1-\widehat p(k)&=\sin^2(k/2)\int_{-\pi}^{\pi}
	\frac{2r\Gamma g^2+4g^4\sin^2q}{\mathcal Z(q,k)}\frac{\dd q}{2\pi}.
	\label{eq:endpointsymbol}
\end{align}
For a clean auxiliary return, averaging the two ground endpoints multiplies the auxiliary-center transform by $\cos(k/2)$. In position space, the same endpoint choice adds $1/4$ to the variance about the auxiliary center. The full second moment therefore contains an additional $\eta/4$, as well as the ground-reset and auxiliary-center contributions. Taking $-\widehat p''(0)$ in Eq.~\eqref{eq:endpointsymbol} gives
\begin{equation}
	\nu=\frac{4\kappa g^2}{(\gamma+\kappa)\mathcal D_\gamma}
	+\frac{\gamma\kappa+8g^2-\mathcal D_\gamma}{2(\gamma+\kappa)^2}.
\end{equation}
Rationalizing the second term gives the manifestly nonnegative expression in Eq.~\eqref{eq:constants} and avoids numerical cancellation when $\gamma\kappa\gg g^2$. These coefficients are obtained from the full generator with all auxiliary states retained.

\section{Physical random potential and transport}\label{app:potential}
The spatial limit and the rate of motion require separate arguments. The negative logarithm of the smoothed invariant weight must converge in distribution to a Brownian potential. The reset motion must also acquire the variance specified by the time scale in the main text. The first three subsections construct that potential at a sufficiently small fixed disorder amplitude. The explicit binary-amplitude bounds extend the construction to the parameters used in the simulations. The final subsection combines the spatial weights with the variance accumulated along the path to obtain the transport limit. Section~\ref{sec:clock} then connects the reset process to laboratory time and the complete terminal measurement.
\subsection{Positive invariant weight and the physical increment}
The potential is constructed from positive weights that are unchanged by one reset. On the infinite chain, these weights need not sum to one. We need them to be finite at each site and to grow slowly enough for the convolutions below to exist. Recurrence of the reset chain supplies the weights; their growth is controlled after that construction.

Let $\Omega$ be the probability space of the i.i.d. frozen field. We use $\ell^p$ for spatial sequence norms and $L^q=L^q(\Omega)$ for norms over the field law. The order of these norms is retained in every mixed space. Constants may depend on fixed $g,\kappa$ and the indicated exponents, but are uniform for sufficiently small positive $\gamma$. Our convolution convention is $(Cf)_j=\sum_n C_nf_{j-n}$. Let $\mathsf T f_j=f_{j+1}$ and $\Delta=\mathsf T-I$ denote the lattice shift and difference. We also write $\mathsf Qf=q*f$ for convolution with $q$ and $(Bf)_j=b_jf_j$ for multiplication by the field. The outgoing reset kernel, recalled from Eq.~\eqref{eq:reset}, is
\begin{equation}
	K_b(j,j+n)=p_n+\tfrac12[b_{j+n-1}q_{n-1}-b_{j+n}q_n],
	\label{eq:Bkernel}
\end{equation}
and $(1-d)p_n\le K_b(j,j+n)\le(1+d)p_n$, where $d=\varepsilon M<1$. For the clean endpoint law, the smoothing kernel and its Fourier transform satisfy
\begin{align}
	A_n&=\sum_{m>|n|}(m-|n|)p_m,\nonumber\\
	\widehat A&=\frac{1-\widehat p}{4\sin^2(k/2)},\qquad \sum_nA_n=\nu/2.
\end{align}
We construct the invariant weight on the infinite chain before introducing killing on a finite interval.

Telescoping over destinations gives
\begin{equation}
	K_b(j,[k,\infty))=\sum_{n\ge k-j}p_n+\tfrac12b_{k-1}q_{k-1-j}.
\end{equation}
The increase in this tail probability when the starting site moves one step to the right can be written explicitly. With $m=k-j$,
\begin{equation}
\begin{aligned}
K_b(j+1,[k,\infty))&-K_b(j,[k,\infty))\\
={}&\gamma Q_{m-1}+\frac{1+b_{k-1}}2q_{m-2}\\
&+\frac{1-b_{k-1}}2q_{m-1}\ge0.
\end{aligned}
\label{eq:reset-tail-order}
\end{equation} Quantile updates therefore preserve the order of two starting positions. The probability of escape to the right is nondecreasing in the starting site. It is also stationary under shifts of the field, so its nonnegative increments have zero mean and vanish almost surely. The escape probability is consequently independent of the starting site. Conditional expectation along a path identifies this constant with the indicator of eventual right escape. Shift ergodicity and bond reflection then exclude escape to either infinity. 

Escape to infinity is thus excluded, but the jumps can skip sites. We must also exclude paths that cross zero repeatedly without returning to it. The temporal resolvent of Eq.~\eqref{eq:bloch} expresses $p_n$, for $n\ge1$, as a positive mixture of $\rho^{n-1}$ with $\rho\le\theta_\gamma<1$. Each crossing then lands at zero with conditional probability at least $(1-d)(1-\theta_\gamma)/(1+d)$. This fixed positive bound prevents infinitely many crossings from all missing zero. The chain is therefore irreducible and recurrent at every fixed positive rate, for almost every frozen field.

Count the visits to site $j$ during an excursion from zero, before the first return to zero. Define the invariant weight by
\begin{equation}
r_j=\E_0^b\sum_{m<T_0^+}\ind_{\{J_m=j\}},\qquad T_0^+=\inf\{m\ge1:J_m=0\}.
\end{equation}
This weight is positive, finite at each site, and unique up to scale \cite{Norris}. Let $c_\gamma=(1-d)p_1>0$. Invariance and the two nearest-neighbor transition bounds give $r_{j+1}\ge c_\gamma r_j$ and $r_j\ge c_\gamma r_{j+1}$. Hence
\begin{equation}
c_\gamma\le\frac{r_{j+1}}{r_j}\le c_\gamma^{-1}.
\label{eq:invariant-ratio-bound}
\end{equation}
The logarithms of these ratios are therefore integrable at each fixed positive rate. Uniqueness makes the ratio field stationary under lattice shifts, and reflection makes its mean zero. Applying the ergodic theorem to $\log(r_n/r_0)=\sum_{j=0}^{n-1}\log(r_{j+1}/r_j)$ gives $\log r_n=o(|n|)$ in both directions. This step does not require a lower bound on $c_\gamma$ uniform in $\gamma$. Thus both $r$ and $1/r$ grow subexponentially. All convolutions used below are absolutely convergent when their exponential weights lie inside a smaller fixed-rate strip. The expected absolute crossing count in a return excursion is finite by these bounds. Zero net cut crossings then give Eq.~\eqref{eq:zerocurrent}.

Write $x=2Ar/\nu$ and $w=\mathsf Qr/\eta$ for the two smoothed weights. The smoothed weight used in the main text is $v_j=(x_j+x_{j+1})/2$, normalized by $v_0=1$. We use $\zeta_j=b_jw_j$ for the bias-weighted activity and $\vartheta_j=w_j/v_j$ for its local ratio. Then $a_j=b_j\vartheta_j$, and the exact convolution identities are
\begin{align}
	\Delta x&=\alpha\zeta,&w&=v+\mathcal T_\gamma\zeta,\nonumber\\
	\Delta v&=\alpha(I+\mathsf T)\zeta/2,&
	\mathcal T_\gamma\Delta A&=\mathsf Q/2-\alpha(I+\mathsf T)A/2.
	\label{eq:Bphysical}
\end{align}
The relative increment $u_j=\alpha a_j/2$ and its telescoping logarithm are given in Sec.~\ref{sec:mechanism}, Eqs.~\eqref{eq:relativeincrement} and~\eqref{eq:logincrement}.

\subsection{Bounds on the source and local weights}\label{app:source}
The physical increment depends on local ratios of the invariant weight. We compare it with a stationary auxiliary source whose spatial fluctuations can be controlled directly. Stationary here means that its statistics are unchanged by a lattice translation. The auxiliary source is not a population; it is introduced to estimate the accumulated physical bias. Constructing this source requires uniform bounds on the convolution operator below. A local comparison with the physical weights will then prepare the cumulative estimate in the next subsection.

Write $(\mathcal T_\gamma f)_j=\sum_nt_nf_{j-n}$. This real-valued, odd convolution has the multiplier
\begin{equation}
	\widehat{\mathcal T}_\gamma=-i\frac{\widehat q_c/\widehat A-2\alpha\cos(k/2)}{4\sin(k/2)},\qquad
	\widehat q=\ee^{ik/2}\widehat q_c,
	\label{eq:Bsymbol}
\end{equation}
with the removable value zero at $k=0$. Put $\delta=\sqrt{\gamma/\kappa}$, so $\alpha\asymp\delta$. The Sylvester equation for the full ground--auxiliary block gives
\begin{align}
|t_n|&\le\frac{C\ee^{-c\delta|n|}}{1+|n|},&
|t_{n+1}-t_n|&\le\frac{C\ee^{-c\delta|n|}}{1+n^2},\nonumber\\
q_n+|(A^{-1})_n|&\le\frac{C\ee^{-c\delta|n|}}{1+n^2},&
\sum_n n^2\ee^{c\delta|n|}p_n&\le C\delta^{-1}.
\label{eq:kernelbounds}
\end{align}
The third and fourth weighted moments are $O(\delta^{-2})$ and $O(\delta^{-3})$. To specify the kernel input, let $\ell=\gamma/2$, $r_d=\kappa/2$, $\Gamma_d=\ell+r_d$, $c_k=\cos(k/2)$, and $C_q=\cos q$. The exact spatial symbols are
\begin{align}
4\widehat A&=\int_{-\pi}^{\pi}\frac{2r_d\Gamma_dg^2+4g^4(1-C_q^2)}{\mathcal Z(q,k)}\frac{\dd q}{2\pi},\nonumber\\
\widehat q_c&=\kappa\int_{-\pi}^{\pi}\frac{\Gamma_dg^2(c_k-C_q)}{\mathcal Z(q,k)}\frac{\dd q}{2\pi},\nonumber\\
\mathcal Z(q,k)&=\ell r_d\Gamma_d^2+2\Gamma_d^2g^2(1-c_kC_q)\nonumber\\
&\quad+4g^4\sin^2(k/2)(1-C_q^2).
\label{eq:Bsymbols}
\end{align}
View the denominator in Eq.~\eqref{eq:Bsymbols} as a polynomial in $C_q$, and let $X_*$ be its root near $C_q=1$. Then $F:=X_*^2-1=\kappa\gamma/(4g^2)+k^2/4+O((\gamma+|k|^2)^2)$. Polynomial division separates the analytic terms from those containing $\sqrt F$ and $\gamma/\sqrt F$. On a smaller complex strip, their second derivatives have uniformly bounded integrals. The singular contribution is controlled by $\int_{\R}\gamma(\gamma+t^2)^{-3/2}dt=2$. Contour displacement and integration by parts give Eq.~\eqref{eq:kernelbounds}. The difference estimate for $t_n$ also uses Eq.~\eqref{eq:Bphysical}. Bounds on the positive moments follow from Cauchy estimates on $\widehat p-1=-4\sin^2(k/2)\widehat A$ and Tonelli on the nonnegative $\cosh$ series. This proves actual exponential moments without identifying analytic continuation with a convergent moment-generating series.

The kernel estimates bound the convolution uniformly as the dephasing rate decreases. For scalar sequences, split the sine series at distance $|k|^{-1}$. Oddness controls the cancellation in the kernel, while the difference estimate controls its change under a spatial shift. Together they give the uniform $\ell^p$ bounds.

For sequences that depend on the field, apply the scalar $\ell^q$ bound at each realization and then use Fubini. This gives the bound when the spatial and field exponents coincide. A vector-valued interval decomposition extends it to spatial exponent two. In that decomposition, the scalar bound controls the contribution near each interval and the kernel difference controls the contribution far from it. The resulting estimate is $\|\mathcal T_\gamma z\|_{\ell^2(\Z;L^q(\Omega))}\le C_q\|z\|_{\ell^2(\Z;L^q(\Omega))}$, with a constant independent of the small rate. Duality covers $q<2$. This estimate does not identify $\ell^2(\Z;L^q(\Omega))$ with $L^q(\Omega;\ell^2(\Z))$. In the first norm the field average is taken before the spatial sum; in the second their order is reversed.

For stationary functions of the field, averaging the spatial estimate over long shift orbits gives the stationary $L^p$ bound. Abel regularization defines the zero-rate operator by damping distant kernel terms and then removing that damping. To establish strong convergence on the field space, start with a bounded centered function $F$ that depends on finitely many field coordinates. Such functions are called cylinder functions. The covariance of $F$ with a lattice translate vanishes once the two sets of coordinates are disjoint. Consequently,
\begin{equation}
\|(\mathcal T_\gamma-\mathcal T_0)F\|_{L^2}^2
\le C_F\sum_n|t_{\gamma,n}-t_{0,n}|^2\longrightarrow0.
\label{eq:source-strong-convergence}
\end{equation}
The last step uses pointwise kernel convergence and the uniform bound $C/(1+|n|)$. Oddness makes both operators vanish on constants. Uniform stationary $L^p$ bounds, interpolation, and density of cylinder functions then extend the strong limit to the required field norms.

 The field acts by bounded multiplication. Choosing its amplitude small enough makes the inverse series converge in the stationary and mixed spaces. We choose $d$ so that these inverses exist through exponent $64$, and define
\begin{equation}
s_\gamma=(I-\mathcal T_\gamma B)^{-1}1,\qquad Z_{\gamma,j}=b_js_{\gamma,j},\qquad \E Z_{\gamma,j}=0.
\end{equation}
The auxiliary source $s$ need not be positive. To estimate its fluctuations, we compare two fields that differ at just one site. Let $c=b^{(i)}$ be the field obtained by replacing $b_i$ with an independent draw. Subtracting the source equations and transposing the resulting inverse gives
\begin{equation}
\sum_jz_j[Z_j(b)-Z_j(c)]=(b_i-c_i)s_i(b)[(I+\mathcal T_\gamma B_c)^{-1}z]_i.
\label{eq:resampling}
\end{equation}
The untransposed difference is $(I-B_c\mathcal T)^{-1}[(b_i-c_i)e_is_i(b)]$. We establish the identity for finite Neumann polynomials and pass to the infinite-chain limit using contraction in the mixed space. The relation $\mathcal T^*=-\mathcal T$ fixes the multiplication order under transposition. The rank-one resolvent identity also gives $s_i(b)s_i(-c)=s_i(c)s_i(-b)$.

The source values are correlated in space because each depends on the field beyond its own site. We reveal the field one site at a time and project the centered source onto the corresponding product-space martingale differences. Each difference has zero conditional mean given the previously revealed sites. Applying H\"older's inequality and the mixed-space inverse bound gives
\begin{equation}
	\left\|\sum_jz_jZ_{\gamma,j}\right\|_q\le C_qd\|z\|_2,\qquad 2\le q\le32.
	\label{eq:Blinear}
\end{equation}
Restricting a convolution to an interval does not commute with applying it. The commutator measures this boundary error, including the influence of sites outside the interval. For an interval $I$, the commutator $[\mathcal T,\ind_I]$ has squared coefficient sum $2\sum_n\min(|I|,|n|)t_n^2\le C\log(2+|I|)$. We apply Eq.~\eqref{eq:Blinear} rowwise and the mixed inverse to obtain
\begin{equation}
\left\|\sum_{j\in I}Z_{\gamma,j}-\sum_{i\in I}D_{\gamma,i}\right\|_2^2\le C\log(2+|I|),
\label{eq:source-remainder}
\end{equation}
where $D_{\gamma,i}=\E[(b_i-c_i)s_i(b)s_i(-c)\mid\mathcal F_i]$ and $\mathcal F_i=\sigma(\xi_j:j\le i)$. The commutator estimate includes replacements outside $I$. Exchanging the original and replacement variables makes $D_{\gamma,i}$ a stationary centered martingale difference. Strong convergence of the source gives $D_{\gamma,0}\to D_{0,0}$ in $L^2$. Doob's inequality then controls the partial sums of $D_{\gamma,i}-D_{0,i}$. For the limiting stationary sequence, ergodicity gives the conditional variance limit, and the available higher moment gives the Lindeberg condition \cite{Brown}. The fourth-moment increment estimate in Eq.~\eqref{eq:Blinear} gives tightness of the interpolated source sums. Together these steps yield
\begin{align}
	\alpha\sum_{j=0}^{\lfloor x/\alpha^2\rfloor-1}Z_{\gamma,j}&\Rightarrow\Sigma W(x),\nonumber\\
	\Sigma^2=\E D_{0,0}^2&=\varepsilon^2\sigma^2+O(\varepsilon^4)>0.
	\label{eq:BsourceCLT}
\end{align}

The source sum now has the required Brownian limit. It remains to identify that limit with the logarithm of the smoothed physical weight. We begin locally, fixing the overall normalization at one site. Exponentially decaying spatial weights let us apply the inverse to the physical solution on the infinite chain.

To compare the physical weights near site $j$, use the primitive $\mathcal H_j$ anchored by $(\mathcal H_jz)_j=0$. Its increments are $\Delta \mathcal H_jz=(z+\mathsf Tz)/2$. Set $S_j=\mathcal T+\alpha\mathcal H_j$ and $\phi_j(k)=\ee^{-\beta\delta|k-j|}$, with $0<\beta<c/16$. Conjugation by this weight changes $\mathcal T$ by a norm bounded by $C\sum_{n\ge1}\ee^{-c\delta n}(\ee^{\beta\delta n}-1)/n$. For the primitive, the geometric Hardy estimate gives
\begin{equation}
\|\phi_j\alpha\mathcal H_j\phi_j^{-1}\|_p\le
\frac{\alpha(1+\ee^{\beta\delta})}{2(1-\ee^{-\beta\delta})}\le H_*.
\end{equation}
Let $L_{\rm wt}$ be a bound for the fully conjugated $S_j$ for $6/5\le p\le64$. If $dL_{\rm wt}<1/2$, both $(I-S_jB)^{-1}$ and $(I-BS_j)^{-1}$ have norm at most two on these weighted spaces. With $v_j=1$, the physical equations read
\begin{equation}
	w=1+S_jBw,\qquad v=1+\alpha\mathcal H_jBw.
	\label{eq:Banchor}
\end{equation}
The subexponential physical solutions already belong to the weighted space, so the inverse can be applied to them. This gives $\|\phi_jw/v_j\|_p\le2\|\phi_j\|_p$. On $|k-j|\le\delta^{-1}$, the exact gradient also gives $|v_k/v_j-1|\le K_pd$. Choose $K_2d\le1/2$ to obtain a positive local lower bound. Chaining these intervals gives $v_k/v_j\le Ce^{2K_2d\delta|k-j|}$. Stationary averaging over the local interval then gives uniform moments of $\vartheta_0$ through exponent 64.

For $\mathfrak x=w-s$, $\mathfrak y=v-1$, exact subtraction gives
\begin{equation}
\mathfrak x=S_jB\mathfrak x+\alpha\mathcal H_jZ,\qquad \mathfrak y=\alpha\mathcal H_jZ+\alpha\mathcal H_jB\mathfrak x.
\end{equation}
A primitive row has squared coefficient norm at most $|k-j|$. Equation~\eqref{eq:Blinear} and weighted inversion imply
\begin{align}
\E\|\phi_j\mathfrak x\|_q^q+\E\|\phi_j\mathfrak y\|_q^q&\le C_q\delta^{q/2-1},\nonumber\\
\|\vartheta_0-s_0\|_r&\le C_r\sqrt\delta,\quad 2\le r\le16.
\label{eq:Blocal}
\end{align}
For the second estimate, write $\vartheta-s=(\mathfrak x-s\mathfrak y)/v$ and average on the interval where $v\ge1/2$. H\"older's inequality uses the exponents $r$ and $2r$ on that interval. This local comparison does not yet control the sum of errors over the longer observation window. The next subsection keeps the signs of those errors and bounds their cumulative sum.

\subsection{Positive cut quantity and signed replacement}
A local error bound alone does not control its sum over the growing observation window. Summing absolute errors would discard cancellations that are needed on that window. We instead derive the change in the logarithmic weight when one field variable is replaced. For this purpose, we compare two admissible fields using a bilinear sum over pairs of sites on opposite sides of cut $j$. The weighted tails make the sum absolutely convergent. Define
\begin{align}
	\mathcal Q_j(F,G)&=\sum_{\ell<j,k\ge j}t_{\ell-k}(F_\ell G_k+G_\ell F_k),\nonumber\\
	\mathcal W_j(b,c)&=x_j(b)x_j(-c)+\alpha\mathcal Q_j(\zeta(b),\zeta(-c)),\nonumber\\
	\mathcal F_j(b,c)&=\mathcal W_j(b,c)/[v_j(b)v_j(-c)].
	\label{eq:Bcutdefinition}
\end{align}
Equation~\eqref{eq:Bcutdefinition} fixes the additive constant in the cut quantity. Across the cut, the kernel entries are bounded by a weighted Hilbert matrix with entries $1/(m+n+1)$. Apply the weighted Schur test with row weights $(n+1)^{-r}$ to obtain
\begin{equation}
|\mathcal Q_j(F,G)|\le C_p^Q\|\phi_jF\|_p\|\phi_jG\|_{p'},\qquad p'=p/(p-1).
\label{eq:Bcutbound}
\end{equation}
The remaining exponential decay also permits the weight $\phi_j^2$. Moving the cut gives $\Delta\mathcal Q_j=F_j(\mathcal TG)_j+G_j(\mathcal TF)_j$. The midpoint product rule gives
\begin{equation}
\begin{aligned}
\Delta[x(b)x(-c)]_j={}&\alpha\zeta_j(b)v_j(-c)\\
&+\alpha\zeta_j(-c)v_j(b).
\end{aligned}
\label{eq:midpoint-product-rule}
\end{equation}
Adding $\alpha\Delta\mathcal Q_j$ replaces each $v$ in this expression by $w=v+\mathcal T\zeta$. Since $\zeta_j(b)=b_jw_j(b)$ and $\zeta_j(-c)=-c_jw_j(-c)$, we obtain
\begin{equation}
\Delta\mathcal W_j(b,c)=\alpha(b_j-c_j)w_j(b)w_j(-c).
\label{eq:cut-jump}
\end{equation}
Moreover $x_j/v_j=1-u_j\in[1-d,1+d]$, and the weighted square bounds give $|\alpha\mathcal Q_j|/[v_j(b)v_j(-c)]\le K_Wd^2$. Imposing both $d\le1/4$ and $K_Wd^2\le1/16$ proves
\begin{equation}
\tfrac12\le\mathcal F_j(b,c)\le\tfrac{13}8.
\label{eq:Bpositivecut}
\end{equation}
When $c=b$, the cut quantity is conserved. Its normalization therefore gives $v_j(b)v_j(-b)=\mathcal F_0(b,b)/\mathcal F_j(b,b)$. The positive bounds come from Eqs.~\eqref{eq:Bcutdefinition} and~\eqref{eq:Bpositivecut}, rather than from the difference equation alone.

For $c=b^{(i)}$, we normalize the compared smoothed weights at cut $j$ and put $\delta\zeta=\zeta(b)-\zeta(c)$. The physical response is
\begin{equation}
(I-B_bS_j)\delta\zeta=(b_i-c_i)e_iw_i(c).
\label{eq:Bphysicalinverse}
\end{equation}
Both physical solutions lie in the weighted domain before the inverse is applied. Their cut difference is
\begin{equation}
-\alpha(\delta\zeta)_jx_j(-c)/2+\alpha\mathcal Q_j(\delta\zeta,\zeta(-c)).
\label{eq:cut-weight-difference}
\end{equation} Choose $d$ small enough that $2K_2d<\beta/2$, so exponential decay remains after allowing for growth of the smoothed weights. The weighted inverse carries the change at the replaced site to the cut. Equation~\eqref{eq:Bcutbound} bounds the resulting bilinear term, and the positive cut bounds permit taking the difference of logarithms. The spatial norm of the unperturbed factor supplies the scale dependence in the cut estimate. The physical moment bounds then control the norm over fields. For $\mathfrak r_{i,j}=\log \mathcal F_j(c,c)-\log \mathcal F_j(b,c)$, these steps give
\begin{align}
\|\mathfrak r_{i,j}\|_q&\le C_{p,q}\alpha\delta^{-1/p'}\ee^{-\beta\delta|i-j|/2},\nonumber\\
\sum_{i\in\Z}\|\mathfrak r_{i,j}\|_q^2&\le C_{p,q}\delta^{2/p-1}.
\label{eq:Bboundary}
\end{align}
Here $p$ and $p'$ lie in $[6/5,64]$ and $2\le q\le64$; the sum includes every exterior site. To obtain the second estimate, square the first and sum the exponential over the replacement site. This sum is $\sum_i\ee^{-\beta\delta|i-j|}=O(\delta^{-1})$. Since $\alpha\asymp\delta$, the remaining factor is $\alpha^2\delta^{-2/p'-1}=O(\delta^{2/p-1})$.

Set
\begin{equation}
\begin{aligned}
L_{m,n}&=\log(v_n/v_m),\\
\rho_i&=\log[\mathcal W_{i+1}(b,c)/\mathcal W_i(b,c)].
\end{aligned}
\label{eq:cut-bilinear-definitions}
\end{equation} Only the cut at the replaced site contributes an interior term; changes at the interval ends remain as boundary terms. The single jump in Eq.~\eqref{eq:cut-jump} gives the exact finite-change identity
\begin{equation}
L_{m,n}(b)-L_{m,n}(c)=\ind_{[m,n)}(i)\rho_i+\mathfrak r_{i,n}-\mathfrak r_{i,m}.
\label{eq:Bcocyclechange}
\end{equation}
Exchanging the original and replacement coordinates changes the sign of $\rho_i$; the exterior decay justifies this comparison at fixed rate. Its conditional projection $\Delta_i^{\rm phys}=\E[\rho_i\mid\mathcal F_i]$ is therefore a martingale difference. Projecting over all replacement sites, with $6/5\le p<2$, yields
\begin{equation}
\left\|L_{m,n}-\sum_{i=m}^{n-1}\Delta_i^{\rm phys}\right\|_q\le C_{p,q}\delta^{1/p-1/2}.
\label{eq:Binterval}
\end{equation}
For $p=6/5,q=8$, a union over $O_R(\alpha^{-2})$ endpoints bounds the eighth moment of the maximum by $C_R\delta^{2/3}\to0$.

We next compare the local cut with its limiting value. Equation~\eqref{eq:Blocal} gives $\|\|\phi_j^2(\zeta-Z)\|_2\|_{2q}=O(1)$ for the weighted difference between $\zeta$ and $Z$. Each separate physical or auxiliary weighted norm is $O(\delta^{-1/2})$. Subtracting the bilinear terms in Eq.~\eqref{eq:Bcutbound} and multiplying by $\alpha$ therefore gives $O(\sqrt\delta)$. A source at distance $m$ contributes a row with coefficient norm $Ce^{-c\delta m}/\sqrt m$. Equation~\eqref{eq:Blinear}, followed by summation over these rows, gives the same order. The endpoint product contributes $O(\alpha)$. Thus
\begin{equation}
	\|\mathcal F_j(b,b)-1\|_q+\|\mathcal F_i(b,b^{(i)})-1\|_q\le C_q\sqrt\delta,\quad 2\le q\le16.
	\label{eq:Bcutlocal}
\end{equation}
At the changed site the exact response is
\begin{equation}
\rho_i=\log\!\left[1+\alpha(b_i-c_i)\frac{\vartheta_i(b)\vartheta_i(-c)}{\mathcal F_i(b,c)}\right].
\label{eq:Brho}
\end{equation}
Equation~\eqref{eq:Bpositivecut} and the ratios of neighboring smoothed weights keep the argument of the logarithm away from zero. We can therefore expand Eq.~\eqref{eq:Brho} with a controlled Taylor remainder. The physical eighth moments, local comparison, and Eq.~\eqref{eq:Bcutlocal} give $\Delta_{\gamma,0}^{\rm phys}/\alpha-D_{\gamma,0}\to0$ in $L^2$. Doob's inequality transfers this local response comparison to partial sums over $O(\alpha^{-2})$ sites. Combining it with Eqs.~\eqref{eq:source-remainder} and~\eqref{eq:Binterval} gives
\begin{equation}
\max_{|n|\le R/\alpha^2}\left|\log(v_n/v_0)-\alpha\sum_{j=0}^{n-1}Z_{\gamma,j}\right|\xrightarrow{\Pp}0.
\label{eq:Blogreplacement}
\end{equation}
Over the full observation window, the cubic remainder in Eq.~\eqref{eq:logincrement} has expectation $O(\alpha)$. The fourth-moment bound makes the largest endpoint term vanish. Subtracting these errors from Eq.~\eqref{eq:Blogreplacement} gives
\begin{equation}
\alpha\max_{|n|\le R/\alpha^2}\left|\sum_{j=0}^{n-1}(a_j-Z_{\gamma,j})\right|\xrightarrow{\Pp}0.
\label{eq:signedreplacement}
\end{equation}
This estimate controls signed sums and retains the cancellations between local errors. It proves $V_\gamma(h_\gamma j)=-\log v_j\Rightarrow-\Sigma W$ without a surviving deterministic drift.

All conditions use finitely many norm endpoints. If $L_{\rm aux}$ bounds the stationary and mixed operator constants at exponents 2 and 64, a sufficient fixed choice is
\begin{align}
d_*<\min\{&1/4,(2L_{\rm aux})^{-1},(2L_{\rm wt})^{-1},\nonumber\\
&(2K_2)^{-1},\beta/(8K_2),(4\sqrt{K_W})^{-1},\nonumber\\
&\sigma^2/(32M^2C_2^T)\}.
\label{eq:Bthreshold}
\end{align}
Here $C_2^T$ is the stationary $L^2$ operator bound. The last condition keeps the limiting variance positive. Pairing $D_{0,0}$ with $\xi_0$ gives a leading term $\varepsilon\sigma^2$, with an error at most $\varepsilon\sigma^2/2$. All constants in this choice are independent of $\gamma$ for all sufficiently small $\gamma$.

\subsection{Fixed binary amplitude and explicit weighted constants}\label{app:explicit}
For $g/\kappa=1/10$ and $d=1/5$, the following constants replace the small-amplitude threshold. These bounds must justify the inverses and moment estimates used above at this specified amplitude. The available moment exponent is lower than in the general small-amplitude proof, so the cumulative comparison will require a coarser spatial mesh. We begin by separating the dependence on the random signs from the spatial operator bounds.

Expand a function of independent symmetric signs as $F=\sum_S\widehat F(S)\prod_{j\in S}\xi_j$, where $S$ runs over finite subsets of $\Z$. Each product describes dependence on a finite set of random signs, and the empty product is one. This is the Walsh expansion used below. For $P\ge2$, we define
\begin{equation}
\|F\|_{\HH_P}^2=\sum_{\substack{S\subset\Z\\ |S|<\infty}}(P-1)^{|S|}|\widehat F(S)|^2.
\end{equation}
Bonami's inequality gives $\|F\|_P\le\|F\|_{\HH_P}$ \cite{Bonami}. Multiplication by $\xi_j$ has norm $\sqrt{P-1}$ in this Hilbert space. A deterministic spatial $\ell^2$ operator acts on each Walsh coefficient and retains its spatial operator norm on $\ell^2(\HH_P)$. The contraction estimate must therefore use the weighted norm of the anchored spatial operator. This norm controls the spatial response, while the factor from sign multiplication controls its moments over the disorder.

Put $\lambda^2=4F(\gamma,0)$. In a fixed small-momentum neighborhood, Eq.~\eqref{eq:Bsymbols} gives the following local complex-strip expansion for $|\Im k|<b'\lambda$ with $b'<1$:
\begin{align}
\widehat q_c/\widehat A&=2\sqrt{k^2+\lambda^2}+\varrho_\gamma(k),\nonumber\\
\varrho_\gamma''&=O_{b'}(1),\qquad\lambda=\alpha+O(\alpha^2),
\label{eq:Bnormalform}
\end{align}
where $\varrho_\gamma$ is even and $\varrho_\gamma(0)=2(\alpha-\lambda)=O(\gamma)$. The square root in Eq.~\eqref{eq:Bnormalform} describes only the small-momentum neighborhood. We retain the exact lattice symbols elsewhere. Under $k\mapsto k+2\pi$, the centered auxiliary symbol $\widehat q_c$ changes sign. The physical symbol $\widehat q=\ee^{ik/2}\widehat q_c$ and the combinations in Eq.~\eqref{eq:Bsymbol} are periodic. Outside the local neighborhood, these periodic combinations have uniform analytic bounds on the shrinking complex strip. Split the anchored primitive into its two half-line pieces. Let $P_+$ project onto $j\ge0$ and $P_-$ onto $j<0$. Define $V_+$ to have kernel 1 for $n>0$ and $1/2$ at zero. Define $V_-$ to have kernel $-1$ for $n<0$ and $-1/2$ at zero. Then
\begin{equation}
\mathcal H_0=P_+V_+P_++P_-V_-P_--\tfrac12\mathbf1\otimes\delta_0.
\end{equation}
Here $\mathbf1_j=1$ at every site, and $(\mathbf1\otimes\delta_0)f=f_0\mathbf1$. Thus the last term subtracts half the value at the anchor from every component. After conjugation by $\phi_j=\ee^{-\beta\alpha|j|}$, its diagonal combination with $\mathcal T$ has symbol $-i\widehat q_c/(4\widehat A\sin(k/2))$ on $\Im k=\pm\beta\alpha$. The real pole is not used. The limiting diagonal profile is $-iC(z)$ with $C(z)=\sqrt{1+z^2}/z$; the cross blocks use $-iB(z)$ with $B(z)=(\sqrt{1+z^2}-1)/z$. On $\Im z=\beta<1$, $\sup|B|\le1$ and
\begin{equation}
C_\beta:=\sup|C|=
\begin{cases}\sqrt{1-\beta^2}/\beta,&\beta^2\le1/3,\\
[4\beta^2/(4\beta^2-1)]^{1/4},&\beta^2\ge1/3.
\end{cases}
\end{equation}
The diagonal blocks have norm at most $C_\beta+o(1)$. The combined off-diagonal block contributes $1+o(1)$, while the anchor correction contributes $(\alpha/2)\sqrt{\coth(\beta\alpha)}=o(1)$. The exact symbol at the remaining frequencies converges uniformly to its bounded zero-rate limit. Adding these bounds gives
\begin{equation}
\limsup_{\gamma\downarrow0}\|\phi S_0\phi^{-1}\|_2\le1+C_\beta.
\label{eq:Bweightedconstant}
\end{equation}
At $\beta=19/20$ this is $1+(361/261)^{1/4}<2.085$. For sufficiently small rates, we may therefore take $C_S=21/10$. With $P=25/4$, the actual contraction and inverse are
\begin{align}
dC_S\sqrt{P-1}&=21\sqrt{21}/100<1,\nonumber\\
J_P&=(1-21\sqrt{21}/100)^{-1}<26.6.
\label{eq:Bexplicitconstant}
\end{align}
The physical solution belongs to the weighted $\ell^2$ space before inversion. Pathwise uniqueness therefore identifies the Walsh series with that solution, giving
\begin{equation}
\sum_j \ee^{-1.9\alpha|j|}\|w_j/v_0\|_{\HH_P}^2\le J_P^2\coth(19\alpha/20).
\label{eq:Bexplicitmoment}
\end{equation}
At environmental exponent two, the inverse norm is bounded by $50/29$. For $|j|\le\lfloor(10\alpha)^{-1}\rfloor$, the gradient bound tends to $d(50/29)\ee^{0.095}\sqrt{0.1/0.95}<0.124$. Hence $v_j/v_0\ge1/2$ on this interval for sufficiently small $\alpha$. Stationary averaging then gives uniform $P$th moments of $\vartheta_0$. Applying the forcing equations to each Walsh coefficient gives the local comparison for exponents through $r=P/2$.

At small rates, $\|\mathcal T_\gamma\|_2\le1.1$. The stationary Walsh inverse for the source is then $[1-1.1d\sqrt{q-1}]^{-1}$. Its deterministic-test bound follows from resampling at exponents $2q$. 

The commutator row depends on the same field as its source. A bound for a prescribed deterministic input therefore cannot be applied directly to this row. We keep that field dependence by expanding the source itself. Expand the source in its complex amplitude on the circle of radius $\rho=0.49$. The $n$th coefficient has Walsh degree at most $n$. Cauchy's coefficient estimate and Bonami's inequality give a convergent series through $q=P$, since $d\sqrt{P-1}<\rho$ and $1.1\rho\sqrt3<1$. The convergent series controls the random source that actually enters the commutator row. It does not assume an inverse for arbitrary $\ell^2(L^P)$ inputs.

A changed bias also affects sites outside the observation interval. To bound this distant contribution, we need spatial bounds at exponents $3/2$ and 3. Write $\widehat{\mathcal T}=-i\mathfrak a_\gamma$, where $\mathfrak a_\gamma$ is real and odd. Define $\mu_\gamma(\dd\theta)=[1/(2\pi)+\mathfrak a_\gamma'(\theta)/2]\dd\theta$ on the circle. For $n\ne0$, integration by parts gives its Fourier coefficient $\pi nt_n$. Hence $\mathcal T=\int U_\theta \mathsf H U_\theta^{-1}\mu_\gamma(\dd\theta)$, with $(U_\theta z)_j=\ee^{ij\theta}z_j$ and $\mathsf H_n=1/(\pi n)$ for $n\ne0$. The discrete Hilbert-transform norm \cite{BK} then gives $\|\mathcal T\|_p\le\|\mu_\gamma\|_{\rm TV}\cot[\pi/(2p^*)]$, where $p^* =\max(p,p')$. At $r=g/\kappa\le1$, a sufficient limiting variation bound is
\begin{align}
M_r&=1-2z_*+\frac8\pi\arctan z_*+2[1-(1+8r^2)^{-1/2}],\nonumber\\
z_*&=\sqrt{4/\pi-1}.
\end{align}
To obtain it, set $u=\sin(k/2)$ for $0<k<\pi$. The exact zero-rate symbols are
\begin{align}
q_{c,0}(k)&=\sqrt{\frac{1-u}{1+u+16r^2u^2}},\nonumber\\
\mathfrak a_0(k)&=\frac{u q_{c,0}(k)}{1-\cos(k/2)q_{c,0}(k)}
=\frac{\tan[(\pi-k)/4]}{1+f_r(u)},\nonumber\\
f_r(u)&=\frac{\sqrt{1+16r^2u^2/(1+u)}-1}{u}.
\end{align}
For $r\le1$, differentiation gives $f_r'\ge0$. The last factor in $\mathfrak a_0$ therefore decreases from 1 to $(1+8r^2)^{-1/2}$. Integration by parts bounds its contribution to the total variation. The base factor has one crossing at $a'_{\rm base}=-1/\pi$, which gives the first three terms of $M_r$. The rounded small-momentum layer supplies the initial 1. The conjugation bound for the screened reference kernel of $-iB$ is
\begin{equation}
D_\beta=\frac2\pi+\frac1\beta-\frac{2\sqrt{1-\beta^2}}{\pi\beta}(\pi/2+\arcsin\beta).
\end{equation}
Here the continuum reference kernel and its sampled discrete sequence are
\begin{align}
t(x)&=\frac{\operatorname{sgn}x}{\pi}\int_1^\infty\frac{\sqrt{s^2-1}}s \ee^{-s|x|}\dd s,\nonumber\\
t^{\rm ref}_{\lambda,n}&=\lambda t(\lambda n)\quad(n\ne0),\qquad t^{\rm ref}_{\lambda,0}=0.
\label{eq:discretereference}
\end{align}
The sequence is odd, with positive entries for $n>0$. Write its lattice symbol as $\widehat T_\lambda^{\rm ref}=-i\mathfrak a_\lambda^{\rm ref}$. To construct this periodic symbol, periodize $\lambda^{-1}B'(k/\lambda)$ and subtract its mean $1/\pi$. Integrating the result and fixing its value at the origin to zero gives
\begin{align}
\mathfrak a_\lambda^{\rm ref}(k)={}&B(k/\lambda)-k/\pi\nonumber\\
&+\sum_{m\ne0}[B((k+2\pi m)/\lambda)-B(2\pi m/\lambda)].
\label{eq:periodicreference}
\end{align}
Both the differentiated series and the paired image series converge normally. This regularized construction defines the lattice reference; it does not require an unregularized periodic sum of the nondecaying function $B$.

The reference kernel is useful only if its weighted tail agrees with that of the actual lattice kernel. We prove this agreement from the difference of their symbols. Keeping both symbols periodic is essential for the contour displacement below.

Let $E_\gamma=\mathcal T_\gamma-T_\lambda^{\rm ref}$. To expose the cancellation at zero, put $S(k)=\sqrt{k^2+\lambda^2}$ and $N_\gamma(k)=\varrho_\gamma(k)-\varrho_\gamma(0)$. Equations~\eqref{eq:Bsymbol} and \eqref{eq:Bnormalform} give the exact local identity
\begin{align}
\mathfrak a_\gamma(k)-B(k/\lambda)
={}&[S(k)-\lambda]\left[\frac1{2\sin(k/2)}-\frac1k\right]\nonumber\\
&+\frac{N_\gamma(k)}{4\sin(k/2)}+\frac\alpha2\tan(k/4).
\label{eq:Bquotientcancellation}
\end{align}
Since $\varrho_\gamma$ is even and $\varrho_\gamma''=O(1)$, we have $|N_\gamma(k)|\le C|k|^2$, $|N_\gamma'(k)|\le C|k|$, and $|N_\gamma''(k)|\le C$. Differentiate $N_\gamma/[4\sin(k/2)]$ twice. On $k=t\pm i\sigma\lambda$, with fixed $0<\sigma<1$, each quotient-rule term is bounded by $C_\sigma/(\lambda+|t|)$. The sine correction and tangent term have integrable second derivatives. The periodic image terms in Eq.~\eqref{eq:periodicreference} do as well. For the remaining frequencies, we use the exact periodic lattice symbol. These estimates give
\begin{equation}
	\int_{-\pi}^{\pi}|\widehat E_\gamma''(k\pm i\sigma\lambda)|\dd k
	\le C_\sigma(1+|\log\lambda|).
\end{equation}
Contour displacement, with matching periodic endpoints, gives for $n\ne0$
\begin{equation}
|t_{\gamma,n}-t^{\rm ref}_{\lambda,n}|
\le\frac{C_\sigma(1+|\log\lambda|)\ee^{-\sigma\lambda|n|}}{n^2}.
\end{equation}
Choose fixed $\beta<\sigma<1$. Since $\lambda/\alpha\to1$, exponential decay remains after multiplication by $\ee^{\beta\alpha|n|}$. Split the sum at $|n|=\alpha^{-1}$ to obtain
\begin{equation}
\sum_n|t_{\gamma,n}-t^{\rm ref}_{\lambda,n}|(\ee^{\beta\alpha|n|}-1)
=O(\alpha[1+|\log\alpha|]^2)\longrightarrow0.
\label{eq:weighteddiscretedifference}
\end{equation}
For the positive reference sequence, summing its geometric series and applying dominated convergence gives the same conjugation cost $D_\beta$ defined above
\begin{equation}
\sum_n|t^{\rm ref}_{\lambda,n}|(\ee^{\beta\alpha|n|}-1)
\longrightarrow\frac{2\beta}\pi\int_0^1\frac{\sqrt{1-u^2}}{1-\beta u}\dd u=D_\beta.
\end{equation}
Equation~\eqref{eq:weighteddiscretedifference} transfers this cost to the actual discrete kernel. The finite anchored primitive costs $1/\beta+o(1)$. Hence at $\beta=0.7$,
\begin{equation}
\limsup\|\phi S_0\phi^{-1}\|_{3/2,3}\le M_{0.1}\sqrt3+D_{0.7}+10/7<4.15.
\label{eq:BexplicitLp}
\end{equation}
Here $\|L\|_{3/2,3}:=\max_{p\in\{3/2,3\}}\|L\|_{\ell^p\to\ell^p}$ uses deterministic spatial norms. For sufficiently small rates, the operator norm is therefore at most 4.2. The pathwise inverse has norm at most $25/4$ in either multiplication order. 

The inverse bounds control the size of the solution, but do not by themselves make the mixed cut positive. We obtain positivity from a flux whose individual terms are nonnegative. Let $H_c(0)=0$ and $\Delta H_c=w(-c)$, and define
\begin{align}
C_j(b,c)=\sum_{m<j,n\ge j}\{&r_m(b)K_c(m,n)[H_c(n)-H_c(j)]\nonumber\\
&+r_n(b)K_c(n,m)[H_c(j)-H_c(m)]\}.
\label{eq:Bpositiveflux}
\end{align}
Every summand is nonnegative, and the exact normalization is
\begin{equation}
\mathcal W_j(b,c)=2C_j(b,c)/\nu.
\label{eq:Bfluxnormalization}
\end{equation}
To determine the additive constant, use the weaker weight $\ee^{-0.3\alpha|j|}$. Its operator bound tends to $1+\sqrt{91}/3<4.18$, so the pathwise contraction is below 0.84. Truncate both fields outside $[-N,N]$ and construct their anchored solutions in this weighted space. Set $x=v-\alpha\zeta/2$ and $r=(\nu/2)A^{-1}x$. The exact convolution identities give invariance and harmonicity for these solutions. Their algebraic truncations need not be positive for this step. Each truncated solution has constant tails with exponentially decaying corrections. In the left tail, the flux tends to the product of the two left constants times $\sum_{n\ge1}n^2p_n=\nu/2$. The cut correction tends to zero there, fixing Eq.~\eqref{eq:Bfluxnormalization} for every truncation.

Strong convergence of the multiplication operators and inverse series then passes to the original fields. After summing the harmonic increments, the cross-cut flux matrix decays as $Ce^{-0.8\alpha n}/(1+n)$. Dividing by the two $0.3\alpha$ weights leaves a square-summable matrix at fixed rate. Both bilinear forms therefore converge with the same additive constant.

Positivity alone does not give a lower bound uniform in the dephasing rate. We also need a definite amount of invariant weight on each side of the cut. This follows by bounding the mass in blocks of the coherent propagation length. The same inverse and the exact raw-weight relation
\begin{align}
r&=v+U\zeta,\qquad \widehat U(0)=-\alpha/2,\nonumber\\
\widehat U&=\frac\eta2\frac{\widehat A^{-1}-(1+\ee^{ik})/\nu}{\ee^{ik}-1}
\label{eq:Braw}
\end{align}
give weighted square norms $O(\alpha^{-1/2})$ for $w/v_j$, $v/v_j$, and $r/v_j$. To bound the mass in a block around $j$, pair the weights with $\chi_k=\ee^{-0.6\alpha|k-j|}$. Kernel cancellation gives $|\mathcal T^*\chi|+|U^*\chi|\le C\chi$ and $\mathsf Q^*\chi\le C\chi$. Using $v=w-\mathcal T\zeta$, $v=r-U\zeta$, $w=\mathsf Qr/\eta$, and $|\zeta|\le dw$, we obtain $\langle\chi,v\rangle\le C\langle\chi,z\rangle$ for $z=r,w$. The local positive lower bound gives $\langle\chi,v\rangle\ge cv_j/\alpha$. Weighted Cauchy--Schwarz bounds the contribution outside the block by $Cv_j\alpha^{-1}\ee^{-0.3R}$. Subtracting this tail gives the lower mass bound. The weighted square norm gives the upper bound. Thus, for one fixed sufficiently large $R$,
\begin{equation}
c\alpha^{-1}v_j\le\sum_{|k-j|\le R/\alpha}z_k\le C_R\alpha^{-1}v_j,\quad z=r,w.
\label{eq:Bblockmass}
\end{equation}
The clean temporal mixture gives $p_n\ge c_L\alpha^2$ for $|n|\le L/\alpha$. To obtain this bound, restrict its positive frequency integral to $|\omega|\le c\gamma$. In this interval, the interior root satisfies $1-|z|\asymp\alpha$, and the ground coefficient is of order $\alpha^{-1}$. Retaining two separated blocks in Eq.~\eqref{eq:Bpositiveflux} then gives $C_j\ge c/\alpha$. The weighted cross-cut estimate gives $C_j\le C/\alpha$. Hence
\begin{equation}
0<c_F\le\mathcal F_j(b,c)\le C_F<\infty.
\label{eq:Bbinarypositive}
\end{equation}
These constants are qualitative and uniform at small rates.

The weak inverse gives $v_{j+n}/v_j\le C\ee^{0.35\alpha|n|}$. The $0.7\alpha$ inverse in Eq.~\eqref{eq:BexplicitLp} thus leaves $0.35\alpha$ exterior decay. Equation~\eqref{eq:Bphysicalinverse} and the cut bound yield
\begin{align}
\|\mathfrak r_{i,j}\|_q&\le C_q\alpha^{2/3}\ee^{-0.35\alpha|i-j|},\nonumber\\
\left\|L_{m,n}-\sum_{i=m}^{n-1}\Delta_i^{\rm phys}\right\|_q&\le C_q\alpha^{1/6},\quad 2\le q\le P.
\label{eq:Bbinaryboundary}
\end{align}
The available moment exponent is too small for a union bound over all microscopic endpoints. Each endpoint error is small, but there are too many endpoints to sum their probability bounds directly. We instead use a mesh with spacing $L=\lfloor\alpha^{-1}\rfloor$, containing $O_R(\alpha^{-1})$ endpoints. At exponent $P$, the mesh-point error is $O(\alpha^{P/6-1})=O(\alpha^{1/24})$. Between adjacent mesh points, the source maximal estimate and local comparison bound the oscillations of both the physical sum and its martingale by $O(\sqrt\alpha)$ in $L^{r_0}$, where $r_0=P/2$. A union bound over these blocks gives $O(\alpha^{r_0/2-1})=O(\alpha^{9/16})$. The full-window cubic and endpoint errors vanish by the earlier estimates.

It is sufficient to show that the cut quantity approaches one locally. Subtract the bilinear terms and apply the local comparison bound. The source estimate for each row, used as in the derivation of Eq.~\eqref{eq:Bcutlocal}, then gives $\|\mathcal F_i(b,b)-1\|_1=O(\sqrt\alpha)$. Equation~\eqref{eq:Bbinaryboundary} transfers this result to a mixed replacement site. Boundedness in Eq.~\eqref{eq:Bbinarypositive} gives convergence of the cut quantity and its inverse in every finite $L^q$. Equation~\eqref{eq:Brho}, local comparison, and uniform moments for $P/2>2$ imply $\Delta_{\gamma,0}^{\rm phys}/\alpha\to D_{0,0}$ in $L^2$. Doob's inequality and the mesh argument prove Eq.~\eqref{eq:Blogreplacement} at this amplitude.

It remains to bound the limiting potential variance from below. Since $\|\mathcal T_0\|_2\le1$ and $\E s_0=1$, we have $\E|s_0-1|^2\le d^2/(1-d^2)$. Let $\mathcal J_0$ flip every field sign except the sign at zero. This transformation preserves the binary-field law. The independent replacement flips $\xi_0$ with conditional probability $1/2$; on that event, $-b^{(0)}=\mathcal J_0b$. Therefore
\begin{align}
\E[\xi_0D_{0,0}]&=d\E[s_0(b)s_0(\mathcal J_0b)]\ge d\left(1-\frac{d^2}{1-d^2}\right),\nonumber\\
\Sigma^2&\ge d^2\left(\frac{1-2d^2}{1-d^2}\right)^2=\frac{529}{14400},\quad d=1/5.
\label{eq:binaryvariance}
\end{align}
The physical replacement has already identified this source variance with the physical-potential variance. No source positivity is assumed. The next subsection uses these spatial estimates to determine the rate at which the reset particle moves through the limiting potential.

\subsection{Harmonic scale, mobility, and path convergence}\label{app:transport}
The potential limit describes the accumulated bias in space. To obtain the motion, we also need the variance acquired as the particle visits different sites. The harmonic coordinate has zero mean increment after each reset, so its accumulated variance sets the rate of motion. The exact factorization is
\begin{equation}
K_b-I=\mathsf T^{-1}[\Delta A+\mathsf QB/2]\Delta.
\label{eq:factor}
\end{equation}
Apply Eq.~\eqref{eq:zerocurrent} to the opposite field $-b$ and use $\mathsf Q^*=\mathsf T^{-1}\mathsf Q$. Together with Eq.~\eqref{eq:factor}, this gives $K_bH_\gamma=H_\gamma$, as claimed in Eq.~\eqref{eq:harmonic}. The calculation interchanges convolution operators only. Multiplication by the frozen field stays in its original order.

Positivity and recurrence make both scale endpoints infinite. Cut conservation relates the smoothed weights for opposite frozen fields. In the general small-amplitude case, the eighth-moment bound controls their maximum deviation from reciprocity. In the explicit binary case, bounded cut factors and local $L^1$ convergence suffice after averaging the activity. The harmonic increments use the weight for the opposite field. Cut conservation makes its smoothed weight on the long scale the reciprocal of the original one. Write $v_j^-=v_j(-b)$. For $n=\lfloor x/h_\gamma\rfloor$, separate the harmonic sum into its smoothed part and its local fluctuation:
\begin{equation}
\begin{aligned}
h_\gamma H_\gamma(n)={}&h_\gamma\sum_{j=0}^{n-1}v_j^-
+h_\gamma\sum_{j=0}^{n-1}[w_j(-b)-v_j^-],\\
v_j^-={}&\frac{\mathcal F_0(b,b)}{\mathcal F_j(b,b)v_j}.
\end{aligned}
\label{eq:harmonic-coordinate-decomposition}
\end{equation}
The source block bounds and local comparison make the second sum vanish on compact intervals. The cut relation turns the first sum into a Riemann sum for $\int_0^x\ee^{V_\gamma(y)}\dd y$. In the binary case the cut errors are controlled after averaging, as above. These estimates give
\begin{align}
\sup_{|x|\le R}\left|h_\gamma H_\gamma(x/h_\gamma)-\int_0^x\ee^{V_\gamma(y)}\dd y\right|&\xrightarrow{\Pp}0,\nonumber\\
(V_\gamma,h_\gamma H_\gamma(\cdot/h_\gamma))&\Rightarrow(V,S_V).
\label{eq:scalemass}
\end{align}
Here $S_V(x)=\int_0^x\ee^{V(y)}\dd y$. The second statement is joint convergence in distribution in the locally uniform topology. It does not assert convergence in probability to a Brownian potential on the original field space.

We also need the local mass of the unsmoothed invariant weight. The smoothing kernel $2A/\nu$ has first moment $O(\alpha^{-1})$, shorter than the macroscopic length $h_\gamma^{-1}$. Continuity of the smoothed weight controls the contribution from nearby sites. The distant tail decays exponentially faster than this weight can grow. Positivity then gives the local-mass convergence $h_\gamma\sum_jr_j\delta_{h_\gamma j}\Rightarrow \ee^{-V(x)}\dd x$. Equation~\eqref{eq:Braw} and the strong weighted bound also give $\sup_\gamma\|r_0/v_0\|_P<\infty$, with $P=25/4$ for binary disorder.

For a coordinate with zero mean increment, the second moment of each increment determines the variance growth. We must compare that local quantity with the limiting diffusion coefficient, and then control the error accumulated along the path.

Define $\mathcal A_\gamma(j)=\sum_nK_b(j,j+n)[H_\gamma(j+n)-H_\gamma(j)]^2$. The harmonic coordinate $H_\gamma$ is defined globally; its increments are estimated through the local ratios
\begin{align}
D_j(n)&=\frac{H_\gamma(j+n)-H_\gamma(j)}{v_j^-},\nonumber\\
W_j(n)&=\frac{w_{j+n}(-b)}{v_j^-},\qquad v_j^-=v_j(-b).
\end{align}
For binary disorder, choose the conjugate exponents $q=P/(P-1)=25/21$ and $r_b=2q=50/21$. The weak weighted Walsh inverse exists because $(0.2\times4.2)^2(r_b-1)=609/625<1$. Its norm is bounded by $(1-\sqrt{609}/25)^{-1}<78$. The corresponding forcing estimate gives $\|D_j(n)-n\|_{r_b}\le C\sqrt{|n|}\ee^{0.3\alpha|n|}$. Hence the squared-increment error is bounded in $L^q$ by $C[|n|^{3/2}\ee^{0.3\alpha|n|}+|n|\ee^{0.6\alpha|n|}]$. Positive weighted moments of $p$ make the normalized clean error $O(\nu^{-1/4})$. The exact routing correction is
\begin{equation}
\mathfrak R_j=\tfrac12\sum_n b_{j+n}q_nW_j(n)[D_j(n+1)+D_j(n)].
\end{equation}
The physical moment bounds allow growth with exponent $0.35\alpha|n|$ for each factor. This leaves decay against the $0.8\alpha$ kernel strip. The inverse-square $q$ tail gives $\|\mathfrak R_j\|_q\le C[1+|\log\alpha|]$. Thus
\begin{equation}
e_j=\left|\frac{\mathcal A_\gamma(j)}{\nu(v_j^-)^2}-1\right|,\qquad \|e_0\|_q\le C\alpha^{1/4}.
\label{eq:Bbracketerror}
\end{equation}
The general small-amplitude argument gives the same estimate with higher finite moments. Equation~\eqref{eq:Bbracketerror} controls an environmental moment at one site. The passage to the process limit below uses occupation times, not a maximum over microscopic sites. A small error averaged over fields does not by itself control the time spent at sites with larger errors. The next estimate bounds that time through the invariant weights.

To estimate visits to each site, we terminate the reset path when it leaves a finite interval. Its Green function counts the visits before exit, with a time discount. Bounding this function will convert local field estimates into an error accumulated along the actual trajectory.

Partition a lattice interval $\Lambda$ into complete consecutive blocks $B_m$ of size $L\asymp\alpha^{-1}$. Extend each test function by zero outside $\Lambda$. Let $K_b^\Lambda$ retain only destinations inside $\Lambda$, and let $\tau_\Lambda=\inf\{m:J_m\notin\Lambda\}$ be the exit time. Write $v_{\min}=\min_{j\in\Lambda}v_j$ and use the invariant-weight inner product $\langle z,w\rangle_r=\sum_jr_j\bar z_jw_j$. For a fixed $a>0$, set $\theta=\ee^{-a/n_\gamma}$. Exact invariance gives
\begin{align}
\mathcal E_\theta(z)&:=\Re\langle z,(I-\theta K_b^\Lambda)z\rangle_r\nonumber\\
&=\tfrac\theta2\sum_{i,j}r_iK_b(i,j)|z_i-z_j|^2+(1-\theta)\sum_i r_i|z_i|^2.
\label{eq:Benergy}
\end{align}
Equation~\eqref{eq:Benergy} retains pairs crossing the boundary in both directions. For each block, let $W_m=\sum_{i\in B_m}r_i$ be its mass and $U_m=W_m^{-1}\sum_{i\in B_m}r_iz_i$ its weighted average. Set $\mathsf Z_m=\sum_{i\in B_m}|z_i-U_m|^2$. The same-block double sum is $W_m\mathsf Z_m+L\sum_{i\in B_m}r_i|z_i-U_m|^2$. For neighboring blocks, introduce the unweighted average $\bar z_m=L^{-1}\sum_{i\in B_m}z_i$. Jensen's inequality gives
\begin{equation}
\begin{aligned}
\sum_{i\in B_m,j\in B_{m+1}}r_i|z_i-z_j|^2
&\ge W_mL|U_m-\bar z_{m+1}|^2,\\
L|\bar z_{m+1}-U_{m+1}|^2&\le\mathsf Z_{m+1}.
\end{aligned}
\label{eq:block-jensen-bound}
\end{equation}
The same-block term therefore controls the error in replacing $\bar z_{m+1}$ by $U_{m+1}$. The triangle inequality then controls the difference of neighboring weighted averages. The block mass lower bound fixes its coefficient. Together with $K_b(i,j)\ge c\alpha^2$ at these distances, this gives
\begin{align}
\mathcal E_\theta(z)\ge cv_{\min}\big[&\alpha\sum_m\mathsf Z_m+\sum_m|U_{m+1}-U_m|^2\nonumber\\
&+\alpha^2\sum_m|U_m|^2\big].
\end{align}
The zero blocks outside $\Lambda$ impose Dirichlet conditions at both ends. The coarse one-dimensional Green diagonal is $1/[\alpha\sqrt{\alpha^2+4}]$, so $|z_i|^2\le C\mathcal E_\theta(z)/(v_{\min}\alpha)$. Now take $z=(I-\theta K_b^\Lambda)^{-1}\delta_j$, whose energy is $r_jz_j$. The resolvent is nonnegative. Applying the pointwise bound first at $j$ gives $z_j^2\le Cr_jz_j/(v_{\min}\alpha)$ and hence $z_j\le Cr_j/(v_{\min}\alpha)$. Applying it again at $i$ yields
\begin{equation}
|z_i|^2\le\frac{Cr_jz_j}{v_{\min}\alpha}
\le C\left(\frac{r_j}{v_{\min}\alpha}\right)^2.
\label{eq:resolvent-point-bound}
\end{equation}
This gives $G_\theta(i,j)\le Cr_j/(v_{\min}\alpha)$; no symmetry of the Green function is required. On a finite time interval, removing the discount changes only the constant. Hence
\begin{equation}
\frac1{n_\gamma}\E_i^b\sum_{m<n_\gamma U\wedge\tau_\Lambda}f(J_m)
\le\frac{C_U\alpha^2}{v_{\min}}\sum_{j\in\Lambda}r_jf_j,\quad f\ge0.
\label{eq:Boccupation}
\end{equation}
The definitions $n_\gamma=(\nu\alpha^4)^{-1}$ and $\nu\alpha=\eta$ give the exact identity $1/(n_\gamma\alpha)=\nu\alpha^3=\eta\alpha^2$. Killing has not been replaced by reflection.

We apply the occupation bound where the potential stays bounded throughout the observation region. On this event the smoothed weight has a positive lower bound, so the spatially averaged error controls the trajectory average.

On $\mathcal C_\gamma(R,M)=\{\sup_{|x|\le R+1}|V_\gamma(x)|\le M\}$, take an interval made of complete blocks that contains $[-R,R]/h_\gamma$ and lies within $[-R-1,R+1]/h_\gamma$. Define
\begin{equation}
\mathfrak e_\gamma^b(R,U)=\E_0^b\sum_{m<n_\gamma U\wedge\tau_\Lambda}\frac1{n_\gamma}
\left|\frac{\mathcal A_\gamma(J_m)}\nu-\ee^{2V_\gamma(h_\gamma J_m)}\right|.
\label{eq:Bcumulative}
\end{equation}
Cut conservation bounds the smoothed weight $v^-$ for the opposite field. In Eq.~\eqref{eq:Boccupation}, write $r_j=v_j(r_j/v_j)$ and use the bounded smoothed weight on $\mathcal C_\gamma(R,M)$. Use the conjugate exponents $P=25/4$ and $q=25/21$ available for the binary amplitude. The higher moments in the general small-amplitude case also allow this choice. H\"older's inequality gives
\begin{equation}
\E_\xi[(r_j/v_j)e_j]
\le\|r_j/v_j\|_P\|e_j\|_q=O(\alpha^{1/4}).
\label{eq:occupation-holder-bound}
\end{equation}
For the remaining error, write $\mathcal F_j=\mathcal F_j(b,b)$. Cut conservation gives $|(v_j^-)^2-v_j^{-2}|\le C_M(|\mathcal F_0-1|+|\mathcal F_j-1|)$. The cut factors are uniformly bounded, and their $L^1$ error is $O(\sqrt\alpha)$. Therefore
\begin{equation}
\|\mathcal F_j-1\|_q
\le C\|\mathcal F_j-1\|_1^{1/q}
=O(\alpha^{21/50}).
\label{eq:cut-error-lq-bound}
\end{equation}
H\"older's inequality applies to these cut errors and $r_j/v_j$ even when they are correlated. Finally, the factor $\alpha^2$ in the occupation bound compensates for the $O(\alpha^{-2})$ sites in the observation window. Combining the two errors yields
\begin{equation}
\E_\xi[\ind_{\mathcal C_\gamma(R,M)}\mathfrak e_\gamma^b(R,U)]\le C_{R,U,M}(\alpha^{1/4}+\alpha^{21/50})\to0.
\label{eq:Bcumulativebound}
\end{equation}
No division by the event probability or independence of path and field is used.

 The weak deterministic inverse gives
\begin{equation}
|D_j(n)|^2\le C\alpha^{-1}|n|\ee^{0.6\alpha|n|}.
\label{eq:deterministic-weak-inverse}
\end{equation} For displacements $|n|>a/\alpha^2$, the contribution to the accumulated conditional variance is bounded by a polynomial factor times $\ee^{-ca/\alpha}$. For smaller displacements, Eq.~\eqref{eq:scalemass} controls the modulus of the harmonic coordinate. Taking $\gamma\downarrow0$ before $a\downarrow0$ gives the compactness and Lindeberg estimates, including jumps that leave the interval.

To see why cumulative control suffices, let $Y_\gamma(t)=h_\gamma H_\gamma(J_{\lfloor n_\gamma t\rfloor})$, stopped on a fixed interval in the harmonic coordinate. In this coordinate, called natural scale, the conditional drift vanishes. For any stopping time $0\le\vartheta\le U$ and $0\le s\le1$, optional sampling gives
\begin{equation}
	\E_0^b|Y_{\vartheta+s}^{\rm stop}-Y_\vartheta^{\rm stop}|^2\le C_M(s+2/n_\gamma)+\mathfrak e_\gamma^b(R,U+1).
	\label{eq:Bstoppingmodulus}
\end{equation}
 The expected contribution of any stopped random subinterval is bounded by the total nonnegative error $\mathfrak e_\gamma^b(R,U+1)$. This stopping-time estimate and the Lindeberg condition give tightness with continuous limits and vanishing exit overshoot. Expand a smooth test function along the harmonic-coordinate increments. Harmonicity cancels the linear term. For small jumps, the remainder is bounded by the modulus of continuity of the second derivative times the accumulated conditional variance. The Lindeberg estimate removes the large-jump remainder. Replacing that variance costs the cumulative error in Eq.~\eqref{eq:Bcumulative}. The final incomplete time cell contributes at most $C_M/n_\gamma+\mathfrak e_\gamma^b$. The limiting process is identified by its zero conditional drift and its local variance. In mathematical terms, each localized limit solves the natural-scale martingale problem with variance $\ee^{2V(S_V^{-1}(y))}$.

The limiting law is unique. Time-change Brownian motion using the inverse of $\int_0^u \ee^{-2V(S_V^{-1}(B_s))}\,\dd s$, and then map it back with $S_V^{-1}$. The resulting diffusion has scale density $\ee^V$ and speed density $2\ee^{-V}$. For a Brownian potential, both scale endpoints are infinite, and the diffusion is recurrent and nonexplosive. A bounded positive continuation after exit identifies the localized exit laws. Letting the interval grow then gives compact containment and removes the stopping.

The path argument applies once the spatial quantities converge and the accumulated errors vanish. For the original random environments, however, the potential and harmonic coordinate converge in distribution. We must therefore carry the errors along with them when passing to a limiting conditional path law.

The eventwise estimate also gives unconditional convergence in probability of each accumulated error. For fixed $R,U$ and $\epsilon>0$,
\begin{equation}
\begin{aligned}
\Pp_\xi(\mathfrak e_\gamma^b(R,U)>\epsilon)
\le{}&\Pp_\xi(\mathcal C_\gamma(R,M)^c)\\
&+\epsilon^{-1}\E_\xi[\ind_{\mathcal C_\gamma(R,M)}\mathfrak e_\gamma^b(R,U)].
\end{aligned}
\label{eq:field-exit-probability}
\end{equation}
At fixed $M$, the second term vanishes as $\gamma\downarrow0$. Tightness of the potential on compact intervals makes the limit superior of the first term vanish as $M\to\infty$.

To pass to random conditional laws, retain the raw i.i.d. field together with $(V_\gamma,h_\gamma H_\gamma)$. Adjoin the cumulative errors for integer radii and time horizons, and the Lindeberg errors for rational thresholds. These are measurable countable path sums or monotone limits. For any subsequence, take a further subsequence and represent these joint quantities on a probability space where they converge almost surely. This representation preserves their joint laws and the relations that define the errors. The error coordinates then vanish almost surely on that space. 

Choose rational $M$ above the limiting potential bound on the compact interval. The event used in Eq.~\eqref{eq:Bcumulativebound} then holds eventually. The deterministic martingale argument therefore applies to almost every represented environment. Uniqueness identifies the limit and yields convergence in distribution of the random conditional path laws. Section~\ref{sec:clock} transfers this convergence to laboratory time and the complete terminal Born measurement, retaining duration--displacement correlations.

\section{Packet centering in the original frozen field}\label{app:clock}
\subsection{Duration, displacement, and the unfinished packet}
The slow coherent mode in Eq.~\eqref{eq:bloch} gives the joint law used in Sec.~\ref{sec:packet}. We derive its two contributions before treating the frozen return biases. Throughout this appendix, $g,\kappa>0$ are fixed and $\gamma\downarrow0$.

At zero dephasing, the slow eigenvalue is $-D_*k^2+O(k^4)$, with $D_*=2g^2/\kappa$. Its ground component tends to one, while its auxiliary component is $-2gk/\kappa+O(k^3)$ in the centered Fourier convention. Fourier inversion and the exact identity $\mathcal U_{\gamma,\kappa}(t)=\ee^{-\gamma t/2}\mathcal U_{0,\kappa-\gamma}(t)$ therefore give the leading ground amplitude
\begin{equation}
G(x,t)\simeq\frac{\ee^{-\gamma t/2}}{\sqrt{4\pi D_*t}}
\exp\!\left[-\frac{x^2}{4D_*t}\right].
\label{eq:packet-ground-amplitude}
\end{equation}
This asymptotic holds on the diffusive spatial scale as $t\to\infty$, with $\gamma t$ fixed. The auxiliary amplitude is proportional to the spatial derivative of $G$. Consequently its jump density satisfies
\begin{equation}
\kappa p^E(x,t)\simeq\frac{x^2}{2D_*t^2}p^G(x,t).
\label{eq:packet-aux-density}
\end{equation}
The auxiliary half-site offset vanishes after multiplying position by $\sqrt\gamma$.

Put $z=\sqrt\gamma x$ and $u=\gamma t$. Squaring Eq.~\eqref{eq:packet-ground-amplitude} and summing over positions yields
\begin{align}
\frac{S_\gamma(u/\gamma)}{\sqrt\gamma}
&\longrightarrow \frac{c_*\ee^{-u}}{\sqrt{\pi u}},\nonumber\\
\mathcal L(z\mid u,\text{survival})&\longrightarrow\mathcal N(0,D_*u),
\label{eq:packet-survival-limit}
\end{align}
where $c_*=\sqrt\kappa/(4g)$. The auxiliary survival weight is smaller by a factor of order $\gamma$ at fixed $u>0$. Multiplying the ground population by its reset rate $\gamma$ gives the $u^{-1/2}$ term in Eq.~\eqref{eq:levymarks}. Equation~\eqref{eq:packet-aux-density} gives its $z^2/(2D_*u^{5/2})$ term. Both terms are needed even though the surviving auxiliary weight vanishes: auxiliary population decays at the fixed rate $\kappa$.

These limits first apply on duration intervals bounded away from zero. To extend them to the accumulated process, the short segments must contribute neither an omitted drift nor an additional diffusion term. For a clean segment with duration $W$ and endpoint displacement $J$, the sector estimates in Eq.~\eqref{eq:sectorbounds} give, for $0<a\le1$,
\begin{equation}
\limsup_{\gamma\downarrow0}\frac1{\sqrt\gamma}
\E\!\left[(\gamma W+\gamma J^2)\ind_{\{\gamma W\le a\}}\right]
\le C(\sqrt a+a^{3/2}).
\label{eq:packet-small-segments}
\end{equation}
For the duration term, use $\E[W\ind_{\{W\le t\}}]\le\int_0^t S_\gamma(v)\dd v$. For the displacement term, integrate $\gamma M_{2,G}(v)+\kappa M_{2,E}(v)$ up to $a/\gamma$. Choosing an auxiliary endpoint adds at most $1/4$ to the clean second moment. These integrals give the two powers of $a$ in Eq.~\eqref{eq:packet-small-segments}. The factor $\ee^{-u}$ controls large durations, and clean reflection symmetry removes the displacement drift. This establishes the joint limit specified by Eq.~\eqref{eq:levymarks}.

At measurement, the unfinished packet contributes its survival weight, given by Eq.~\eqref{eq:packet-survival-limit}. Its Laplace--Fourier transform is
\begin{equation}
\frac{c_*}{\sqrt\pi}\int_0^\infty
a^{-1/2}\ee^{-(1+s+D_*k^2/2)a}\dd a
=\frac{c_*}{\sqrt{1+s+D_*k^2/2}}.
\label{eq:packet-survival-transform}
\end{equation}
Combining this factor with the completed-segment process~\cite{StrakaHenry} gives Eq.~\eqref{eq:packet-transform}. It remains to show that the endpoint choices in a frozen field add no drift on this scale.

\subsection{Invariant environment at zero dephasing}\label{app:packet}
The exact zero-rate endpoint symbol is
\begin{equation}
\widehat p_0(k)=(1-s_k)
\sqrt{\frac{1+s_k}{1+s_k+16(g/\kappa)^2s_k^2}}.
\end{equation}
Here $s_k=|\sin(k/2)|$. Consequently
\begin{equation}
\sup_{0<|k|\le\pi}\frac{|\ee^{ik}-1|}{1-\widehat p_0(k)}=2.
\label{eq:resonance-multiplier}
\end{equation}
Indeed, the square-root factor is at most one, so $1-\widehat p_0(k)\ge s_k$; the ratio tends to two as $k\to0$.

Let $\mathbb P$ be the i.i.d. law of the field $b$, with $|b_j|\le d<1$. The field viewed from reset site $j$ is $\theta_jb$, where $(\theta_jb)_m=b_{j+m}$. On $L^2(\mathbb P)$, define the shift $\mathsf T F=F\circ\theta_1$, multiplication $\mathsf B F=b_0F$, and
\begin{equation}
\mathsf P_0=\sum_np_{0,n}\mathsf T^n,\qquad
\mathsf Q_0=\sum_nq_{0,n}\mathsf T^n.
\end{equation}
The shift is unitary and $\mathsf P_0$ is self-adjoint. Also, $\|\mathsf Q_0\|\le1$ because $\sum_nq_{0,n}=1$. Equation~\eqref{eq:reset} gives the environment transition operator and its adjoint:
\begin{align}
\mathsf K_0&=\mathsf P_0+\tfrac12\mathsf Q_0\mathsf B(\mathsf T-I),\nonumber\\
\mathsf K_0^*&=\mathsf P_0+\tfrac12(\mathsf T^{-1}-I)\mathsf B\mathsf Q_0^*.
\label{eq:environment-operator}
\end{align}
An invariant density $\pi$ of unit mean must satisfy $(I-\mathsf P_0)\pi=\tfrac12(\mathsf T^{-1}-I)\mathsf B\mathsf Q_0^*\pi$. The composite operator
\begin{equation}
\mathsf R=(I-\mathsf P_0)^{-1}(\mathsf T^{-1}-I),\qquad
\mathsf D=\tfrac12\mathsf R\mathsf B\mathsf Q_0^*
\end{equation}
is defined by its shift spectral multiplier, with value zero on constants. Equation~\eqref{eq:resonance-multiplier} gives $\|\mathsf R\|\le2$, hence $\|\mathsf D\|\le d$. The inverse of $I-\mathsf P_0$ alone need not be bounded. Since $\mathsf D$ has mean-zero range, the unique unit-mean $L^2$ solution is
\begin{equation}
\pi=(I-\mathsf D)^{-1}1=\sum_{m=0}^\infty\mathsf D^m1.
\label{eq:environment-density}
\end{equation}

Positivity requires a separate argument because $\mathsf D$ does not preserve positive functions. For $0<z<1$, replace $\mathsf D$ by $\mathsf D_z=\tfrac z2(I-z\mathsf P_0)^{-1}(\mathsf T^{-1}-I)\mathsf B\mathsf Q_0^*$. The same spectral bound gives $\|\mathsf D_z\|\le d$. The density $\pi_z=(I-\mathsf D_z)^{-1}1$ obeys
\begin{align}
(I-z\mathsf K_0^*)\pi_z&=(1-z)1,\nonumber\\
\pi_z&=(1-z)\sum_{m=0}^\infty z^m(\mathsf K_0^*)^m1\ge0.
\end{align}
The last equality follows from the positive Markov resolvent on $L^1(\mathbb P)$. The bounded spectral multipliers converge strongly as $z\uparrow1$, so $\pi_z\to\pi$ in $L^2$. Thus $\pi\ge0$ and $\E_{\mathbb P}\pi=1$.

Equation~\eqref{eq:kernelcomparison} implies $\mathsf K_0^*\pi\ge(1-d)\mathsf P_0\pi$. All $p_{0,n}$ are positive. If $\pi$ vanishes at a field, it therefore vanishes at every shifted field, up to null sets. The zero set is shift invariant. Ergodicity of the i.i.d. field law and unit normalization then imply $\pi>0$ almost surely. The invariant law $\pi\,\dd\mathbb P$ is equivalent to $\mathbb P$. The same positivity makes every invariant event of the environment process shift invariant, so this process is ergodic.

Write $q_\gamma(a)$ for the auxiliary-return probabilities. The bounded mean displacement due to the endpoint choice is
\begin{equation}
\beta_\gamma(j)=\tfrac12\sum_aq_\gamma(a)b_{j+a}.
\label{eq:endpoint-choice-drift}
\end{equation}
Reflection sends $b_j$ to $-b_{-j-1}$ and uses $q_0(a)=q_0(-a-1)$. It preserves the environment dynamics and $\mathbb P$, and reverses $\beta_0$. Uniqueness of the normalized density in Eq.~\eqref{eq:environment-density} makes its invariant mean zero. The Markov ergodic theorem therefore gives vanishing time averages of $\beta_0(J_m)$ for almost every original field, without redrawing that field on revisits.

\subsection{Comparison with positive dephasing}
The comparison must hold for the same field and for $O(\gamma^{-1/2})$ reset steps. We give the row estimate needed for this transfer. Put $\delta=\sqrt{\gamma/\kappa}$ and $m_n=1+|n|$. The zero-rate symbol has the expansion $\widehat p_0(k)=1-|k|/2+O(k^2)$. Its Fourier coefficients satisfy $p_{0,n}\sim(2\pi n^2)^{-1}$; positivity at the remaining finite sites gives
\begin{equation}
p_{0,n}\ge c\,m_n^{-2}.
\label{eq:zero-rate-row-lower}
\end{equation}
All constants in this subsection may depend on $g,\kappa,d$, but not on the field, site, or sufficiently small $\gamma$.

For completeness, let $G_n(t;\kappa')$ and $E_n(t;\kappa')$ be the zero-dephasing amplitudes with $\kappa/2\le\kappa'\le\kappa$, and put $v=\kappa t$. Fourier inversion of Eq.~\eqref{eq:bloch} yields
\begin{align}
|G_n(t;\kappa')|&\le C(1+v)^{-1/2}\ee^{-c\rho_n(v)},\nonumber\\
|E_n(t;\kappa')|+\kappa|\partial_{\kappa'}E_n(t;\kappa')|
&\le C(1+v)^{-1}\ee^{-c\rho_n(v)},\nonumber\\
\rho_n(v)&=\min\{n^2/(1+v),|n|\}.
\label{eq:pointwise-segment-bounds}
\end{align}
Near $k=0$, the slow eigenvalue is quadratic and the auxiliary amplitude has one factor of $k$. Their Gaussian Fourier integrals give the two powers of $1+v$. A derivative with respect to $\kappa'$ preserves these bounds: differentiating the exponential adds $vk^2$, which is bounded after rescaling $k$ by $(1+v)^{-1/2}$. Shifting the Fourier contour by an amount proportional to $\min\{|n|/(1+v),1\}$ gives the spatial exponential. Away from $k=0$, a uniform decay gap absorbs the remaining terms. The full matrix exponential controls coincident eigenvalues, so these estimates do not require a gap between its two eigenvalues.

The exact factor $\ee^{-\gamma t/2}$ and the change $\kappa\mapsto\kappa-\gamma$ now bound the difference of the integrated kernels. For the ground reset, Eq.~\eqref{eq:pointwise-segment-bounds} gives
\begin{equation}
\gamma Q_{\gamma,n}\le C\delta^2\int_0^\infty
\frac{\ee^{-\delta^2v-c\rho_n(v)}}{1+v}\dd v.
\end{equation}
For the auxiliary return, changing the exponential contributes $\min(\delta^2v,1)$, and changing $\kappa'$ contributes $O(\delta^2)$. Hence
\begin{equation}
|q_{\gamma,n}-q_{0,n}|\le C\int_0^\infty
\frac{\min(\delta^2v,1)+\delta^2}{(1+v)^2}
\ee^{-c\rho_n(v)}\dd v.
\end{equation}
Splitting the integrals at $v=m_n^2$ and $v=\delta^{-2}$ gives
\begin{align}
D_{\gamma,n}&:=\gamma Q_{\gamma,n}+|q_{\gamma,n}-q_{0,n}|\le C h_\delta(m_n),\nonumber\\
h_\delta(m)&=\begin{cases}
\delta^2[1+\log(1/(\delta m))],&m\le\delta^{-1},\\
m^{-2},&m>\delta^{-1}.
\end{cases}
\label{eq:row-difference-envelope}
\end{align}
The logarithm comes from the interval $m_n^2\lesssim v\lesssim\delta^{-2}$. Beyond the spatial cutoff, the zero-rate $m_n^{-2}$ tail bounds the difference. Direct summation gives
\begin{equation}
\sum_nD_{\gamma,n}\le C\delta,\qquad
\sum_nm_n^2D_{\gamma,n}^2\le C\delta.
\label{eq:row-envelope-sums}
\end{equation}
For the second sum below the cutoff, substitute $x=\delta m$ in $\delta^4\sum m^2[1+\log(1/(\delta m))]^2$; the integral of $x^2(1+|\log x|)^2$ is finite. Above the cutoff, $\sum_{m>\delta^{-1}}m^{-2}=O(\delta)$.

Equation~\eqref{eq:reset} bounds each row difference by $C(D_{\gamma,n}+D_{\gamma,n-1})$. Its denominator is bounded below by $(1-d)p_{0,n}$. Equations~\eqref{eq:zero-rate-row-lower} and \eqref{eq:row-envelope-sums} therefore give
\begin{equation}
\sup_{b,j}\sum_n\frac{[K_{\gamma,b}(j,j+n)-K_{0,b}(j,j+n)]^2}
{K_{0,b}(j,j+n)}\le C\sqrt\gamma.
\label{eq:row-chi-square}
\end{equation}
The first sum in Eq.~\eqref{eq:row-envelope-sums} also bounds row total variation and $\|\beta_\gamma-\beta_0\|_\infty$ by $C\sqrt\gamma$.

Let $\mathbb Q_{\gamma,n}^b$ and $\mathbb Q_{0,n}^b$ be the laws of the first $n$ reset steps from site zero in this same field. Their likelihood ratio is the product of the corresponding transition ratios. Conditional iteration of Eq.~\eqref{eq:row-chi-square} gives, for $n\le U/\sqrt\gamma$,
\begin{equation}
\E_0^b L_{\gamma,n}^2\le(1+C\sqrt\gamma)^n\le\ee^{CU+o(1)}.
\end{equation}
Here $L_{\gamma,n}=\dd\mathbb Q_{\gamma,n}^b/\dd\mathbb Q_{0,n}^b$ and the expectation is under the zero-rate path law. For any event $A$ determined by these steps, Cauchy--Schwarz gives
\begin{equation}
\mathbb Q_{\gamma,n}^b(A)\le\ee^{CU/2+o(1)}
[\mathbb Q_{0,n}^b(A)]^{1/2}.
\label{eq:path-law-transfer}
\end{equation}
Thus the zero-rate ergodic centering transfers to positive dephasing for almost every original field. Vanishing time averages imply that the maximal centered partial sum divided by $n$ also vanishes. Multiplication by $\sqrt\gamma$ therefore removes the accumulated $\beta_0$ on every bounded interval of rescaled jump count. Replacing it by $\beta_\gamma$ adds $O(U\sqrt\gamma)$. The endpoint-choice fluctuation about its conditional mean is a bounded-increment martingale; its scaled mean-square maximum is $O(U\sqrt\gamma)$.

The no-jump durations and ground or auxiliary-center displacements are independent of the field. Only the bounded choice of an auxiliary endpoint uses $b$. The preceding drift and martingale bounds therefore leave the joint limit in Eq.~\eqref{eq:levymarks} unchanged. Its increasing duration process gives a tight rescaled number of jumps at any fixed positive observation time, allowing the bound on $U$ to be removed. Adding the unfinished packet gives Eq.~\eqref{eq:packet-transform}, with variance $4R$ in the coordinate $a_\gamma X$. This proves the single-time terminal limit for almost every original field. It does not assign a classical unmeasured position trajectory to a coherent segment.

\section{Uniform response limits and Brownian-potential perturbation}\label{app:response}
\subsection{Strong spatial convergence of the response kernels}
The covariance depends on the spatial integral of the squared response. Convergence of the Fourier transform at each fixed momentum is therefore insufficient. Starting from Eqs.~\eqref{eq:renewalquotient}--\eqref{eq:occupation-symbols}, we derive bounds over the full momentum range and near both ends of the time integral.

For the long limit, let $(W,J)$ be the duration and endpoint displacement of one clean segment, with joint density given by Eq.~\eqref{eq:jointlaw}. Expanding its transform gives
\begin{equation}
1-\widetilde{\mathcal K}_\gamma(k,\lambda)
=\lambda\tau+k^2\nu/2+\mathcal \varrho_\gamma.
\end{equation}
The remainder is bounded by $\lambda^2\E W^2/2+k^4\E J^4/24+\lambda k^2\E(WJ^2)/2$. At the long-scale arguments of Eq.~\eqref{eq:occupation-symbols}, this remainder is smaller than the leading term $\nu h_\gamma^2$. We bound $\E(WJ^2)$ by Cauchy--Schwarz. The calculation retains the correlation between duration and displacement.

The covariance requires strong spatial convergence of the response kernels, beyond pointwise convergence of their symbols. We obtain a uniform lower bound for the renewal denominator from a positive integral. The Laplace weight adds $\lambda$ to both decay rates. Write $\Gamma_\lambda=(\gamma+\kappa+2\lambda)/2$ and, for the spatial integral, $c_k=\cos(k/2)$, $s_k=\sin(k/2)$, and $C_q=\cos q$. The cross-mode Sylvester equation gives
\begin{equation}
\begin{aligned}
D_\lambda(q,k)={}&(\gamma+\lambda)(\kappa+\lambda)\Gamma_\lambda^2
+8\Gamma_\lambda^2g^2(1-c_kC_q)\\
&+16g^4s_k^2(1-C_q^2),\\
N_\lambda(q,k)={}&\lambda\Gamma_\lambda[(\kappa+\lambda)\Gamma_\lambda
+8g^2(1-c_kC_q)]\\
&+s_k^2[4\kappa\Gamma_\lambda g^2+16g^4(1-C_q^2)],\\
1-\widetilde{\mathcal K}_\gamma(k,\lambda)
={}&\int_{-\pi}^{\pi}\frac{N_\lambda(q,k)}{D_\lambda(q,k)}\frac{\dd q}{2\pi}.
\end{aligned}
\label{eq:response-resolvent-definitions}
\end{equation}
Both numerator and denominator are positive. Fix a small $q_0>0$. On $|q|\le q_0$, for all $|k|\le\pi$ and sufficiently small positive $\gamma,\lambda$,
\begin{equation}
\begin{aligned}
D_\lambda(q,k)&\le C_1(\gamma+\lambda+k^2+q^2),\\
N_\lambda(q,k)&\ge c_1(\lambda+k^2).
\end{aligned}
\label{eq:response-resolvent-local-bounds}
\end{equation}
Here $C_1,c_1>0$ depend on fixed $g,\kappa,q_0$. The bounds use $1-c_kC_q\le C(k^2+q^2)$ and $s_k^2\ge k^2/\pi^2$. Integrating just over this small interval gives
\begin{equation}
1-\widetilde{\mathcal K}_\gamma(k,\lambda)
\ge c\frac{\lambda+k^2}{\sqrt{\gamma+\lambda+k^2}}.
\label{eq:response-resolvent-lower-bound}
\end{equation}
Taking $\lambda=\gamma z$ or $z/T_\gamma\asymp\gamma^2$, with fixed $z>0$, gives the two bounds over the entire lattice momentum interval:
\begin{align}
	1-\widetilde{\mathcal K}_\gamma(k,\gamma z)&\ge c_z(\sqrt\gamma+|k|),\nonumber\\
	1-\widetilde{\mathcal K}_\gamma(k,z/T_\gamma)&\ge c_z\frac{\gamma^2+k^2}{\sqrt\gamma+|k|}.
\end{align}
The absolute value of the single-segment auxiliary numerator is bounded by its zero-momentum integral, at most $1/\kappa$. Combining this fact with the denominator bounds and the exact scale factors gives $|\mathcal A^p_{\gamma,z}(p)|\le C_z/(1+|p|)$ and $|\mathcal A^l_{\gamma,z}(p)|\le C_z(1+\sqrt\gamma|p|)/(1+p^2)$ on their respective lattice momentum intervals. Extend each amplitude by zero outside that interval. These bounds give an integrable envelope for its square, so dominated convergence yields spatial $L^2$ convergence of the Laplace integrals at the lattice level.

The step-density transforms also contain the cell factors specified after Eq.~\eqref{eq:occupation-symbols}. Let $\delta_\gamma$ denote the relevant cell width, $a_\gamma$ or $h_\gamma$. On every bounded frequency interval, $\operatorname{sinc}(\delta_\gamma p/2)\to1$. Before zero extension, the centered lattice amplitude changes sign under $p\mapsto p+2\pi/\delta_\gamma$, while its magnitude is periodic. All Fourier copies are retained in the identity
\begin{equation}
\sum_{m\in\Z}\operatorname{sinc}^2(\delta_\gamma p/2+m\pi)=1.
\label{eq:sinc-copy-identity}
\end{equation}
By discrete Plancherel, the norm of the full step function equals the norm over the lattice interval. The bounded-frequency limit and convergence of these norms give strong $L^2$ convergence of the complete step-density transforms, including the copies outside that interval.

The occupations are positive, which lets us recover time-integrated estimates from their Laplace transforms. Approximate a continuous time weight by a finite sum of decaying exponentials. A positive Laplace integral bounds the approximation error. Approximate interval indicators from above and below in the same way. Finally, $\|e_r\|_2\le C_Rr^{-3/4}$ and $\|p_r\|_2=(4\pi r)^{-1/4}$ make both time-endpoint contributions integrable. This gives convergence of the time-integrated occupations without requiring a bounded microscopic density at the initial instant.

For the packet scale, the clean terminal law gives $d_j/a_\gamma\to\mathsf P_{2s}f'$ uniformly away from the time endpoints. For the long scale it gives $d_j/h_\gamma\to P_sf'$. Substitution into Eq.~\eqref{eq:gradient}, with the preceding norm estimates, yields $G_j/a_\gamma\to H_R$ and $G_j/\alpha\to K_t/2$ strongly in spatial $L^2$. The exact long prefactor is
\begin{equation}
\frac\kappa2T_\gamma f_{E,\gamma}h_\gamma^2=\frac\alpha2.
\end{equation}
Together with the cell widths, this gives the prefactors in Eq.~\eqref{eq:ordered}.

\subsection{Brownian-potential response}\label{app:brownian}
Write $P_t^Vf(x)=\E_x^Vf(Z_t^V)$ for the expectation under the diffusion defined by Eq.~\eqref{eq:scale-speed}. A Brownian potential is continuous but is neither differentiable nor bounded on the whole line. We therefore begin with a response bound for bounded potentials that uses no potential derivative. Its constants are independent of the support length, so the bound can later be applied on growing intervals. Let $w$ be bounded and continuous on $\R$, and set $M=\|w\|_\infty$. Here $\lambda$ is the potential amplitude, and we require $\theta_{\rm pot}=|\lambda|M\le1/10$. For $f\in C_c^\infty(\R)$, let $\delta_wf$ denote the first variation at zero potential. Then
\begin{align}
|P_t^{\lambda w}f(0)-P_tf(0)-\lambda \delta_wf(t,0)|
&\le50t^{-1/4}\lambda^2M^2\|f\|_2,\nonumber\\
|\delta_wf(t,0)|&\le11t^{-1/4}M\|f\|_2.
\label{eq:boundedpotentiallemma}
\end{align}
The right-hand norm is taken with respect to Lebesgue measure.

To derive the weak equation without differentiating $w$, use the harmonic coordinate
\begin{equation}
y=S_\lambda(x)=\int_0^x\ee^{\lambda w(r)}\dd r.
\label{eq:brownian-scale-coordinate}
\end{equation}
In this coordinate the generator is $a(y)\partial_y^2/2$, where $a(y)=\ee^{2\lambda w(S_\lambda^{-1}(y))}$. Put $\rho_\lambda=\ee^{-\lambda w}$. The mass weight $a(y)^{-1}\dd y$ becomes $\rho_\lambda(x)\dd x$. If $\widetilde u(y)=u(S_\lambda^{-1}(y))$, the energy transforms as
\begin{equation}
\frac12\int|\partial_y\widetilde u|^2\dd y
=\frac12\int\rho_\lambda(x)|u'(x)|^2\dd x.
\label{eq:brownian-energy-transform}
\end{equation}
Thus both the mass and the energy vary with the potential, even though no derivative of the potential is used.

For $\Re z>0$, define $u_\lambda(x)=\int_0^\infty\ee^{-zt}P_t^{\lambda w}f(x)\dd t$. This is the Laplace transform of the mean readout; $z$ is its complex Laplace parameter. We continue this resolvent to the sector used below. Let $H^1(\R)$ denote the space of square-integrable functions with a square-integrable first weak derivative. With $u=u_\lambda$ and a test function $v\in H^1(\R)$, the weak equation is
\begin{equation}
b_\lambda(u,v):=\int\rho_\lambda(zu\bar v+\tfrac12u'\bar v')\dd x
=\int\rho_\lambda f\bar v\dd x=:F_\lambda(v).
\label{eq:roughweakform}
\end{equation}
For $r=|z|>0$ and $|\arg z|\le3\pi/4$, use the energy norm $\|u\|_r^2=r\|u\|_2^2+\|u'\|_2^2/2$. Rotate the clean form by $\ee^{-i\arg(z)/2}$ to obtain coercivity constant $c=\cos(3\pi/8)$. The perturbation of the form is bounded by $\ee^{\theta_{\rm pot}}-1<c$. Its second-order remainder is bounded by $\theta_{\rm pot}^2\ee^{\theta_{\rm pot}}/2$. The right-hand-side perturbations satisfy the same estimates with an extra factor $r^{-1/2}\|f\|_2$. If $b_1$ and $F_1$ are obtained by inserting $-w$ and $u_1=b_0^{-1}(F_1-b_1u_0)$, then
\begin{align}
	\|u_0\|_r&\le c^{-1}r^{-1/2}\|f\|_2,\nonumber\\
	\|u_1\|_r&\le M(c+1)c^{-2}r^{-1/2}\|f\|_2.
\end{align}
For $\varrho_\lambda=u_\lambda-u_0-\lambda u_1$, exact subtraction gives
\begin{align}
b_\lambda \varrho_\lambda={}&F_\lambda-F_0-\lambda F_1\nonumber\\
&-(b_\lambda-b_0-\lambda b_1)u_0
-\lambda(b_\lambda-b_0)u_1.
\end{align}
For $\theta_{\rm pot}\le1/10$, this gives $\|\varrho_\lambda\|_r\le43\theta_{\rm pot}^2r^{-1/2}\|f\|_2$. For example, the constant is bounded by
\begin{equation}
\frac{\tfrac12\ee^{1/10}(1+c^{-1})
+10(\ee^{1/10}-1)(c+1)c^{-2}}
{c-(\ee^{1/10}-1)}<43.
\end{equation}
The Sobolev estimate $|u(0)|\le2^{-1/4}r^{-1/4}\|u\|_r$ gives a pointwise remainder $O(\theta_{\rm pot}^2r^{-3/4}\|f\|_2)$.

Invert the resolvents along $\arg z=\pm3\pi/4$. Their absolute integral is bounded by
\begin{equation}
\pi^{-1}2^{-1/4}43\int_0^\infty \ee^{-tr/\sqrt2}r^{-3/4}\,dr<50t^{-1/4}.
\label{eq:laplace-inversion-bound}
\end{equation} At a fixed connecting radius $a>0$, the spectral theorem gives inversion in $H^1$, so point evaluation is legitimate. The pointwise contribution of the connecting arc is $O(a^{1/4})$ and vanishes as $a\downarrow0$. Weighted self-adjointness identifies the inverse transform with the scale-speed semigroup. No gap at zero frequency is assumed. Applying the same inversion to $u_1$ gives the constant 11 in Eq.~\eqref{eq:boundedpotentiallemma}.

For a smooth $w$, the mass variation cancels the $wu_0''$ term and leaves $-w'u_0'/2$. Thus, with the kernel $K_t^f$ defined as in Eq.~\eqref{eq:longkernel},
\begin{equation}
\delta_wf(t,0)=\tfrac12\int_\R w(y)(K_t^f)'(y)\dd y.
\label{eq:roughderivative}
\end{equation}
For bounded, uniformly continuous $w$, approximate $w$ uniformly by smooth functions. Since $K_t^f$ vanishes at infinity and $\|(K_t^f)'\|_1\le2\sqrt{2t/\pi}\|f'\|_\infty+t\|f''\|_\infty$, the first-variation formula passes to the limit. The clipped Brownian potentials used below are bounded and uniformly continuous.

For $0<|\lambda|<\ee^{-1}$ choose
\begin{equation}
R=[\log(1/|\lambda|)]^2,\qquad
E_R=\{\sup_{[-R,R]}|W|\le R\}.
\label{eq:brownianlocalization}
\end{equation}
Reflection on the two half-lines gives $\Pp(E_R^c)\le8\ee^{-R/2}$. On $[-R,R]$, clip the values of $W$ to $[-R,R]$ and extend the result outside the interval by its endpoint constants. Call this potential $w_R$. It satisfies $\|w_R\|_\infty\le R$ and agrees with $W$ throughout the interval on $E_R$. If the potential is bounded there by $\theta_{\rm pot}$, each scale endpoint has magnitude at least $Re^{-\theta_{\rm pot}}$. The variance rate of the stopped natural-scale process is at most $\ee^{2\theta_{\rm pot}}$. Brownian time change therefore gives
\begin{equation}
\Pp_0^V(\tau_R\le t)\le4\exp[-R^2\ee^{-4\theta_{\rm pot}}/(2t)].
\label{eq:roughexit}
\end{equation}
This uses no bounded-drift assumption or information about the exterior potential.

Choose an even smooth cutoff $\chi$ with $0\le\chi\le1$, equal to one on $[-2,2]$ and zero outside $[-3,3]$. Set $f_R(x)=\chi(x/R)\tanh x$. Then $\|f_R\|_2\le\sqrt{6R}$ and $P_tf_R(0)=0$. On $E_R$, the two diffusions agree up to their exit from $[-R,R]$. Their terminal means differ by at most twice the exit bound in Eq.~\eqref{eq:roughexit}. On $E_R^c$, the difference is at most two. The resulting environmental $L^2$ errors are smaller than every fixed power of $|\lambda|$, even after division by $\lambda$. Since $|\lambda|R\le1/10$ eventually, Eq.~\eqref{eq:boundedpotentiallemma} gives the explicit growing-domain bound
\begin{equation}
\|P_t^{\lambda w_R}f_R(0)-\lambda \delta_{w_R}f_R(t,0)\|_{L^2(W)}
\le C_t\lambda^2R^{5/2}.
\label{eq:growingremainder}
\end{equation}
On $E_R$, the constant exterior extensions in Eq.~\eqref{eq:roughderivative} cancel the two integration-by-parts boundary terms. Hence $\delta_{w_R}f_R=-\tfrac12\int_{-R}^{R}K_t^{f_R}\dd W$. On $E_R^c$, Eq.~\eqref{eq:boundedpotentiallemma} bounds this derivative by $C_tR^{3/2}$. The Wiener integral with deterministic kernel has a bounded fourth moment. These estimates bound the environmental $L^2$ discrepancy by $C_t[R^{3/2}\Pp(E_R^c)^{1/2}+\Pp(E_R^c)^{1/4}]$. The cutoff remains one up to $2R$, so Gaussian tails make $K_t^{f_R}\ind_{[-R,R]}-K_t$ exponentially small in spatial $L^2$. The Wiener isometry and Eq.~\eqref{eq:growingremainder} then give, for $f=\tanh$,
\begin{equation}
	P_t^{\lambda W}f(0)=-\frac\lambda2\int_\R K_t(y)\dd W(y)+o_{L^2}(\lambda).
	\label{eq:brownian-response}
\end{equation}
The $L^2$ norm of the remainder is $O_t(\lambda^2[\log(1/|\lambda|)]^5)$. Reflection makes its odd-readout environmental mean zero. For $V=-\lambda W$, the first variation changes sign and the variance remains $\lambda^2A_t+o(\lambda^2)$.

The short-time covariance requires a test function that changes with the scaling parameter. Set $\lambda=s^{1/4}$ and $f_\lambda(x)=\lambda^{-2}\tanh(\lambda^2x)$. In the environment law, Brownian scaling gives
\begin{equation}
P_s^W[\tanh](0)\overset{d}{=}\lambda^2P_1^{\lambda W}f_\lambda(0).
\label{eq:brownian-readout-scaling}
\end{equation} The global bound $\|f_\lambda\|_\infty=\lambda^{-2}$ grows as $\lambda\to0$. However, $|f_\lambda(x)|\le|x|$ and $\|f_\lambda'\|_\infty\le1$ uniformly. For $f_{\lambda,R}=\chi(x/R)f_\lambda(x)$, these bounds give $\|f_{\lambda,R}\|_2\le\sqrt{18}\,R^{3/2}$ and a uniformly bounded derivative.

Applying Eq.~\eqref{eq:boundedpotentiallemma} and dividing by $\lambda$ leaves a remainder $O(\lambda R^{7/2})$. Before division by $\lambda$, the good-event exit error and the environmental $L^2$ error on $E_R^c$ are bounded by $C\lambda^{-2}[\ee^{-cR^2}+\Pp(E_R^c)^{1/2}]$. They remain smaller than every power of $\lambda$ after division as well.

On the exceptional event, the first variation is bounded by $CR^{5/2}$. Multiplication by $\Pp(E_R^c)^{1/2}$ makes this contribution vanish faster than every power. Thus the growing global test bound and the localized first variation are controlled separately.

On $|y|\le R$, $0\le u\le1$, the derivative of the heat-evolved test function obeys
\begin{equation}
	|(P_uf_{\lambda,R})'(y)-1|\le C\lambda^4(R^2+1)+C\ee^{-R^2/2}.
\end{equation}
On the plateau this follows from $1-f_\lambda'(z)=\tanh^2(\lambda^2z)\le\lambda^4z^2$; the rest is a Gaussian tail. Thus the first-variation kernel tends in $L^2$ to $J(y)=\int_0^1p_u(y)\dd u$. In the rescaled-environment representation,
\begin{equation}
\lambda^{-1}P_1^{\lambda W}f_\lambda(0)\longrightarrow
-\tfrac12\int_\R J(y)\dd W(y)\quad\hbox{in }L^2(W).
\end{equation}
The variance tends to $\tfrac14\int_0^1\int_0^1p_{u+v}(0)\,du\,dv=C_0$. Restoring the Brownian scaling factors gives $C(s)\sim C_0s^{3/2}$. This is convergence of environmental moments, not an almost-sure short-time expansion for one original Brownian sample. The unit-potential covariance $C(s)$ remains distinct from the weak-potential coefficient $A_s$.

\subsection{Quadrature and precision}
The packet cusp is resolved using logarithmic spatial and time coordinates. The two representations
\begin{align}
(\mathsf P_{2s}f')(x)&=\int_0^\infty\frac{k\ee^{-2sk^2}\cos(kx)}{\sinh(\pi k/2)}\dd k,\nonumber\\
(\mathsf P_{2s}f')(x)&=\E\sech^2(x+2\sqrt s\,G)
\end{align}
give consistent ordinary floating-point quadratures. Separate bounds control the omitted region near the origin, the large-$|x|$ tail, the short-time interval, and the Fourier tail. The estimates used in the paper are $B_1\simeq0.355$, $A_1\simeq5.00\times10^{-2}$, and $A_1/B_1\simeq0.141$.

The same positive Gaussian-product representation gives $A_4\simeq0.167$. Setting $u=x^2$ and $v=x^2z$ removes the corner singularity. We use positive Gauss--Legendre and Gaussian quadratures with $(40,80)$, $(64,128)$, and $(96,192)$ nodes. The successive values of $A_4$ differ by less than $2.7\times10^{-7}$ and $5.0\times10^{-9}$. These differences measure numerical convergence, not a rigorous enclosure. Figure~\ref{fig:response}(a) uses the independently evaluated kernel $K_4/2$ on the packet coordinate.

We obtain a separate enclosure for $A_1$ from its positive Gaussian-product representation. Define
\begin{equation}
\mathcal H(a,b,q)=\E[(P_af')(\sqrt qG)(P_bf')(\sqrt qG)].
\label{eq:gaussian-product-definition}
\end{equation}
The two time variables enter symmetrically. On the triangle $v\le u$, the substitution $u=x^2$, $v=x^2z$ has Jacobian $2x^3$. Including both triangles gives
\begin{equation}
\begin{aligned}
A_1={}&\frac1{\sqrt{2\pi}}\int_0^1\!\int_0^1\frac{x^2}{\sqrt{1+z}}\\
&\times\mathcal H\left(1-x^2,1-x^2z,\frac{x^2z}{1+z}\right)\dd z\dd x.
\end{aligned}
\label{eq:A1-square-integral}
\end{equation}
This removes the corner singularity. At $q=0$, derivatives are taken through the heat operator, $\partial_qP_q=P_q\partial_x^2/2$, rather than by differentiating $\sqrt q$.

Direct differentiation of $f=\tanh$ gives the bounds
\begin{equation}
\begin{aligned}
\|f'\|_\infty&\le1,&\|f''\|_\infty&\le1,&\|f^{(3)}\|_\infty&\le2,\\
\|f^{(4)}\|_\infty&\le5,&\|f^{(5)}\|_\infty&\le16.
\end{aligned}
\label{eq:tanh-derivative-bounds}
\end{equation}
Heat evolution preserves these bounds. Hence $|\mathcal H|\le1$, $|\mathcal H_a|,|\mathcal H_b|\le1$, and $|\mathcal H_q|\le3$. The second derivatives satisfy $|\mathcal H_{aa}|,|\mathcal H_{bb}|\le4$, $|\mathcal H_{ab}|\le1$, $|\mathcal H_{aq}|,|\mathcal H_{bq}|\le7.5$, and $|\mathcal H_{qq}|\le24$. For example, the mixed derivative $\mathcal H_{aq}$ is one quarter of a heat average of a product with derivative orders $(4,0)$, twice $(3,1)$, and $(2,2)$ on the two evolved copies of $f'$. Its bound is $(16+10+4)/4=7.5$.

Let $B=x^2/\sqrt{1+z}$ and evaluate $\mathcal H$ at the three arguments in Eq.~\eqref{eq:A1-square-integral}. Their first and second $x$ derivatives are bounded componentwise by $(2,2,1)$. The chain rule gives $|\mathcal H_x|\le7$ and $|\mathcal H_{xx}|\le131$. Since $|B_x|\le2$ and $|B_{xx}|\le2$, we obtain $|\partial_x^2(B\mathcal H)|\le161$. In the other direction, $|\mathcal H_z|\le4$, $|\mathcal H_{zz}|\le49$, $|B_z|\le1/2$, and $|B_{zz}|\le3/4$ give $|\partial_z^2(B\mathcal H)|\le53.75$. Applying the one-dimensional midpoint error in each direction bounds the $N\times N$ rule by
\begin{equation}
\frac{161+53.75}{24\sqrt{2\pi}N^2}<\frac{3.58}{N^2}.
\label{eq:midpoint-error-bound}
\end{equation}

Each Gaussian expectation uses a positive trapezoid rule with step $h=1/5$ and nodes $jh$, $-45\le j\le45$. For an analytic integrand $F$ with $\int|F(y\pm i/2)|\dd y\le M_F$, contour displacement of its Fourier coefficients and Poisson summation give~\cite{TrefethenWeideman}
\begin{equation}
\left|h\sum_{j\in\Z}F(jh)-\int_\R F(y)\dd y\right|
\le\frac{2M_F}{\ee^{\pi/h}-1}.
\label{eq:trapezoid-error-bound}
\end{equation}
For an inner heat average, $|\sech^2(x+iy)|\le\sec^2y$ gives the bound $M_F\le\ee^{1/8}\sec^2(1/2)$. For the outer Gaussian average, $q\le1/2$ gives $M_F\le\ee^{1/8}\sec^4(1/(2\sqrt2))$. Truncating the nodes at $|jh|\le9$ adds less than $3\times10^{-19}$. Each inner and outer error is therefore below $e_G=4.5\times10^{-7}$. The positive quadrature weights sum to less than $1.16$. Propagating the two inner errors through their product and the outer rule gives
\begin{equation}
\frac{e_G+1.16(2e_G+e_G^2)}{\sqrt{2\pi}}<10^{-6}.
\label{eq:gaussian-product-error-bound}
\end{equation}
This bound holds at every midpoint, so it also bounds their positive weighted average.

At $N=64$, outward-rounded binary64 arithmetic encloses the finite quadrature rule in $[5.003,5.004]\times10^{-2}$. Adding the analytic errors in Eqs.~\eqref{eq:midpoint-error-bound} and~\eqref{eq:gaussian-product-error-bound} and rounding outward gives
\begin{equation}
A_1\in[4.91,5.10]\times10^{-2}.
\label{eq:Aenclosure}
\end{equation}
The arithmetic assumption is correctly rounded binary64 basic operations and square root, with outward expansion. The interval evaluation avoids library exponential and hyperbolic functions. For $0\le z\le64$, it evaluates $\ee^{-z/128}$ with a degree-16 Taylor polynomial and remainder below $2.2\times10^{-20}$, followed by seven outward-rounded squarings. The identity $\sech^2x=4r/(1+r)^2$, with $r=\ee^{-2|x|}$, then supplies the integrand. Raw quadrature values and interval endpoints are stored at full machine precision in the data files. The three-significant-figure value is supported by the quadrature convergence checks; the enclosure including analytic errors has the wider interval stated above.

\begin{table}[t]
\caption{Simulation parameters and sample sizes at resonance on the infinite chain. Both terminal times are evaluated on each path.}
\label{tab:design}
\begin{ruledtabular}
\begin{tabular}{@{}lc@{}}
Quantity & Value\\
Rate unit, coupling & $\kappa=1$, $g=0.1$\\
Routing & $b_j=\pm0.2$, equal probabilities\\
Rescaled times & $R=t=1$\\
High rate & $\gamma_H=2.5\times10^{-4}$\\
Middle and low rates & $\gamma_M=\gamma_H/4$, $\gamma_L=\gamma_H/16$\\
Independent fields & $512$\\
Independent batches per field & $2$\\
Paths per batch at $H,M,L$ & $256,256,192$\\
Total production paths & $720\,896$\\
Refinement paths / fields & $1024/8$\\
\end{tabular}
\end{ruledtabular}
\end{table}
\section{Numerical error bounds}\label{app:statistics}

The tabulation grid is uniform in $s=\log(1+\kappa u)$, with linear interpolation in the physical duration $u$. We sample the waiting time by inverting the integrated piecewise-linear total density. At that time, the relative interpolated densities determine the event type and position. Sampling these quantities jointly retains duration--displacement correlations. At the smallest rate, the table uses $\Delta s=2.5\times10^{-5}$, 8192 Fourier modes, and cutoff $\gamma u=40$. We check refinement on 1024 paths from eight fields by halving the step and doubling the modes. The long-scale readout changes by $5.5\times10^{-3}$ in root mean square, with a maximum change of $0.160$. This comparison measures changes in individual trajectories. The following estimates bound the error in the sampled distribution.

For a ground or auxiliary sector, denote its no-jump amplitude by $\Psi$ and its jump rate by $w$, equal to $\gamma$ or $\kappa$. Primes below are duration derivatives, and norms are spatial $\ell^2$ norms. The density interpolation is bounded through
\begin{equation}
\sum_x|\partial_u^2(w|\Psi(x,u)|^2)|
\le2w(\|\Psi'\|^2+\|\Psi\|\|\Psi''\|).
\end{equation}
On an interval of length $\Delta u$, the interpolation error is at most $(\Delta u)^3/12$ times a decreasing upper bound on the preceding expression. For the numerical parameters used here, the two no-jump decay rates are real. Their exponential contributions give positive bounds on $\|\Psi\|^2$, $\|\Psi'\|^2$, and $\|\Psi''\|^2$. We integrate these bounds in momentum using 4096 intervals, taking the largest coefficient and smallest decay rate in each interval. The resulting time envelopes decrease, so their values at the left endpoints bound the interpolation error throughout each time interval. A factor of $1.01$ enlarges these interpolation bounds. Weights obtained by shifting the momentum contour into the complex plane bound spatial truncation and the aliasing caused by a finite Fourier grid. Normalization and cumulative-distribution arithmetic errors are bounded separately. These analytic bounds are evaluated in double precision with safety margins; rounding errors are not enclosed by interval arithmetic.

Let $E$ bound the $L^1$ error in the unnormalized joint density and let $Z$ be its mass. After normalization, the total-variation error obeys
\begin{equation}
e_{\rm TV}\le\frac{E+|Z-1|}{2\min(1,Z)}.
\end{equation}
To propagate this error to time $T$, pair the exact and tabulated waiting times, event types, and positions. The pairs can be chosen so that the probability of any difference in a segment equals $e_{\rm TV}$. While the samples agree, use the same return choices. Summing the mismatch probabilities over segments gives a bound $e_{\rm TV}\E[N(T)+1]$, where
\begin{equation}
\E[N(T)+1]\le\frac T\tau+\frac{\E W^2}{\tau^2},\qquad
\E W^2=-2(\partial_\gamma+\partial_\kappa)\tau.
\end{equation}
The bound $E$ includes interpolation, spatial truncation, the duration cutoff, and Fourier aliasing. Errors from inverting the time and position cumulative distributions are added to $e_{\rm TV}$ after normalization. For the unfinished packet, let $r(a)$ bound the conditional readout error at age $a$, including spatial and arithmetic errors, and let $r_*(a)=\sup_{0\le v\le a}r(v)$. The exact elapsed-time tail then gives the uniform bound
\begin{equation}
\delta_{\rm term}\le r_*(0)+\int_0^\infty\ee^{-\gamma a}\,\dd r_*(a).
\label{eq:terminal-error-average}
\end{equation}
For a piecewise constant envelope, this integral is a sum over its upward increments. The omitted tail beyond $\gamma a=40$ contributes at most $2\ee^{-40}$ for a readout in $[-1,1]$.

Let $\widetilde F_b$ denote the conditional mean obtained from the tabulated dynamics. Before a mismatch, the two paths coincide. After a mismatch, their readouts differ by at most two. Hence, for $f=\tanh$,
\begin{align}
\Delta(T,c)&:=\sup_b|\widetilde F_b(T;c)-F_b(T;c)|\nonumber\\
&\le 2e_{\rm TV}\left(\frac T\tau+\frac{\E W^2}{\tau^2}\right)+\delta_{\rm term}.
\label{eq:terminal-mean-error}
\end{align}
Both conditional means lie in $[-1,1]$. Their second moments differ by at most $2\Delta$, as do the squares of their means over fields. Thus the covariance bias satisfies
\begin{align}
|\Var_b\widetilde F_b-\Var_bF_b|&\le4\Delta(T,c)\nonumber\\
&\le8e_{\rm TV}\left(\frac T\tau+\frac{\E W^2}{\tau^2}\right)+4\delta_{\rm term}.
\label{eq:covariance-error-budget}
\end{align}
Table~\ref{tab:error-components} gives the quantities entering this bound. Substitution at the packet and long observation times gives the upward-rounded bias bounds in Table~\ref{tab:scandata}. Bounds for contrasts are added using the unrounded values before final rounding.

\begin{table}[t]
\caption{Components of the covariance error bound for the parameters of Table~\ref{tab:scandata}. Here $\overline N(T)=T/\tau+\E W^2/\tau^2$, evaluated at $T=1/\gamma$ and $T=T_\gamma$. Entries are rounded upward. The terminal bound applies to both readouts. These are bounds evaluated in floating-point arithmetic with the safety margins described in the text.}
\label{tab:error-components}
\begin{ruledtabular}
\begin{tabular}{ccccc}
Rate & $10^9 e_{\rm TV}$ & $10^9\delta_{\rm term}$ & $\overline N_{\rm p}$ & $\overline N_{\rm l}$\\
$H$ & $3.947$ & $12.81$ & $49.78$ & $326.61$\\
$M$ & $3.957$ & $13.00$ & $100.31$ & $2260.96$\\
$L$ & $0.2747$ & $0.8387$ & $201.47$ & $17023.80$\\
\end{tabular}
\end{ruledtabular}
\end{table}

For the descriptive comparison in Sec.~\ref{sec:data}, the bound on numerical bias $4.8\times10^{-5}$ widens the approximate $90\%$ interval to $[-0.56,0.34]\times10^{-3}$. This interval lies inside the stated tolerance. Table~\ref{tab:design} lists the sample sizes; pilot trajectories are excluded.

\clearpage
\begingroup
\interlinepenalty=10000

\endgroup
\end{document}